\documentclass{article}
\usepackage[utf8]{inputenc}
\usepackage[a4paper, margin=0.8in]{geometry}
\usepackage[style=nature,maxbibnames=99]{biblatex}
\DeclareUnicodeCharacter{2212}{-}
\DeclareUnicodeCharacter{0327}{}

\usepackage{siunitx}
\usepackage{tabularx}
\usepackage{array}
\usepackage{amsmath}
\usepackage{booktabs}
\usepackage{makecell}

\usepackage{graphicx}
\usepackage{caption}
\usepackage{subcaption}
\usepackage{float}
\usepackage{hyperref}
\usepackage[protrusion=true]{microtype}
\usepackage[capitalise,noabbrev]{cleveref}
\usepackage{fancyhdr}
\usepackage{authblk}

\title{Efficient Numerical Schemes for \\ Multidimensional Population Balance Models}
\author[1]{Pavan K. Inguva}
\author[1]{Richard D. Braatz}
\affil[1]{Massachusetts Institute of Technology, 77 Massachusetts Avenue, Cambridge, MA 02139, USA}

\date{}

\begin{document}

\maketitle

\begin{abstract}
\noindent
    Multidimensional population balance models (PBMs) describe chemical and biological processes having a distribution over two or more intrinsic properties (such as size and age, or two independent spatial variables). The incorporation of additional intrinsic variables into a PBM improves its descriptive capability and can be  necessary to capture specific features of interest. As most PBMs of interest cannot be solved analytically, computationally expensive high-order finite difference or finite volume  methods are frequently used to obtain an accurate numerical solution. We propose a finite difference scheme based on operator splitting and solving each sub-problem at the limit of numerical stability that achieves a discretization error that is \textit{zero} for certain classes of PBMs and low enough to be acceptable for other classes. In conjunction to employing specially constructed meshes and variable transformations, the scheme exploits the commutative property of the differential operators present in many classes of PBMs. The scheme has very low computational cost -- potentially as low as just memory reallocation. Multiple case studies  demonstrate the performance of the proposed scheme.
\end{abstract}

\section{Introduction}

Multidimensional population balance models (PBM) are of significant interest due to having the ability to describe population dynamics that vary across multiple intrinsic variables. Examples of such populations include crystals that vary along two independent spatial  dimensions such as length and width \cite{Briesen2006,Ma2007,Zhang2015,Fys2019}; granules that vary in solid, liquid, and gas content \cite{Imm2005,Barr2015} or porosity, binder content, and composition \cite{Iveson2002}; and cell populations that vary in multiple properties such as cell size, age, and intracellular concentrations of species of interest such as enzymes \cite{Mantz2001a, Durr2017, Qued2018}. Another important class of problems where multidimensional PBMs are valuable are systems where, in addition to variations in one or more intrinsic variables, there is spatial variation such as in slug or plug flow in a tube which introduces an additional intrinsic variable in the form of axial position or residence time \cite{Morchain2012,,Rasche2016,Shir2019,Moz2021,Inguva2022}. 

Most PBMs of interest cannot be solved analytically, and so many numerical methods have been developed. Compared to 1D PBMs, multidimensional PBMs are more challenging to solve numerically, as the solution process is much more memory intensive and prone to numerical diffusion and/or dispersion. As such, various numerical schemes have been specifically targeted at multidimensional PBMs based on the finite difference method (FDM) \cite{Mantz2001a,Ma2002}, the finite volume method (FVM) \cite{Gunawan2004,Imm2005,Qamar2007, Pinto2007,Singh2020}, the finite element method (FEM) \cite{Mantz2001b,Ganesan2012}, spectral methods \cite{Mantz2001c}, and Lattice Boltzmann methods \cite{Maju2012}. Often, these methods are computationally costly and/or highly mathematically involved which impacts their adoption and deployment. 

In previous work, we  demonstrated how FDM, when thoughtfully applied using specially constructed meshes and variable transformations, can accurately and efficiently solve 1D PBMs \cite{Inguva2022}. This article describes the extension of those methods to multidimensional PBMs. Many classes of PBMs have differential operators that commute which enables the use of operator splitting techniques with no splitting error \cite{Leveque2002}. These operator splitting techniques transform a multidimensional PBM into a series of 1D sub-problems, each of which can be solved highly accurately and efficiently as previously shown. Although the presented methods and case studies focus on 2D PBMs, the extension to n-dimensional PBMs naturally follows.

\section{Theory and Methods}

This section is structured as a series of cases in which each case outlines the development of the finite difference numerical scheme for a specific class of PBM. Finite difference schemes are benchmarked with the SharpClaw solver \cite{Ket2013}, a high-order weighted essentially non-oscillatory (WENO) solver from the PyClaw package \cite{Ket2012,Mandli2016}. Default solver settings with a maximum CFL number of 1 for all problems were used. For cases where the growth rate in the PBM is variable (in the context of hyperbolic conservation equations, this is also called variable velocity advection), PyClaw requires the equation to be formulated non-conservatively which is done by employing a variable transformation (e.g., see \eqref{eq:size1_model_step1}) or by expanding out the ``spatial" derivatives. 

\subsection{Case 1: PBMs with Constant Growth Rate}

Consider the homogeneous PBM,
\begin{equation}
    \frac{\partial f(t,a_{1},a_{2})}{\partial t} + g_{1}\frac{\partial f(t,a_{1},a_{2})}{\partial a_{1}} + g_{2}\frac{\partial f(t,a_{1},a_{2})}{\partial a_{2}} = 0, \quad f(0,a_{1},a_{2}) = f_{0}(a_{1},a_{2}), 
    \label{eq:con_growth_pbm}
\end{equation}
where $a_{1}$ and $a_{2}$ are the intrinsic variables and $g_{1}$ and $g_{2}$ are the constant positive growth rates. PBMs of this form arise when the driving force for growth is constant which, for example, can occur in crystallization where an external control system maintains a constant supersaturation (see \cite{Inguva2022} and citations therein). The efficient and accurate numerical solution to this class of PBMs using the techniques described in this section is already known (e.g., see \cite{Szy2021,OCW2009,Leveque2002}). The solution of the PBM \eqref{eq:con_growth_pbm} is considered here for pedagogical reasons, as a robust understanding of this section simplifies the presentation of subsequent, more complicated PBMs.

The application of the upwind finite difference scheme to \eqref{eq:con_growth_pbm} gives
\begin{equation}
    \frac{f_{i+1}^{j,k} -f_{i}^{j,k}}{\Delta t} + g_{1}\frac{f_{i}^{j,k} -f_{i}^{j-1,k}}{\Delta a_{1}} + g_{2}\frac{f_{i}^{j,k} -f_{i}^{j,k-1}}{\Delta a_{2}} = 0,
\end{equation}
where $i$ is the time index, $j$ is the index for $a_{1}$, and $k$ is the index for $a_{2}$. The discretized PDE can be specified either in natural variables (such as in the Taylor series expansions) or by employing the index previously used, i.e.,
\begin{align}
    \frac{f_{i+1}^{j,k} -f_{i}^{j,k}}{\Delta t} + g_{1}\frac{f_{i}^{j,k} -f_{i}^{j-1,k}}{\Delta a_{1}} + g_{2}\frac{f_{i}^{j,k} -f_{i}^{j,k-1}}{\Delta a_{2}} \equiv &\, \frac{f(t + \Delta t,a_{1}, a_{2}) - f(t,a_{1}, a_{2})}{\Delta t}  \nonumber \\
    &+ g_{1}\frac{f(t,a_{1}, a_{2}) - f(t,a_{1} - \Delta a_{1}, a_{2})}{\Delta a_{1}} \nonumber \\ 
    &+ g_{2}\frac{f(t,a_{1}, a_{2}) - f(t,a_{1}, a_{2} - \Delta a_{2})}{\Delta a_{2}}. 
\end{align}
Defining $\alpha = \frac{g_{1} \Delta t}{\Delta a_{1}}$ and $\beta = \frac{g_{2} \Delta t}{\Delta a_{2}}$ for compactness, the corresponding upwind scheme is
\begin{equation}
    f_{i+1}^{j,k} = f_{i}^{j,k} - \alpha (f_{i}^{j,k} - f_{i}^{j-1,k}) - \beta (f_{i}^{j,k} - f_{i}^{j,k-1}).
\end{equation}

To characterize the local truncation error of the scheme, consider the Taylor series expansions
\begin{align}
    f(t+\Delta t, a_{1},a_{2}) &= \sum_{n=0}^{\infty} \frac{(\Delta t)^{n}}{n!}\frac{\partial^{n}f}{\partial t^{n}} \bigg|_{t,a_{1},a_{2}}, \nonumber \\
    f(t,a_{1}-\Delta a_{1}, a_{2}) &= \sum_{n=0}^{\infty} (-1)^{n}\frac{(\Delta a_{1})^{n}}{n!} \frac{\partial^{n}f}{\partial a_{1}^{n}} \bigg|_{t,a_{1},a_{2}},\nonumber \\
    f(t,a_{1}, a_{2}-\Delta a_{2}) &= \sum_{n=0}^{\infty} (-1)^{n}\frac{(\Delta a_{2})^{n}}{n!} \frac{\partial^{n}f}{\partial a_{2}^{n}}\bigg|_{t,a_{1},a_{2}}. 
\end{align}
Equation \ref{eq:con_growth_pbm} implies that the higher order derivatives are related by
\begin{align}
    \frac{\partial^{n}f}{\partial t^{n}} &= (-1)^{n} \sum_{p=0}^{n} \frac{n!}{p!(n-p)!} g_{1}^{n-p} g_{2}^{p} \frac{\partial^{n}f}{\partial a_{1}^{n-p} \partial a_{2}^{p}} \nonumber \\ &= (-1)^{n} \!\left( g_{1}^{n}\frac{\partial^{n} f}{\partial a_{1}^{n}} + g_{2}^{n}\frac{\partial^{n} f}{\partial a_{2}^{n}} + \sum_{p=1}^{n-1}  \frac{n!}{p!(n-p)!} g_{1}^{n-p} g_{2}^{p} \frac{\partial^{n}f}{\partial a_{1}^{n-p} \partial a_{2}^{p}} \right)
\end{align}
Correspondingly, the local truncation error is
\begin{align}
    \text{Error} &= \!\bigg[ \frac{\partial f}{\partial t} + g_{1}\frac{\partial f}{\partial a_{1}} + g_{2}\frac{\partial f}{\partial a_{2}} \bigg|_{t,a_{1},a_{2}} \nonumber - \! \left(  \frac{f_{i+1}^{j,k} -f_{i}^{j,k}}{\Delta t} + g_{1}\frac{f_{i}^{j,k} -f_{i}^{j-1,k}}{\Delta a_{1}} + g_{2}\frac{f_{i}^{j,k} -f_{i}^{j,k-1}}{\Delta a_{2}} \right) \nonumber \\
    &= \sum_{n=2}^{\infty}\frac{1}{n!} \bigg[ g_{1}(-1)^{n}(\Delta a_{1})^{n-1} \frac{\partial^{n} f}{\partial a_{1}^{n}} + g_{2}(-1)^{n}(\Delta a_{2})^{n-1} \frac{\partial^{n} f}{\partial a_{2}^{n}} - (\Delta t)^{n-1} \frac{\partial^{n} f}{\partial t^{n}} \bigg|_{t,a_{1},a_{2}} \nonumber \\
    &= \sum_{n=2}^{\infty} \bigg( \frac{(-1)^{n}}{n!} \bigg[ (g_{1}(\Delta a_{1})^{n-1} - g_{1}^{n}(\Delta t)^{n-1})\frac{\partial^{n}f}{\partial a_{1}^{n}} + (g_{2}(\Delta a_{2})^{n-1} - g_{2}^{n}(\Delta t)^{n-1})\frac{\partial^{n}f}{\partial a_{2}^{n}} \nonumber \\ 
    & -(\Delta t)^{n-1} \sum_{p=1}^{n-1}\frac{n!}{p!(n-p)!}g_{1}^{n-p}g_{2}^{p}\frac{\partial^{n}f}{\partial a_{1}^{n-p} \partial a_{2}^{p}} \bigg|_{t,a_{1},a_{2}} \bigg) \nonumber \\
    &=\sum_{n=2}^{\infty} \bigg( \frac{(-1)^{n}}{n!} \bigg[ \frac{1}{g_{1}^{n}}\bigg(\frac{(\Delta a_{1})^{n-1}}{g_{1}^{n-1}} -(\Delta t)^{n-1}\bigg)\frac{\partial^{n}f}{\partial a_{1}^{n}} + \frac{1}{g_{2}^{n}}\bigg(\frac{(\Delta a_{2})^{n-1}}{g_{2}^{n-1}} -(\Delta t)^{n-1}\bigg)\frac{\partial^{n}f}{\partial a_{2}^{n}} \nonumber \\ 
    & -(\Delta t)^{n-1} \sum_{p=1}^{n-1}\frac{n!}{p!(n-p)!}g_{1}^{n-p}g_{2}^{p}\frac{\partial^{n}f}{\partial a_{1}^{n-p} \partial a_{2}^{p}} \bigg|_{t,a_{1},a_{2}} \bigg)  \nonumber \\
    & =\sum_{n=2}^{\infty}  \bigg( \frac{(-1)^{n}(\Delta t)^{n-1}}{n!} \bigg[ \frac{1}{g_{1}^{n}} \left( \frac{1}{\alpha^{n-1}} - 1 \right) \frac{\partial^{n}f}{\partial a_{1}^{n}}  + \frac{1}{g_{2}^{n}} \left( \frac{1}{\beta^{n-1}} - 1 \right) \frac{\partial^{n}f}{\partial a_{2}^{n}} \nonumber \\  &-\sum_{p=1}^{n-1}\frac{n!}{p!(n-p)!}g_{1}^{n-p}g_{2}^{p}\frac{\partial^{n}f}{\partial a_{1}^{n-p} \partial a_{2}^{p}} \bigg|_{t,a_{1},a_{2}}  \bigg)
\end{align}
It can be shown using von Neumann stability analysis that the explicit upwind scheme is conditionally stable for
\begin{equation}
    \alpha, \beta \geq 0,\quad \alpha + \beta \leq 1.
    \label{eq:alpha:beta}
\end{equation}

The above upwind scheme is not able to solve the PBM exactly irrespective of the values of $\alpha$ and $\beta$. The selection of $\alpha$ and $\beta$ to satisfy the second inequality in \eqref{eq:alpha:beta}
 results in excessive numerical diffusion \cite{Leveque2002}.

An alternative approach is to employ operator splitting (also called fractional step methods) \cite{Leveque2002,Hosseini2019} which enables the original PBM \eqref{eq:con_growth_pbm} to be expressed in a manner more amenable to numerical solution.  Furthermore, operator splitting incurs no additional error penalty when the operators commute \cite{Leveque2002}. To illustrate this point, consider the application of first-order order splitting (also known as dimensional splitting in this instance) to \eqref{eq:con_growth_pbm},
\begin{align}
    &\frac{\partial f^{*}}{\partial t} + g_{1}\frac{\partial f^{*}}{\partial a_{1}} = 0 , \quad f^{*}(t,a_{1},a_{2}) = f(t,a_{1},a_{2}), \quad t \in [t,t+\Delta t], \nonumber \\
    &\frac{\partial f^{**}}{\partial t} + g_{2}\frac{\partial f^{**}}{\partial a_{2}} = 0, \quad f^{**}(t,a_{1},a_{2}) = f^{*}(t+\Delta t,a_{1},a_{2}), \quad t \in [t,t+\Delta t], \nonumber \\
    &f(t+\Delta t,a_{1},a_{2}) = f^{**}(t+\Delta t,a_{1},a_{2}). 
    \label{eq:constant_split}
\end{align}
To write the subsequent equations more compactly,  $\mathcal{A}$ and $\mathcal{B}$ are used to represent the operators $g_{1}\frac{\partial}{\partial a_{1}}$ and $g_{2}\frac{\partial}{\partial a_{2}}$ respectively. An expression for the splitting error for first-order splitting can be obtained by considering the Taylor series expansion of $f$ about $t+\Delta t$ \cite{Leveque2002}, 
\begin{align}
    f(t+\Delta t, a_{1}, a_{2}) &= f(t,a_{1},a_{2}) + \sum_{n=1}^{\infty} \frac{(\Delta t)^{n}}{n!} (\mathcal{A} + \mathcal{B})^{n} f(t,a_{1},a_{2}) \nonumber \\
    &= \sum_{n=0}^{\infty} \frac{(\Delta t)^{n}}{n!} (\mathcal{A} + \mathcal{B})^{n} f(t,a_{1},a_{2}) \nonumber \\
    &= e^{\Delta t (\mathcal{A} + \mathcal{B})}f(t,a_{1},a_{2}).
\end{align}
For the dimensional splitting method \eqref{eq:constant_split},
\begin{align}
    f^{**}(t+\Delta t, a_{1},a_{2}) &= e^{\Delta t \mathcal{B}} e^{\Delta t \mathcal{A}} f(t,a_{1},a_{2}) \nonumber \\
    &= \left( \sum_{n=0}^{\infty} \frac{(\Delta t)^{n}}{n!}\mathcal{B}^{n} \right) \left( \sum_{n=0}^{\infty} \frac{(\Delta t)^{n}}{n!}\mathcal{A}^{n} \right) f(t,a_{1},a_{2}) \nonumber \\
    &= \sum_{n=0}^{\infty} \frac{(\Delta t)^{n}}{n!} \left(\sum_{p=0}^{n}\frac{n!}{p!(n-p)!} \mathcal{B}^{p} \mathcal{A}^{n-p} \right) f(t,a_{1},a_{2}).
\end{align}
Therefore the splitting error is
\begin{align}
    \text{Error} &= f(t+\Delta t, a_{1},a_{2}) - f^{**}(t+\Delta t, a_{1},a_{2}) \nonumber \\
    &= \sum_{n=0}^{\infty} \frac{(\Delta t)^{n}}{n!} \left[ (\mathcal{A} + \mathcal{B})^{n} - \left(\sum_{p=0}^{n}\frac{n!}{p!(n-p)!} \mathcal{B}^{p} \mathcal{A}^{n-p} \right) \right]f(t,a_{1},a_{2}) \nonumber \\
    &= \sum_{n=2}^{\infty} \frac{(\Delta t)^{n}}{n!} \left[ (\mathcal{A} + \mathcal{B})^{n} - \left(\sum_{p=0}^{n}\frac{n!}{p!(n-p)!} \mathcal{B}^{p} \mathcal{A}^{n-p} \right) \right]f(t,a_{1},a_{2}) \nonumber \\
    &\approx \frac{(\Delta t)^{2}}{2}(\mathcal{A}\mathcal{B} - \mathcal{B}\mathcal{A})f(t,a_{1},a_{2}) + \mathcal{O}((\Delta t)^{3}).
    \label{eq:binomial}
\end{align}
If the operators commute, i.e., 
\begin{equation}
    \mathcal{A}\mathcal{B} - \mathcal{B}\mathcal{A} = 0,
\end{equation}
then the second-order term is zero. 
The operators do commute for $g_{1}\frac{\partial}{\partial a_{1}}$ and $g_{2}\frac{\partial}{\partial a_{2}}$ for sufficiently smooth $f$,
\begin{equation}
    g_{1}\frac{\partial}{\partial a_{1}} g_{2}\frac{\partial}{\partial a_{2}} = g_{1}g_{2}\frac{\partial^{2}}{\partial a_{1}\partial a_{2}} = g_{2}\frac{\partial}{\partial a_{2}}g_{1}\frac{\partial}{\partial a_{1}}.
\end{equation}
More generally, the binomial formula can be used to show that, for $\mathcal{A}$ and $\mathcal{B}$ that commute, the splitting error \eqref{eq:binomial} is exactly zero \cite{Leveque2002}. Correspondingly, the PBM \eqref{eq:con_growth_pbm} can be solved exactly by solving each 1D PBM sub-problem in \eqref{eq:constant_split} exactly. Each sub-problem can be exactly solved very efficiently using the upwind finite difference scheme when $\text{CFL}=1$ is employed as the scheme simplifies to a form that only requires memory reallocation \cite{Inguva2022,Leveque2002}.   

As we will subsequently demonstrate, many PBMs have operators that commute which enables the use of operator splitting techniques for effective solution. In cases where the operators do not commute, more accurate splitting schemes such as Strang splitting \cite{Strang1968,Speth2013} can be more accurate.

\subsection{Case 2: PBMs with Growth Rate Dependent on Intrinsic Variables Only}

\subsubsection{PBMs with Growth Rate $G_{i} = G_{i}(a_{i})$}

Consider a homogeneous PBM with a growth rate given by $\mathbf{G}(\mathbf{a}) = (G_{1}(a_{1}), G_{2}(a_{2}))$ expressed in conservative form,
\begin{equation}
    \frac{\partial f}{\partial t} + \frac{\partial (G_{1}(a_{1}) f)}{\partial a_{1}} + \frac{\partial (G_{2}(a_{2})f)}{\partial a_{2}}= 0, \quad f(0,a_{1},a_{2}) = f_{0}(a_{1},a_{2}),
    \label{eq:size1_model}
\end{equation}
with $G_{1}(a_{1})$ and $G_{2}(a_{2})$ continuous in $a_{1}$ and $a_{2}$ respectively, bounded, and $G_{1}(a_{1}), G_{2}(a_{2}) > 0, \forall a_{1}, a_{2}$. PBMs of this form  are relevant when the growth rate depends on the intrinsic variable such as size-dependent growth in crystallization and precipitation or age-dependent growth in cell population models (see \cite{Inguva2022} and citations therein). Equation \ref{eq:size1_model} can be transformed by multiplying each term in the equation with $G_{1}(a_{1})G_{2}(a_{2})$ which gives
\begin{equation}
    \frac{\partial (G_{1}(a_{1}) G_{2}(a_{2}) f)}{\partial t} + G_{1}(a_{1}) \frac{\partial (G_{1}(a_{1}) G_{2}(a_{2}) f)}{\partial a_{1}} + G_{2}(a_{2}) \frac{\partial (G_{1}(a_{1}) G_{2}(a_{2}) f)}{\partial a_{2}} = 0.
    \label{eq:size1_model_step1}
\end{equation}
Defining a new variable $\hat{f} = G_{1}(a_{1}) G_{2}(a_{2}) f$ enables \eqref{eq:size1_model_step1} to be written as
\begin{equation}
    \frac{\partial \hat{f}}{\partial t} + G_{1}(a_{1})\frac{\partial \hat{f}}{\partial a_{1}} + G_{2}(a_{2}) \frac{\partial \hat{f}}{\partial a_{2}} = 0.
    \label{eq:size1_model_step2}
\end{equation}
Two variable transformations are introduced,
\begin{equation}
    \tilde{a}_{1} = \int_{0}^{a_{1}} \frac{1}{G_{1}(a_{1})} da_{1}, \quad \tilde{a}_{2} = \int_{0}^{a_{2}} \frac{1}{G_{2}(a_{2})} da_{2},
    \label{eq:size1_var_trans}
\end{equation}
which can be used to transform \eqref{eq:size1_model_step2},
\begin{equation}
    \frac{\partial \hat{f}}{\partial t} + G_{1}(a_{1}) \frac{\partial \hat{f}}{\partial \tilde{a}_{1}} \frac{d\tilde{a}_{1}}{da_{1}}  + G_{2}(a_{2}) \frac{\partial \hat{f}}{\partial \tilde{a}_{2}} \frac{d\tilde{a}_{2}}{da_{2}} = 0,
\end{equation}
\begin{equation}
     \frac{\partial \hat{f}}{\partial t} + \frac{\partial \hat{f}}{\partial \tilde{a}_{1}} + \frac{\partial \hat{f}}{\partial \tilde{a}_{2}} = 0.
     \label{eq:size1_model_step3}
\end{equation}
Under the assumptions, each function $\tilde{a}_{i}(a_{i})$ is invertible \cite{Inguva2022}. Reparametrizing $\hat{f}$ in terms of $\tilde{a}_{i}$,
\begin{equation}
    \tilde{f}(t,\tilde{a}_{1},\tilde{a}_{2}) = \hat{f}(t,a_{1},a_{2}),
\end{equation}
where $\tilde{f}$ denotes that $\hat{f}$ has been reparametrized. This reparametrization enables \eqref{eq:size1_model_step3} to be written in the form
\begin{equation}
    \frac{\partial \tilde{f}}{\partial t} + \frac{\partial \tilde{f}}{\partial \tilde{a}_{1}} + \frac{\partial \tilde{f}}{\partial \tilde{a}_{2}} = 0
\end{equation}
that can be solved exactly as discussed in the previous section. Then all values for $\tilde{f}(t,\tilde{a}_1,\tilde{a}_2)$ are mapped to $\hat{f}(t,a_{1},a_{2})$ by applying the inverses of $\tilde{a}_{i}(a_{i})$.

To demonstrate the effectiveness of the proposed exact method, multiple solution strategies are considered alongside the exact method,
\begin{enumerate}
    \item Apply the upwind scheme to \eqref{eq:size1_model} on a uniform mesh (``Con-Uniform,Upwind")
    \item Apply the WENO scheme to \eqref{eq:size1_model_step1} on a uniform mesh (``Trans-Uniform,WENO")
    \begin{sloppypar}
    \item Apply the upwind scheme to \eqref{eq:size1_model} on a non-uniform mesh to locally enforce CFL = 1 (``Con-Nonuniform,Upwind")
    \end{sloppypar}
    \item Apply the upwind scheme to \eqref{eq:size1_model_step1} on a uniform mesh (``Trans-Uniform,Upwind")
    \begin{sloppypar}
    \item Apply the upwind scheme to \eqref{eq:size1_model_step1} on a non-uniform mesh to locally enforce CFL = 1 (``Trans-Nonuniform,Upwind")
    \end{sloppypar}
    \item Employ the exact method presented in this work (``Exact")
\end{enumerate}
When the uniform meshes are employed in strategies 1 and 4, the value of $\Delta t$ is computed to enforce CFL $\leq 1$ which ensures numerical stability,
\begin{equation}
    \Delta t = \frac{1.0}{\frac{\max G_{1}(a_{1})}{\Delta a_{1}} + \frac{\max G_{2}(a_{2})}{\Delta a_{2}}}.
\end{equation}
To construct the nonuniform mesh for 2D problems for strategies 3 and 5, the mesh is constructed backwards from its end using a specified $\Delta t$,
\begin{equation}
    a_{1}^{j-1} = a_{1}^{j} - \gamma G_{1}(a_{1}^{j})\Delta t, \quad a_{2}^{k-1} = a_{2}^{k} -(1-\gamma)G_{2}(a_{2}^{k})\Delta t, \quad 0 \leq \gamma \leq 1,
\end{equation}
where $\gamma$ is a free parameter which controls the mesh spacing in a particular direction, with $\gamma = 0.5$ used in the case studies. The implementation of the exact method requires computing \eqref{eq:size1_var_trans} and their inverses. These steps can be done offline either analytically or numerically as discussed in previous work \cite{Inguva2022}.

\subsubsection{PBMs with Growth Rate $G_{i} = G_{i}(\mathbf{a})$}

Consider a homogeneous PBM with growth rate $\mathbf{G}(\mathbf{a}) = (G_{1}(a_{1},a_{2}), G_{2}(a_{1},a_{2}))$ expressed in conservative form,
\begin{equation}
    \frac{\partial f}{\partial t} + \frac{\partial (G_{1}(a_{1},a_{2})f)}{\partial a_{1}} + \frac{\partial (G_{2}(a_{1},a_{2})f)}{\partial a_{2}} = 0.
    \label{eq:size2}
\end{equation}
While it is not possible in general to solve this class of PBMs exactly, employing the aforementioned strategies can improve the numerical solution. Applying first-order splitting to \eqref{eq:size2} gives
\begin{align}
    &\frac{\partial f^{*}}{\partial t} + \frac{\partial (G_{1}(a_{1},a_{2})f^{*})}{\partial a_{1}} = 0, \quad f^{*}(t,a_{1}, a_{2}) = f(t,a_{1},a_{2}), \nonumber \\
    &\frac{\partial f^{**}}{\partial t} + \frac{\partial (G_{2}(a_{1},a_{2})f^{**})}{\partial a_{2}} = 0, \quad f^{**}(t,a_{1}, a_{2}) = f^{*}(t+\Delta t,a_{1},a_{2}), \nonumber \\
    &f(t+\Delta t, a_{1},a_{2}) = f^{**}(t+\Delta t, a_{1},a_{2}).
    \label{eq:size2_split}
\end{align}
As the operators do not commute, this splitting incurs an error proportional to $(\Delta t)^{2}$. Although it might be advantageous to use a higher order splitting scheme, they are not considered in the present work as the intention is to demonstrate how comparatively simple techniques can be employed to effectively solve these PBMs. Defining the variable transformation $\hat{f} = G_{i}(a_{1},a_{2})f$ for each sub-problem in \eqref{eq:size2_split} results in
\begin{align}
    &\frac{\partial \hat{f}^{*}}{\partial t} + G_{1}(a_{1},a_{2})\frac{\partial \hat{f}^{*}}{\partial a_{1}} = 0, \quad \hat{f}^{*}(t,a_{1}, a_{2}) = G_{1}(a_{1},a_{2})f(t,a_{1},a_{2}), \nonumber \\
    &\frac{\partial \hat{f}^{**}}{\partial t} + G_{2}(a_{1},a_{2})\frac{\partial \hat{f}^{**}}{\partial a_{2}} = 0, \quad \hat{f}^{**}(t,a_{1}, a_{2}) = \hat{f}^{*}(t+\Delta t,a_{1},a_{2})\frac{G_{2}(a_{1},a_{2})}{G_{1}(a_{1},a_{2})}, \nonumber \\
    &f(t+\Delta t, a_{1},a_{2}) = \frac{\hat{f}^{**}(t+\Delta t, a_{1},a_{2})}{G_{2}(a_{1},a_{2})}.
    \label{eq:size2_split_trans}
\end{align}
Each sub-problem in \eqref{eq:size2_split_trans} can be solved exactly by defining the variable transformation,
\begin{equation}
    \tilde{a}_{i} = \int_{0}^{a_{i}} \frac{1}{G_{i}(a_{1},a_{2})} \partial a_{i},
\end{equation}
which results in
\begin{align}
    &\frac{\partial \tilde{f}^{*}}{\partial t} + \frac{\partial \tilde{f}^{*}}{\partial \tilde{a}_{1}} = 0, \quad \tilde{f}^{*}(t,\tilde{a}_{1}, a_{2}) = G_{1}(a_{1},a_{2})f(t,a_{1},a_{2}), \nonumber \\
    &\frac{\partial \tilde{f}^{**}}{\partial t} + \frac{\partial \tilde{f}^{**}}{\partial \tilde{a}_{2}} = 0, \quad \tilde{f}^{**}(t,a_{1}, \tilde{a}_{2}) = f^{*}(t+\Delta t,a_{1},a_{2})\frac{G_{2}(a_{1},a_{2})}{G_{1}(a_{1},a_{2})}, \nonumber \\
    &f(t+\Delta t, a_{1},a_{2}) = \frac{\hat{f}^{**}(t+\Delta t, a_{1},a_{2})}{G_{2}(a_{1},a_{2})},
    \label{eq:size2_split_exact}
\end{align}
where the $\tilde{f}$ indicates that $f$ has been reparameterized in terms of the transformed variables $\tilde{a}_{i}$. 

Five solution strategies are considered alongside the proposed ``exact" scheme,
\begin{enumerate}
    \item Apply the upwind scheme to \eqref{eq:size2} on a uniform mesh (``Con-Uniform,Upwind)
    \item Apply the WENO scheme to \eqref{eq:size2} on a uniform mesh (``Expanded-Uniform,WENO")
    \item Apply the upwind scheme to \eqref{eq:size2_split} on a uniform mesh (``Split-Con-Uniform,Upwind")
    \item Apply the upwind scheme to \eqref{eq:size2_split_trans} on a uniform mesh (``Split-Trans-Uniform,Upwind")
    \item Apply the upwind scheme to \eqref{eq:size2_split_trans} on a non-uniform mesh to locally enforce CFL = 1 for each sub-problem (``Split-Trans-Nonuniform,Upwind)
    \item Apply the scheme developed in this section (``Split-Exact"). 
\end{enumerate}
For the ``Con-Uniform,Upwind" scheme, the time step $\Delta t$ is evaluated to enforce CFL $\leq 1$ for numerical stability,
\begin{equation}
    \Delta t = \frac{1}{\frac{\max G_{1}(a_{1},a_{2})}{\Delta a_{1}} + \frac{\max G_{2}(a_{1},a_{2})}{\Delta a_{2}}}.
\end{equation}
In comparison, the methods using operator splitting on a uniform mesh permit a larger $\Delta t$,
\begin{equation}
    \Delta t = \min\! \left\{ \frac{\Delta a_{1}}{\max G_{1}(a_{1},a_{2})}, \frac{\Delta a_{2}}{\max G_{2}(a_{1},a_{2})}\right\}.
\end{equation}

\begin{figure}[htbp]
    \centering
    \includegraphics[width=0.55\linewidth]{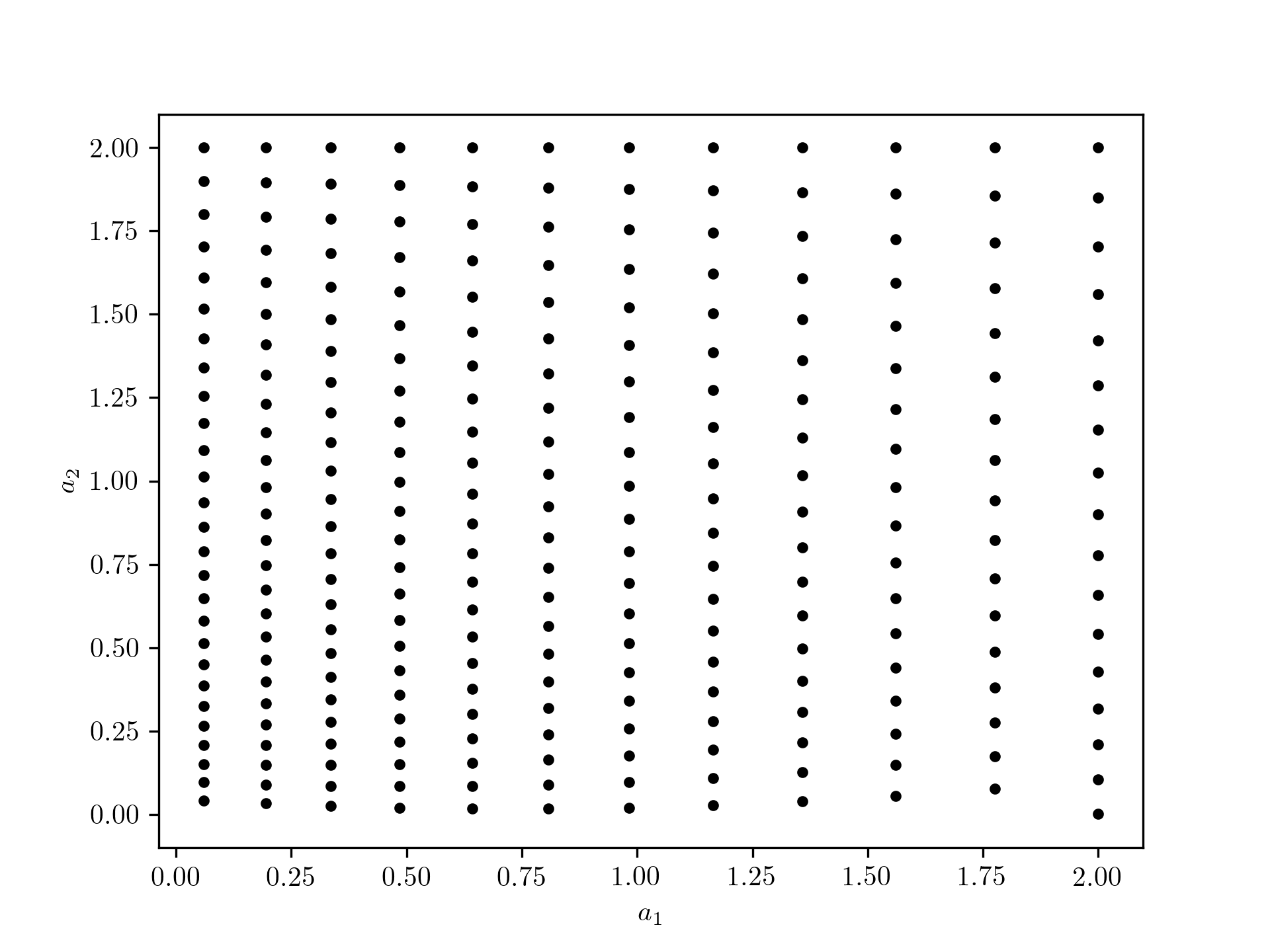}
    \caption{Exemplar jagged mesh generated as part of the ``Split-Trans-Nonuniform,Upwind" scheme.}
    \label{fig:case3_mesh_ex}
\end{figure}

The implementation of the ``Split-Trans-Nonuniform,Upwind" scheme is comparatively involved. One challenge is the mesh construction for each sub-problem in \eqref{eq:size2_split_trans} as each sub-problem solved on its own non-uniform mesh. To outline the mesh construction process for this scheme, suppose that the problem domain is specified to be $a_{1} \in [0,L_{1}]$, $a_{2} \in [0,L_{2}]$, where $L_{1},L_{2}$ are positive constants. For the mesh for the first sub-problem (in $a_{1}$), we first evaluate grid points in the $a_{2}$ direction of the mesh at $a_{1} = L_{1}$ as these points are then used as the basis for stepping backwards to generate the rest of the mesh using the formula,
\begin{equation}
    a_{1}^{j-1,k} = a_{1}^{j,k} -\Delta t \,G_{1}(a_{1}^{j,k}, a_{2}^{k}).
\end{equation}
This task is underspecified and can be implemented in multiple ways such as using a linearly spaced array of $a_{2} \in [0,L_{2}]$ at $a_{1} = L_{1}$. We have elected to use the other growth rate $G_{2}(a_{1},a_{2})$ to construct this array using the formula
\begin{equation}
    a_{2}^{k-1} = a_{2}^{k} - \Delta t \, G_{2}(L_{1}, a_{2}^{k}). 
\end{equation}
The mesh for the second sub-problem (in $a_{2}$) is similarly created. These steps can and should be done offline. The mesh construction process for this scheme involves the formation of jagged arrays (an array of arrays in which member arrays can have different number of elements) and requires due care. An example of such a mesh created as part of the scheme can be seen in Figure~\ref{fig:case3_mesh_ex}. The Awkward Array Python library \cite{Pivarski2020} can be used to handle these jagged arrays.

Both the ``Split-Trans-Nonuniform,Upwind" and ``Split-Exact" schemes require interpolation of the solution from one mesh to the other to solve each sub-problem at each time step. The interpolation step can become computationally costly and potentially constrain accuracy the solution. The SciPy Python Library \cite{Virtanen2020} was used to perform the interpolation and Delaunay computation was performed offline to speed up the interpolation step. Another potential issue is failure of the interpolation step which can result in the population of ``Not a Number" (NaN) values in the solution. This issue is mitigated through the use of a low-order interpolation scheme. 

\subsection{Case 3: PBMs with Time-Dependent Growth Rates}

\subsubsection{PBMs with Growth Rate $G_{i} = G_{i}(t)$}
\label{sec:time}
Consider a homogeneous PBM with $\mathbf{G}(t) = (G_{1}(t),G_{2}(t))$ expressed in conservative form,
\begin{equation}
    \frac{\partial f}{\partial t} + \frac{\partial (G_{1}(t) f)}{\partial a_{1}} + \frac{\partial (G_{2}(t) f)}{\partial a_{2}} = 0, \quad f(0,a_{1},a_{2}) = f_{0}(a_{1},a_{2}).
\end{equation}
Bringing $G_{i}(t)$ out of the ``spatial" derivatives gives
\begin{equation}
    \frac{\partial f}{\partial t} + G_{1}(t)\frac{\partial f}{\partial a_{1}} + G_{2}(t)\frac{\partial f}{\partial a_{2}} = 0.
\end{equation}
The operators $G_{1}(t)\frac{\partial}{\partial a_{1}}$ and $G_{2}(t)\frac{\partial}{\partial a_{2}}$ commute for sufficiently smooth $f$,
\begin{equation}
    G_{1}(t)\frac{\partial}{\partial a_{1}} G_{2}(t)\frac{\partial}{\partial a_{2}} = G_{1}(t)G_{2}(t) \frac{\partial^{2}}{\partial a_{1} \partial a_{2}} = G_{2}(t)\frac{\partial}{\partial a_{2}}G_{1}(t)\frac{\partial}{\partial a_{1}}.
\end{equation}
Therefore, first-order splitting with no splitting error can be applied which gives
\begin{align}
    &\frac{\partial f^{*}}{\partial t}  + G_{1}(t)\frac{\partial f^{*}}{\partial a_{1}} = 0, \quad f^{*}(t,a_{1},a_{2}) = f(t,a_{1},a_{2}), \quad t\in[t,t+\Delta t], \nonumber \\
    &\frac{\partial f^{**}}{\partial t} + G_{2}(t)\frac{\partial f^{**}}{\partial a_{2}} = 0, \quad f^{**}(t,a_{1},a_{2}) = f^{*}(t+\Delta t, a_{1},a_{2}), \quad t \in[t,t+\Delta t], \nonumber \\
    &f(t+\Delta t, a_{1},a_{2}) = f^{**}(t+\Delta t, a_{1},a_{2}).
    \label{eq:time_dep_split}
\end{align}
As shown in previous work \cite{Inguva2022}, each sub-problem can be solved exactly by employing a variable transformation for $t$,
\begin{equation}
    \tilde{t}_{i} = \int_{0}^{t}G_{i}(t') dt'.
\end{equation}
Introducing this variable transformation for $t$ transforms \eqref{eq:time_dep_split} into
\begin{align}
    &\frac{\partial \tilde{f}^{*}}{\partial \tilde{t}_{1}} + \frac{\partial \tilde{f}^{*}}{\partial a_{1}} = 0, \quad \tilde{f}^{*}(\tilde{t}_{1},a_{1},a_{2}) = f(t, a_{1},a_{2}) \nonumber \\
    &\frac{\partial \tilde{f}^{**}}{\partial \tilde{t}_{2}} + \frac{\partial \tilde{f}^{**}}{\partial \tilde{a}_{2}} = 0, \quad \tilde{f}^{**}(\tilde{t}_{2},a_{1},a_{2}) = \tilde{f}^{*}(\tilde{t}_{1} + \Delta \tilde{t}_{1}, a_{1},a_{2}) \nonumber \\
    &f(t+\Delta t, a_{1},a_{2}) = \tilde{f}^{**}(\tilde{t}_{2}+\Delta \tilde{t}_{2}, a_{1},a_{2}),
\end{align}
where $\tilde{f}$ denotes that $f$ has been reparametrized in terms of $\tilde{t}_{i}$. Since each sub-problem can be solved exactly and there is no error arising from operator splitting, this strategy solves the PBM exactly. 

\subsubsection{PBMs with Growth Rate $G_{i} = G_{i,t}(t)G_{i,a_{i}}(a_{i})$}
\label{sec:time_space_sep}

Consider a homogeneous PBM with a separable time- and size-dependent growth rate $\mathbf{G}(t,\mathbf{a}) = (G_{1,t}(t)G_{1,a_{1}}(a_{1}), \allowbreak G_{2,t}(t)G_{2,a_{2}}(a_{2}))$ in conservative form,
\begin{equation}
    \frac{\partial f}{\partial t} + \frac{\partial (G_{1,t}(t)G_{1,a_{1}}(a_{1}) f)}{\partial a_{1}} + \frac{\partial (G_{2,t}(t)G_{2,a_{2}}(a_{2}) f)}{\partial a_{2}} = 0, \quad f(0,a_{1},a_{2}) = f_{0}(a_{1},a_{2}).
    \label{eq:time_space_sep}
\end{equation}
Many physical systems can be described by such a separable growth rate as the growth rate can typically be split into an ``environmental" part which is a function of $t$ and a part that is only a function of the intrinsic variable \cite{Hulburt1964}. This model can be solved exactly by employing the variable transformation and splitting strategies explored in previous cases. Factoring $G_{i,t}(t)$ from the spatial derivatives and multiplying each term in \eqref{eq:time_space_sep} by $G_{1,a_{1}}(a_{1})G_{2,a_{2}}(a_{2})$ transforms \eqref{eq:time_space_sep} into
\begin{align}
    \frac{\partial (G_{1,a_{1}}(a_{1}) G_{2,a_{2}}(a_{2}) f)}{\partial t} &+ G_{1,t}(t)G_{1,a_{1}}(a_{1})\frac{\partial (G_{1,a_{1}}(a_{1}) G_{2,a_{2}}(a_{2}) f)}{\partial a_{1}} \nonumber \\ &+ G_{2,t}(t)G_{2,a_{2}}(a_{2}) \frac{\partial (G_{1,a_{1}}(a_{1}) G_{2,a_{2}}(a_{2}) f)}{\partial a_{2}} = 0.
    \label{eq:time_space_sep_1}
\end{align}
Defining a new variable $\hat{f} = G_{1,a_{1}}(a_{1})G_{2,a_{2}}(a_{2})f$ and introducing the variable transforms for $a_{i}$, i.e., $\tilde{a}_{i} = \int_{0}^{a_{i}} \frac{1}{G_{i,a_{i}}(a_{i})} da_{i}$ transforms \eqref{eq:time_space_sep_1} to
\begin{equation}
    \frac{\partial \hat{f}}{\partial t} + G_{1}(t)\frac{\partial \hat{f}}{\partial \tilde{a}_{1}} + G_{2}(t)\frac{\partial \hat{f}}{\partial \tilde{a}_{2}} = 0.
    \label{eq:time_space_sep_2}
\end{equation}
Reparametrizing $\hat{f}$ in terms of $\tilde{a}_{i}$,
\begin{equation}
    \tilde{f}(t,\tilde{a}_{1},\tilde{a}_{2}) = \hat{f}(t,a_{1},a_{2}),
\end{equation}
where $\tilde{f}$ denotes that $\hat{f}$ has been reparametrized. This reparameterization enables \eqref{eq:time_space_sep_2} to be expressed in a form that can be exactly solved as shown in the previous case,
\begin{equation}
    \frac{\partial \tilde{f}}{\partial t} + G_{1,t}(t)\frac{\partial \tilde{f}}{\partial \tilde{a}_{1}} + G_{2,t}(t)\frac{\partial \tilde{f}}{\partial \tilde{a}_{2}} = 0.
\end{equation}

\subsection{Case 4: Nonhomogeneous PBMs}

\subsubsection{Nonhomogeneous PBMs with Constant Growth Rates}

Consider a PBM with a growth rate $\mathbf{G} = (g_{1},g_{2})$, where $g_{1}$ and $g_{2}$ are positive constants, and a nonhomogeneous term $h(t,a_{1},a_{2})$,
\begin{equation}
    \frac{\partial f}{\partial t} + g_{1}\frac{\partial f}{\partial a_{1}} + g_{2}\frac{\partial f}{\partial a_{2}} = h(t,a_{1},a_{2}), \quad f(0,a_{1},a_{2}) = f_{0}(a_{1},a_{2}). 
    \label{eq:con_nonhomo}
\end{equation}
It is not possible to generally transform \eqref{eq:con_nonhomo} into a form that can be solved exactly. However, significant improvement can be achieved by employing operator splitting and solving each sub-problem while enforcing CFL = 1. Applying first-order splitting gives
\begin{align}
    &\frac{\partial f^{*}}{\partial t} + g_{1}\frac{\partial f^{*}}{\partial a_{1}} = 0, \quad f^{*}(t,a_{1},a_{2}) = f(t,a_{1},a_{2}), \quad t \in [t,t+\Delta t] \nonumber \\
    &\frac{\partial f^{**}}{\partial t} + g_{2}\frac{\partial f^{**}}{\partial a_{2}} = 0, \quad f^{**}(t,a_{1},a_{2}) = f^{*}(t+\Delta t, a_{1},a_{2}), \quad t\in [t,t+\Delta t] \nonumber \\
    &\frac{\partial f^{***}}{\partial t} = h(t,a_{1},a_{2}), \quad f^{***}(t,a_{1},a_{2}) = f^{**}(t+\Delta t, a_{1},a_{2}), \quad t\in [t,t+\Delta t] \nonumber \\
    &f(t+\Delta t,a_{1},a_{2}) = f^{***}(t+\Delta t, a_{1},a_{2}). 
    \label{eq:con_nonhomo_split}
\end{align}
The first two sub-problem can be solved exactly, while the ;ast sub-problem can be solved efficiently using the forward Euler time stepping scheme. Many of the cases previously considered can be transformed into a form comparable to \eqref{eq:con_nonhomo_split}. To demonstrate this, consider a nonhomogeneous PBM with a growth rate $\mathbf{G} = (G_{1,t}(t)G_{1,a_{1}\!}(a_{1}), G_{2,t}(t)G_{2,a_{2}\!}(a_{2}) )$,
\begin{equation}
     \frac{\partial f}{\partial t} + \frac{\partial (G_{1,t}(t)G_{1,a_{1}\!}(a_{1}) f)}{\partial a_{1}} + \frac{\partial (G_{2,t}(t)G_{2,a_{2}\!}(a_{2}) f)}{\partial a_{2}} = b(t,a_{1},a_{2}), \quad f(0,a_{1},a_{2}) = f_{0}(a_{1},a_{2}).
     \label{eq:time_space_sep_nonhomo}
\end{equation}
Employing the steps developed in \cref{sec:time_space_sep} transforms \eqref{eq:time_space_sep_nonhomo} into
\begin{equation}
    \frac{\partial \tilde{f}}{\partial t} + G_{1,t}(t) \frac{\partial \tilde{f}}{\partial \tilde{a}_{1}} + G_{2,t}(t) \frac{\partial \tilde{f}}{\partial \tilde{a}_{2}} = \tilde{G}_{1,\tilde{a}_{1}\!}(\tilde{a}_{1})\tilde{G}_{2,\tilde{a}_{2}\!}(\tilde{a}_{2})\tilde{b}(t,\tilde{a}_{1},\tilde{a}_{2}),
    \label{eq:time_space_nonhomo_2}
\end{equation}
where the superscript $\sim$ over the various functions denotes the reparameterization in terms of the transformed variables $\tilde{a}_{i}$. Recognizing that  $\tilde{h}(t,\tilde{a}_{1},\tilde{a}_{2}) \equiv\tilde{G}_{1,\tilde{a}_{1}\!}(\tilde{a}_{1})\tilde{G}_{2,\tilde{a}_{2}\!}(\tilde{a}_{2})\tilde{b}(t,\tilde{a}_{1},\tilde{a}_{2})$, and applying first-order splitting with a variable transformation for $t$ to \eqref{eq:time_space_nonhomo_2} as in \cref{sec:time} yields
\begin{align}
    &\frac{\partial \tilde{f}^{*}}{\partial \tilde{t}_{1}} + \frac{\partial \tilde{f}^{*}}{\partial \tilde{a}_{1}} = 0, \quad \tilde{f}^{*}(\tilde{t}_{1},\tilde{a}_{1},\tilde{a}_{2}) = \tilde{f}(\tilde{t},\tilde{a}_{1},\tilde{a}_{2}), \quad \tilde{t}_{1} \in [\tilde{t}_{1},\tilde{t}_{1}+\Delta \tilde{t}_{1}] \nonumber \\
    &\frac{\partial \tilde{f}^{**}}{\partial \tilde{t}_{2}} + \frac{\partial \tilde{f}^{**}}{\partial \tilde{a}_{2}} = \tilde{h}(t,\tilde{a}_{1},\tilde{a}_{2}), \quad f^{**}(\tilde{t}_{2},\tilde{a}_{1},\tilde{a}_{2}) = f^{*}(\tilde{t}_{1}+\Delta \tilde{t}_{1}, \tilde{a}_{1},\tilde{a}_{2}),\quad \tilde{t}_{2}\in [\tilde{t}_{2},\tilde{t}_{2}+\Delta \tilde{t}_{2} ] \nonumber \\
    &f(t+\Delta t,a_{1},a_{2}) = f^{**}(t+\Delta t, a_{1},a_{2}).
\end{align}

\subsection{PBMs with a Linear Nonhomogeneous Term}

Consider a PBM with a constant growth rate $\mathbf{G} = (g_{1},g_{2})$ and a linear nonhomogeneous term,
\begin{equation}
    \frac{\partial f}{\partial t} + g_{1}\frac{\partial f}{\partial a_{1}} + g_{2}\frac{\partial f}{\partial a_{2}} = -\lambda(t,a_{1},a_{2})f, \quad f(0,a_{1},a_{2}) = f_{0}(a_{1},a_{2}).
    \label{eq:linear_nonhomo}
\end{equation}
By employing a variable transform $\hat{f} = \mu f$, where the functional form of $\mu$ is to be determined, \eqref{eq:linear_nonhomo} can be transformed into a form that can be solved exactly,
\begin{equation}
    \frac{\partial \hat{f}}{\partial t} + g_{1}\frac{\partial \hat{f}}{\partial a_{1}} + g_{2}\frac{\partial \hat{f}}{\partial a_{2}} = 0.
    \label{eq:linear_nonhomo_transform}
\end{equation}
This equation can be expanded to yield a PDE for $\mu$,
\begin{equation}
    \frac{\partial \mu}{\partial t} + g_{1}\frac{\partial \mu}{\partial a_{1}} + g_{2}\frac{\partial \mu}{\partial a_{2}} = \lambda \mu.
    \label{eq:mu_pde}
\end{equation}
An expression for $\mu$ can be derived from the solution of \eqref{eq:mu_pde}.  

\begin{table}[h!]
\centering
\caption{Summary of functional forms of $\mu$ for different forms of $\lambda$}\renewcommand{\arraystretch}{1.4}
    \begin{tabular}{ccc}
    \toprule
     Form of $\lambda$& Form of $\mu$ & Functional form of $\mu$ \\
     \hline 
     Constant & $\mu(a_{1}) \lor \mu(a_{2}) \lor \mu(t)$ & $e^{\frac{\lambda a_{1}}{g_{1}}} \lor e^{\frac{\lambda a_{2}}{g_{2}}} \lor e^{\lambda t}$  \\
     $\lambda(a_{1})$& $\mu(a_{1})$ &  $\exp\! \left(\frac{1}{g_{1}}\int_{0}^{a_{1}}\lambda(a_{1}') da_{1}' \right)$ \\
     $\lambda(a_{2})$& $\mu(a_{2})$ &  $\exp\! \left(\frac{1}{g_{2}}\int_{0}^{a_{2}}\lambda(a_{2}') da_{2}' \right)$ \\
     $\lambda(t)$& $\mu(t)$ &  $\exp\!{\big(\!\int_{0}^{t}\lambda(t') dt' \big)}$ \\
     $\lambda(t,a_{1})$& $\mu(t,a_{1})$ & $\exp\!{\left(\!\int_{0}^{t}\lambda(t', g_{1}t'+a_{1,0}) dt' \right)}$, $a_{1,0} = a_{1}-g_{1}t$ \\
     $\lambda(t,a_{2})$& $\mu(t,a_{2})$ & $\exp\!{\left(\!\int_{0}^{t}\lambda(t', g_{2}t'+a_{2,0}) dt' \right)}$, $a_{2,0} = a_{2}-g_{2}t$ \\
     $\lambda(a_{2},a_{2})$& $\mu(a_{1},a_{2})$ & $\exp\!{\left(\!\int_{0}^{a_{1}}\frac{1}{g_{1}}\lambda(a_{1}', \frac{g_{2}}{g_{1}}a_{1}'+a_{2,0}) da_{1}' \right)}$, $a_{2,0} = a_{2}-\frac{g_{2}}{g_{1}}a_{1}$ \\
     $\lambda(t,a_{1},a_{2})$ & $\mu(t,a_{1},a_{2})$ & $\exp\!{\left(\!\int_{0}^{t}\lambda(t', g_{1}t'+a_{1,0},g_{2}t + a_{2,0}) dt' \right)}$, $a_{1,0} = a_{1}-g_{1}t$, $a_{2,0} = a_{2} - g_{2}t$ \\
     \bottomrule
    \end{tabular}
    \renewcommand{\arraystretch}{1}
    \label{tbl:muform}
\end{table}

\section{Results and Discussion}
\label{sec:results}

This section is structured as a series of cases which correspond to the various classes of PBMs explored previously. The error of the various numerical schemes is compared via the Root Mean Square Error (RMSE) and Maximum Absolute Error (MAE),
\begin{equation}
    \text{RMSE} = \sqrt{\frac{\sum_{i=1}^{n} (y_{i} - y_{\text{analytical}, i})^{2}}{n}},
\end{equation}
\begin{equation}
    \text{MAE} = \max_{i} |y_{i} - y_{\text{analytical}, i}|.
\end{equation}

\subsection{Case 1: PBMs with Constant Growth Rates}

\begin{figure}[htbp]
    \centering
    \begin{subfigure}{0.45\textwidth}
        \centering
        \includegraphics[width=0.9\textwidth]{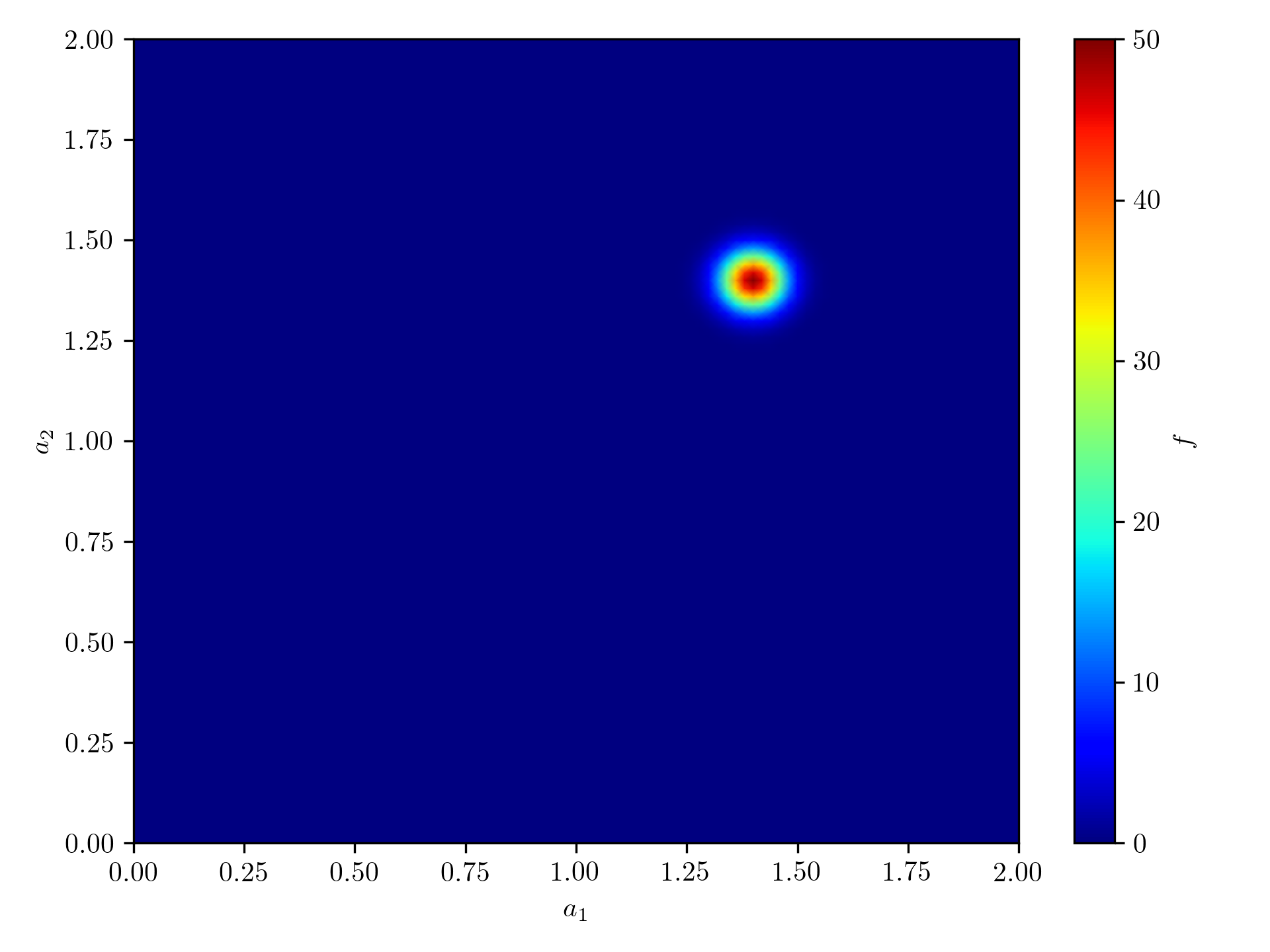} \vspace{-0.2cm}
        \caption{Analytical Solution}
    \end{subfigure}
    \hfill
    \begin{subfigure}{0.45\textwidth}
        \centering
        \includegraphics[width=0.9\textwidth]{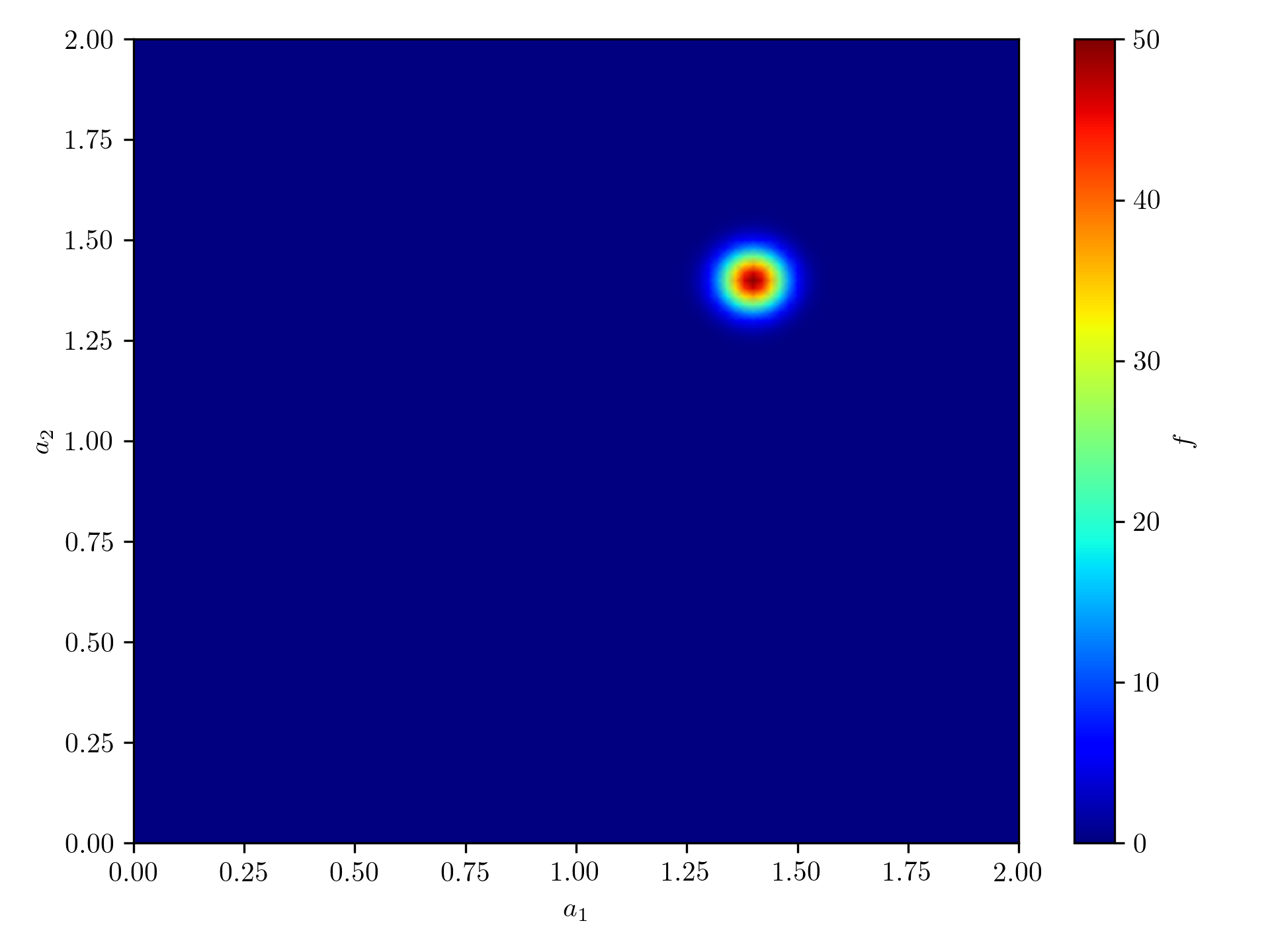}
    \vspace{-0.2cm} \caption{Exact Scheme}
    \end{subfigure}
    \hfill
    \begin{subfigure}{0.45\textwidth}
        \centering
        \includegraphics[width=0.9\textwidth]{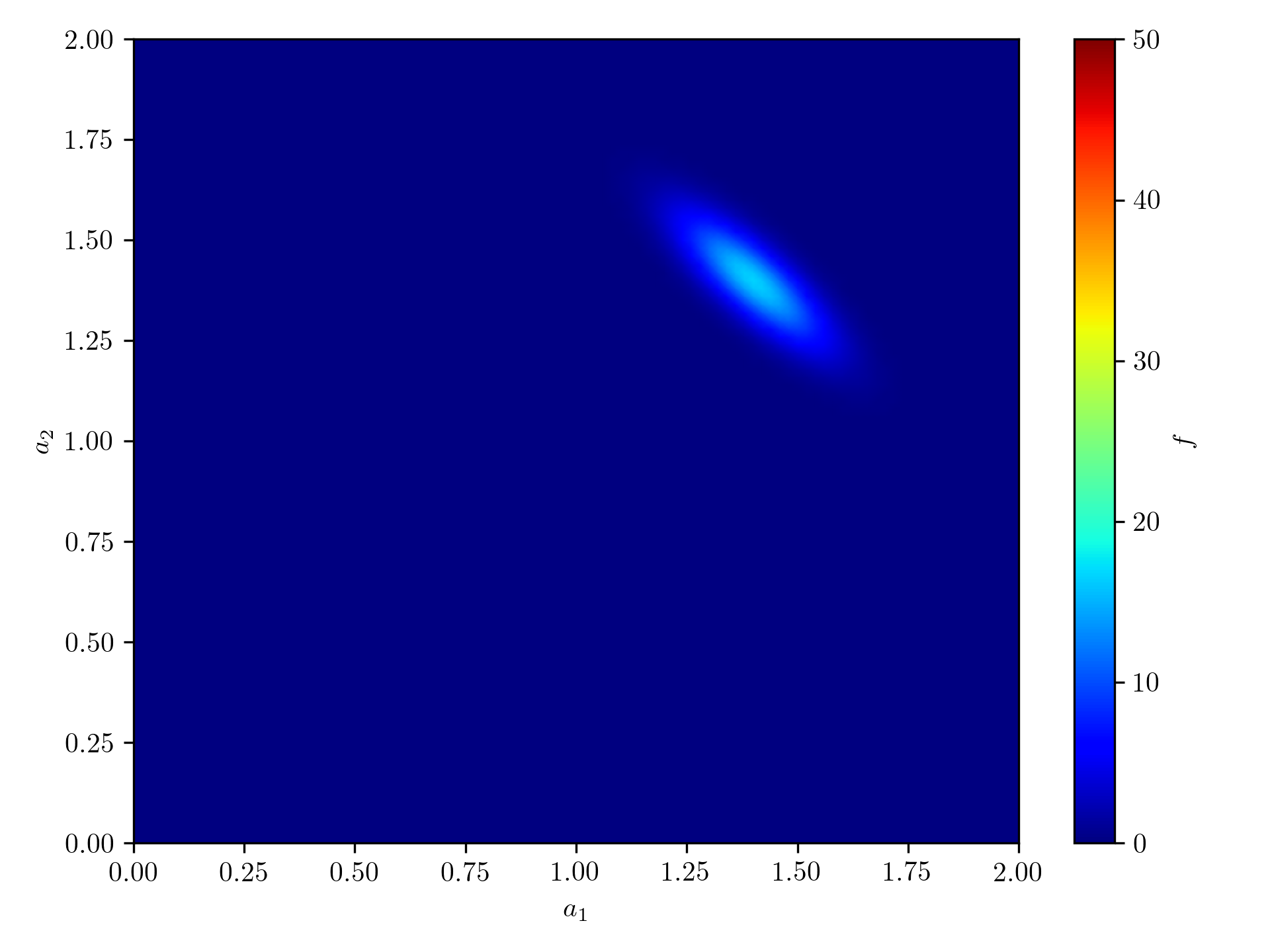}
    \vspace{-0.2cm} \caption{Upwind Scheme}
    \end{subfigure}
    \hfill
    \begin{subfigure}{0.45\textwidth}
        \centering
        \includegraphics[width=0.9\textwidth]{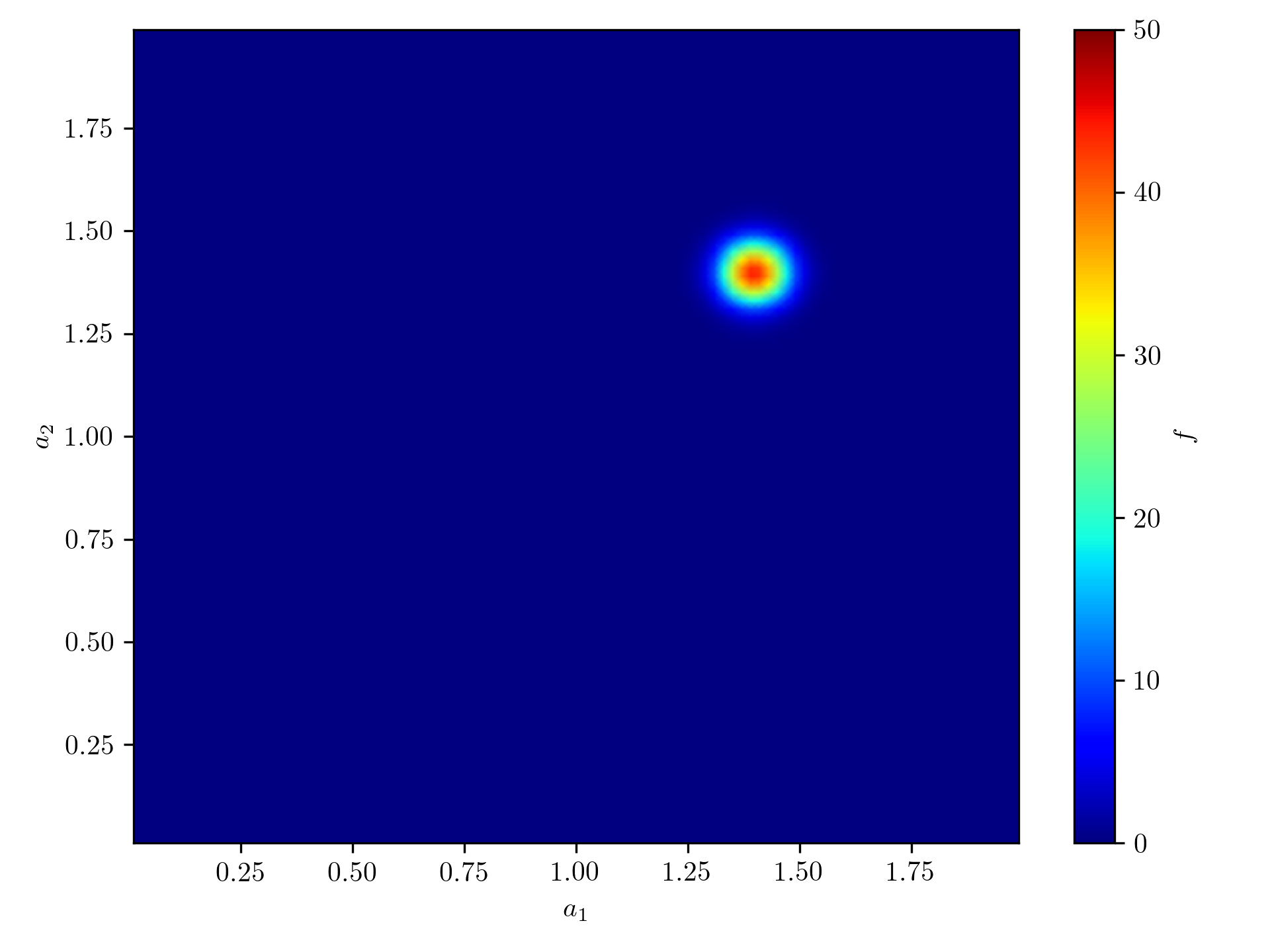}
     \vspace{-0.2cm}\caption{WENO Scheme}
    \end{subfigure}
    \caption{Simulation results for Case 1 at $t=1.0$ using the various schemes. 101 grid points are used in both the $a_{1}$ and $a_{2}$ directions for the Upwind and Exact schemes while 100 cells in both the $a_{1}$ and $a_{2}$ directions for the WENO scheme.}
    \label{fig:case1simulation}
\end{figure}

\begin{figure}[htbp]
    \centering
    \begin{subfigure}{0.45\textwidth}
        \centering
        \includegraphics[width=\textwidth]{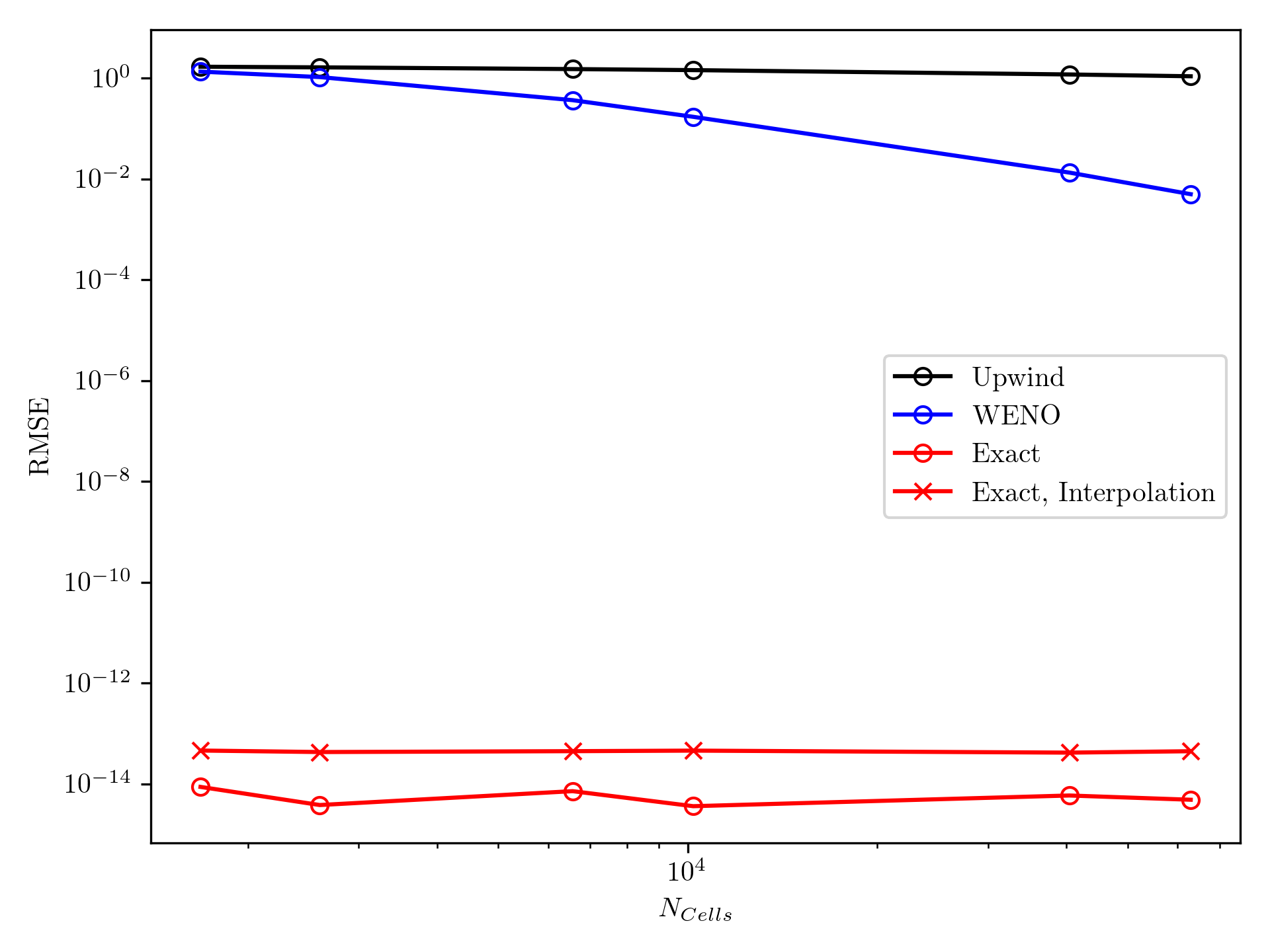}
        \caption{RMSE}
    \end{subfigure}
    \hfill
    \begin{subfigure}{0.45\textwidth}
        \centering
        \includegraphics[width=\textwidth]{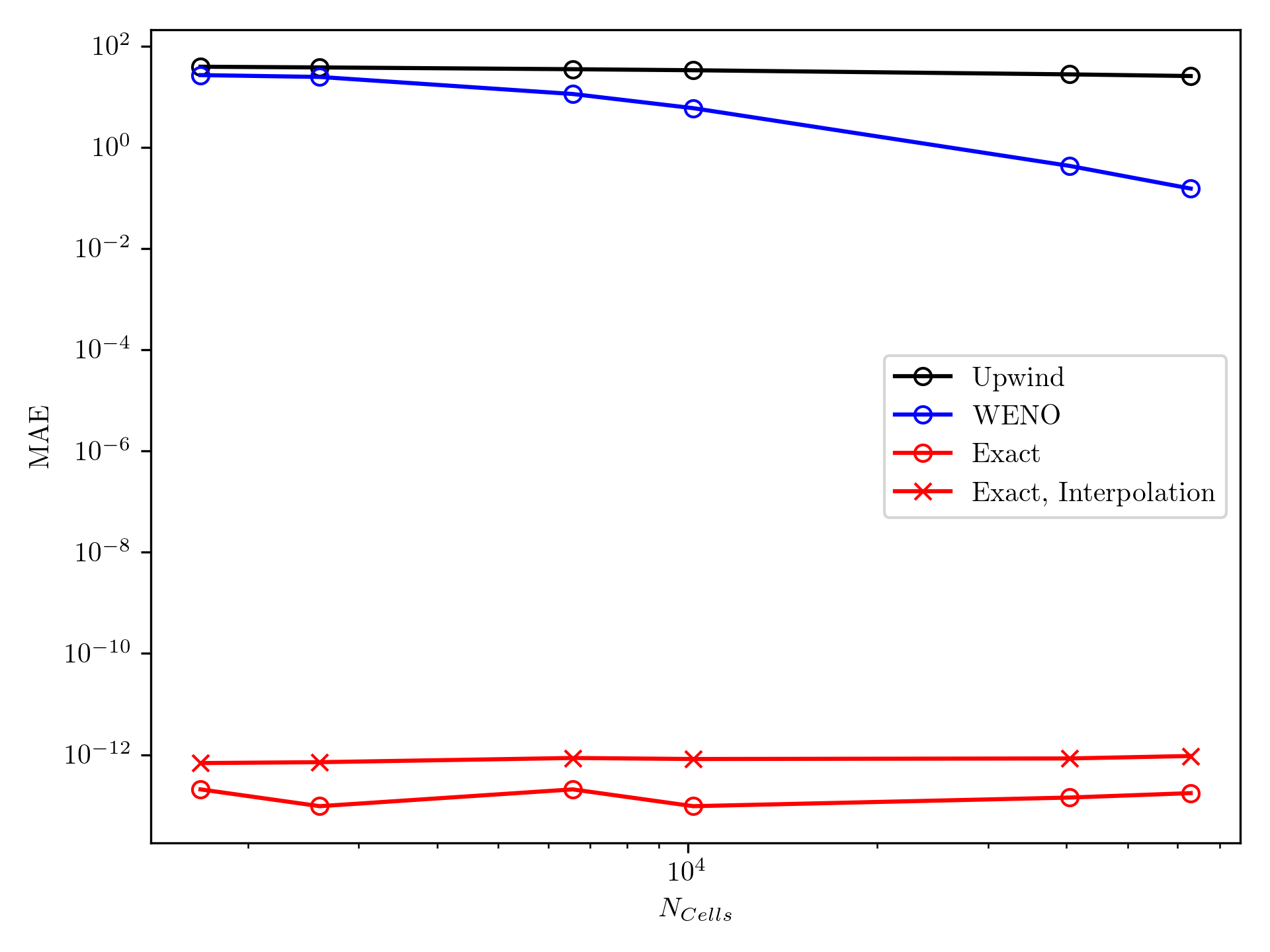}
     \caption{MAE}
    \end{subfigure}
    \caption{Error analysis for Case 1. The use of dimensional splitting and employing $\text{CFL} = 1$ for each subproblem solves the PBM to machine precision as expected.}
    \label{fig:case1error}
\end{figure}

Consider the PBM,
\begin{equation}
    \frac{\partial f}{\partial t} + \frac{\partial f}{\partial a_{1}} + \frac{\partial f}{\partial a_{2}} = 0, \quad f_{0}(a_{1},a_{2}) = 50\exp\!\left( -\frac{(a_{1}-0.4)^{2}}{0.005} - \frac{(a_{2}-0.4)^{2}}{0.005}  \right), \quad a_{i} \in [0,2].
\end{equation}
A no-flux boundary condition is applied at the top and right ends of the domain (i.e.,  $a_{1} = a_{2} = 2$), and a modified Dirichlet boundary which enforces the value of $f$ at the ghost nodes are zero \cite{Gunawan2004} is applied on the left and bottom ends of the domain, i.e., $a_{1} = a_{2} = 0$. The PBM has the analytical solution,
\begin{equation}
    f(t,a_{1},a_{2}) = f_{0}(a_{1}-t,a_{2}-t) = 50\exp\!\left( -\frac{(a_{1}-0.4 -t)^{2}}{0.005} - \frac{(a_{2}-0.4-t)^{2}}{0.005}  \right).
\end{equation}
Exemplar simulation results can be found in \cref{fig:case1simulation} and the convergence analysis can be found in \cref{fig:case1error}. An additional numerical scheme ``Exact,Interpolation", which adapts the ``Exact" scheme by replacing the memory reallocation step with an interpolation function call to compute the updated values at $t + \Delta t$, was also implemented to demonstrate that the use of an interpolation step to compute the values of $f$ at $t + \Delta t$ does not adversely impact error performance. 

\subsection{Case 2: PBMs with Growth Rate $G_{i} = G_{i}(a_{i})$}

\begin{figure}[htbp]
    \centering
    \begin{subfigure}{0.45\textwidth}
        \centering
        \includegraphics[width=0.9\textwidth]{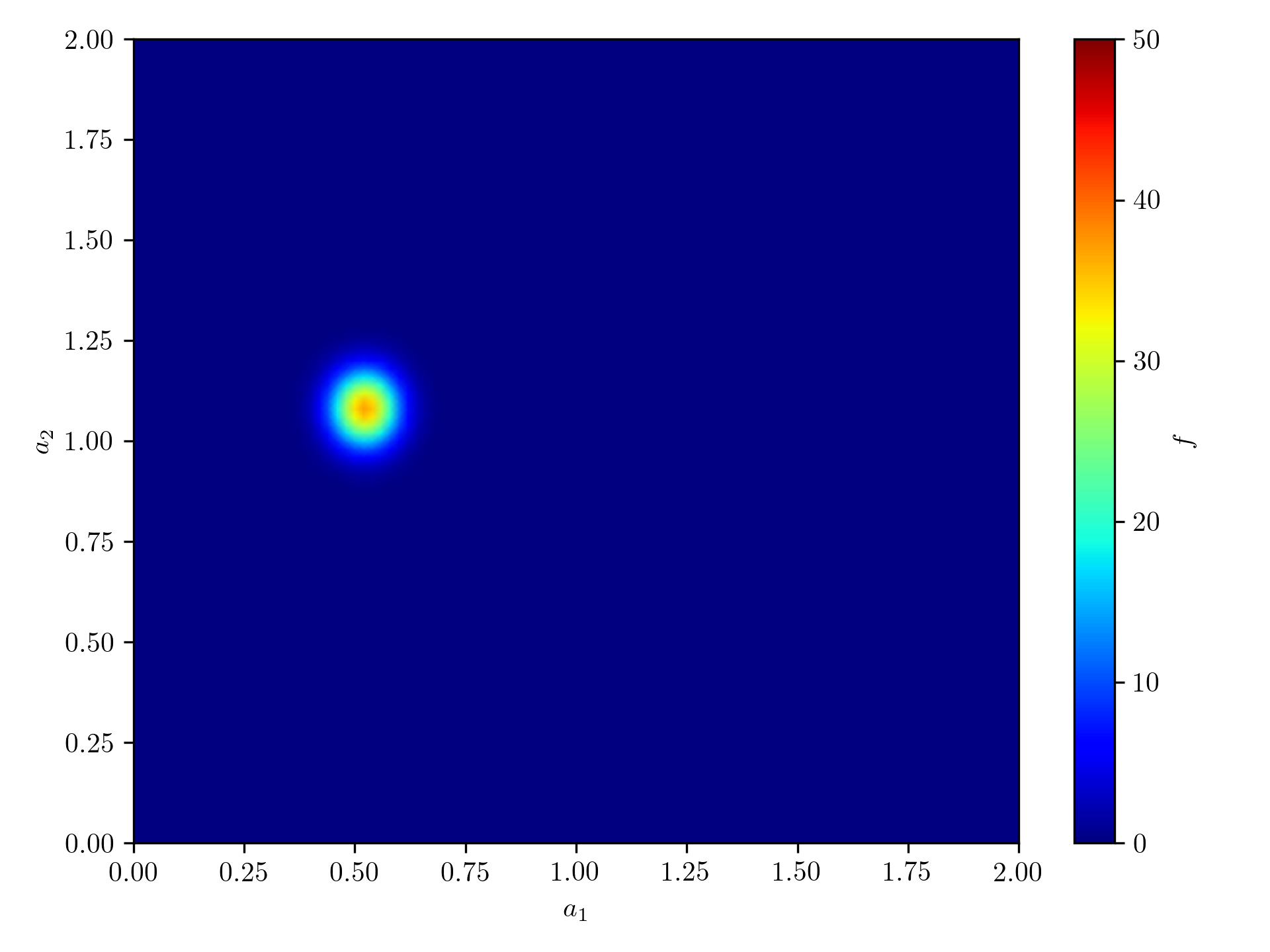}
       \vspace{-0.2cm} \caption{Analytical Solution}
    \end{subfigure}
    \hfill
    \begin{subfigure}{0.45\textwidth}
        \centering
        \includegraphics[width=0.9\textwidth]{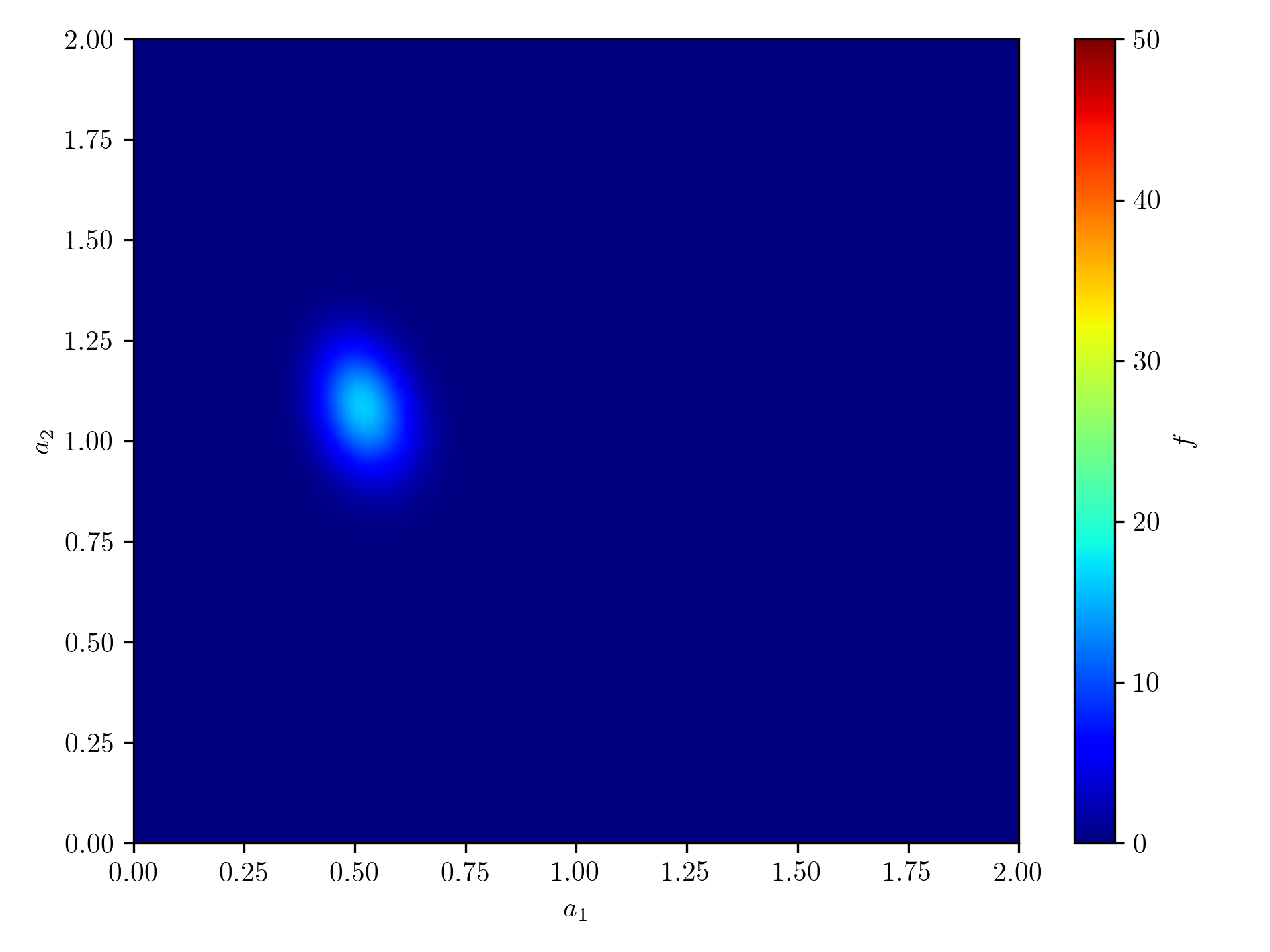}
    \vspace{-0.2cm} \caption{Con-Uniform,Upwind}
    \end{subfigure}
    \hfill
    \begin{subfigure}{0.45\textwidth}
        \centering
        \includegraphics[width=0.9\textwidth]{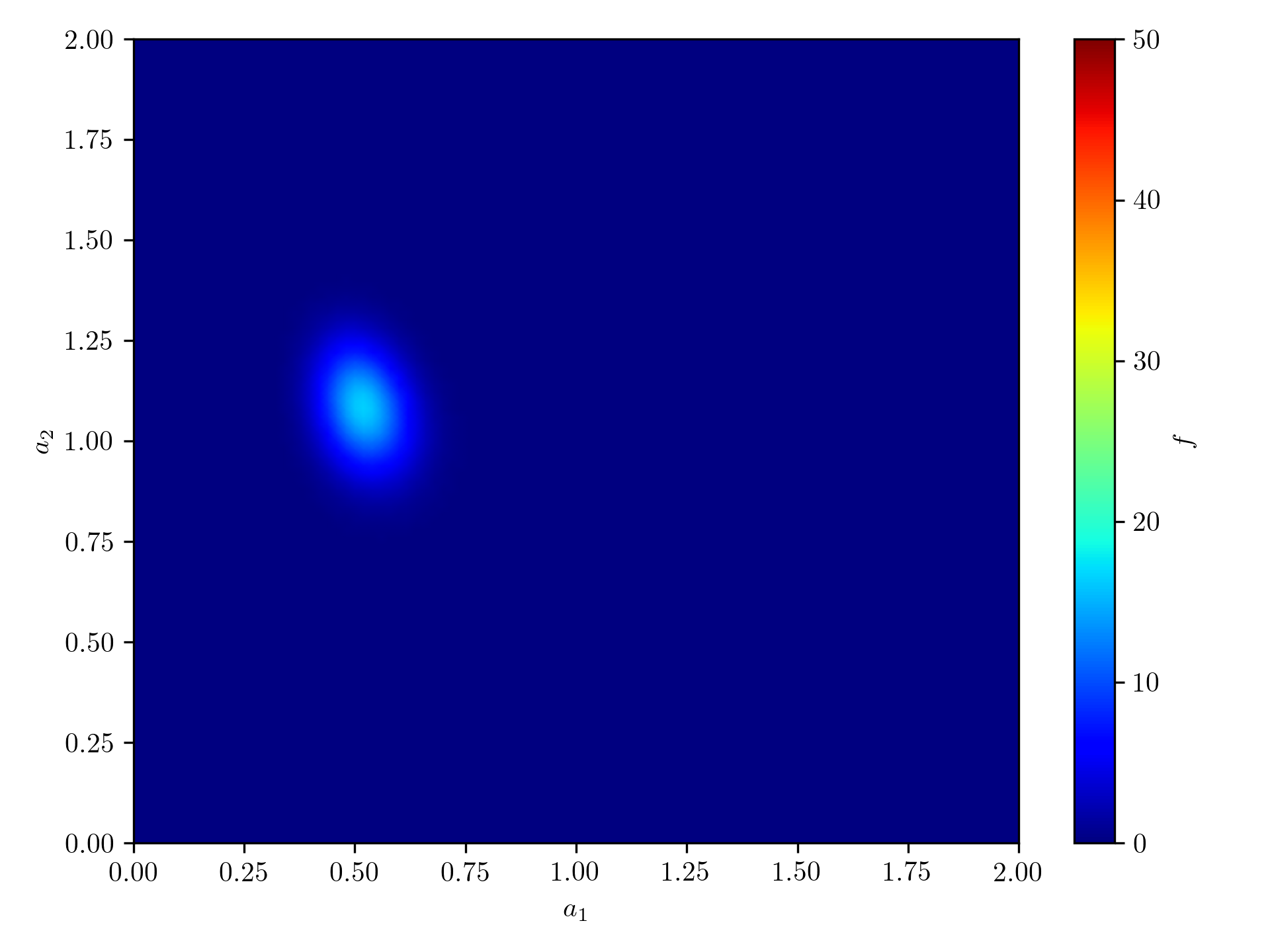}
    \vspace{-0.2cm} \caption{Trans-Uniform,Upwind}
    \end{subfigure}
    \hfill
    \begin{subfigure}{0.45\textwidth}
        \centering
        \includegraphics[width=0.9\textwidth]{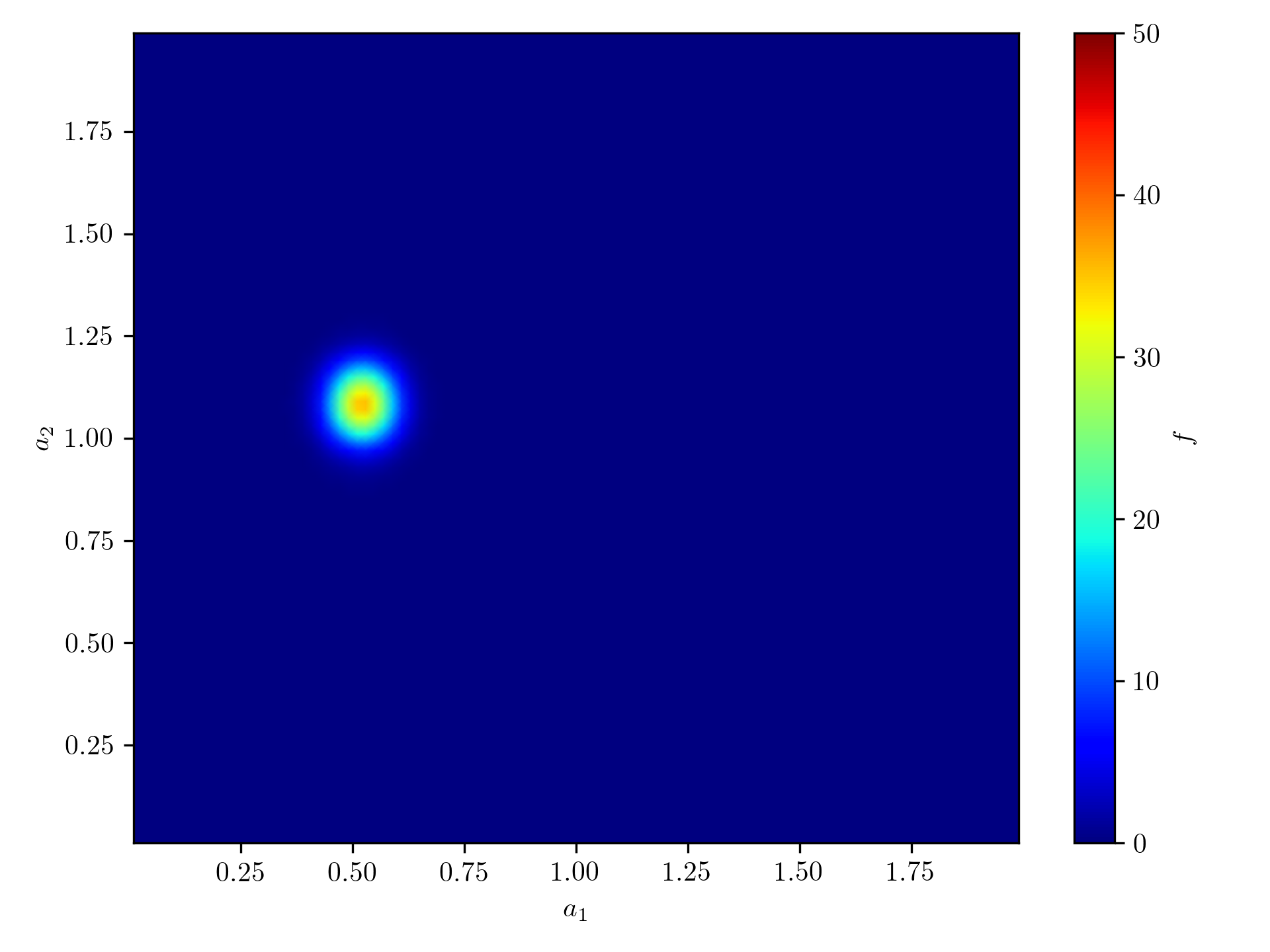}
    \vspace{-0.2cm} \caption{Trans-Uniform,WENO}
    \end{subfigure}
    \hfill
    \begin{subfigure}{0.45\textwidth}
        \centering
        \includegraphics[width=0.9\textwidth]{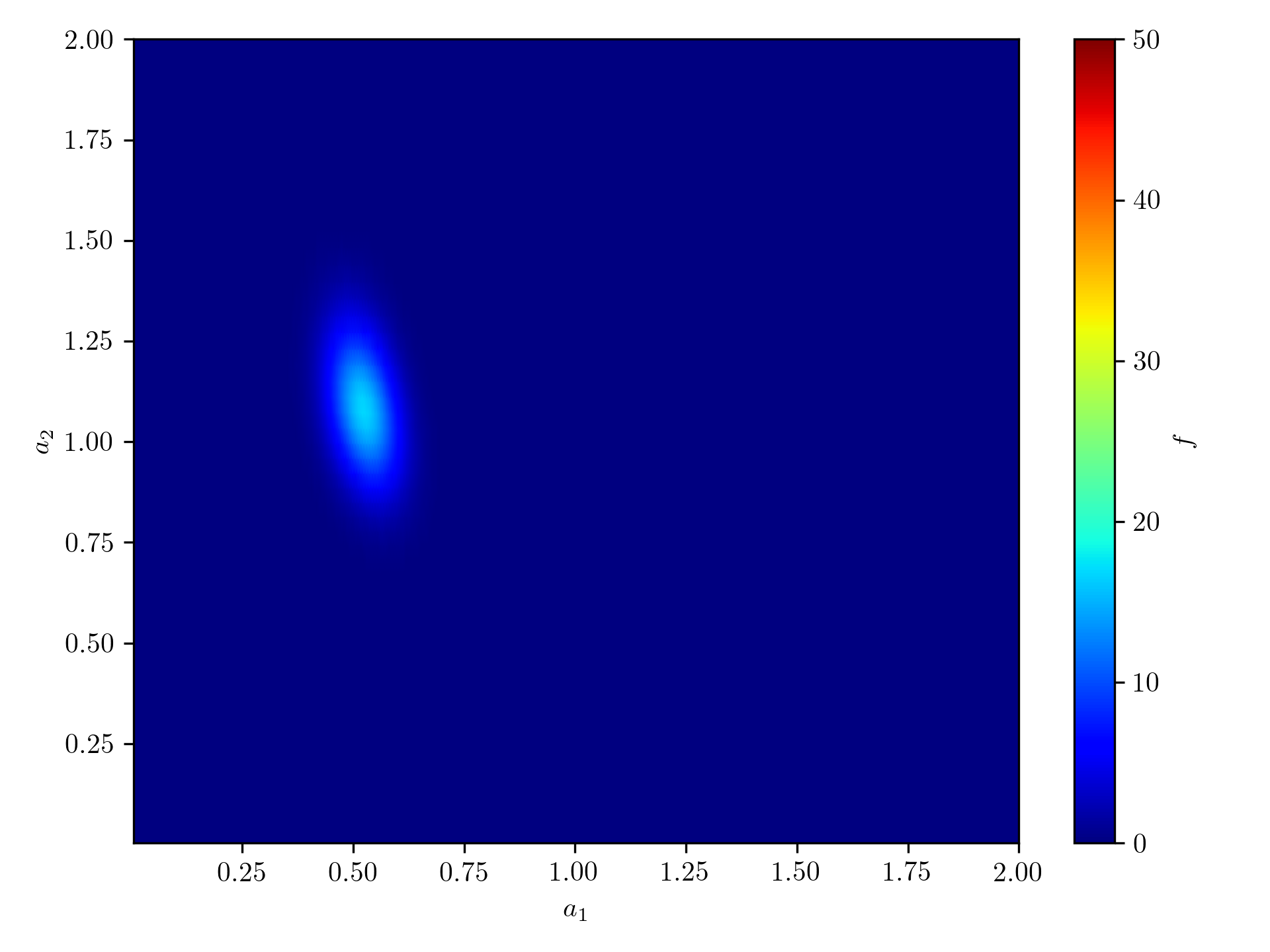}
    \vspace{-0.2cm} \caption{Con-Nonuniform,Upwind}
    \end{subfigure}
    \hfill
    \begin{subfigure}{0.45\textwidth}
        \centering
        \includegraphics[width=0.9\textwidth]{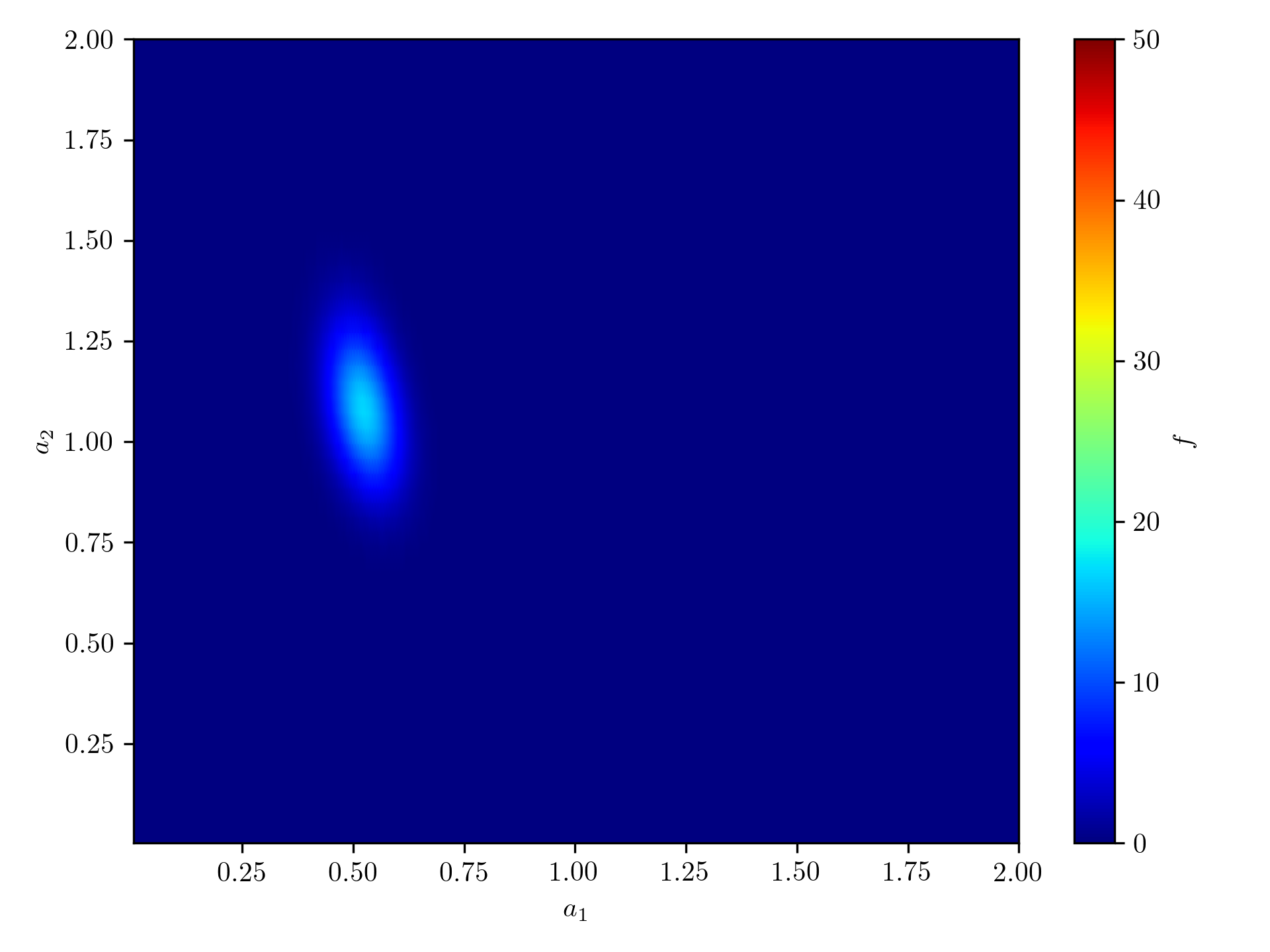}
     \caption{Trans-Nonuniform,Upwind}
    \end{subfigure}
    \hfill
    \begin{subfigure}{0.45\textwidth}
        \centering
        \includegraphics[width=0.9\textwidth]{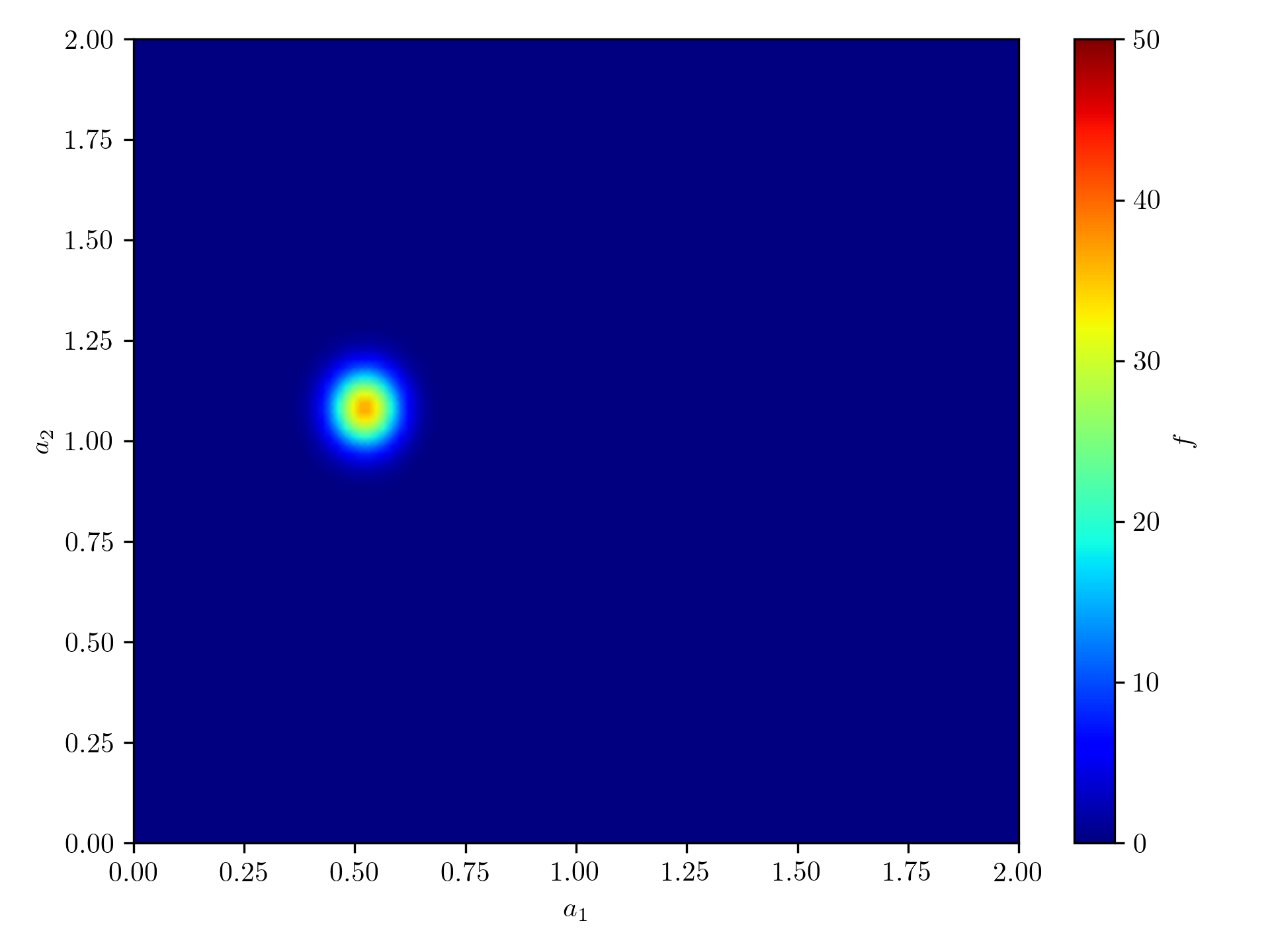}
    \vspace{-0.2cm} \caption{Exact(Analytical)}
    \end{subfigure}
    \hfill
    \begin{subfigure}{0.45\textwidth}
        \centering
        \includegraphics[width=0.9\textwidth]{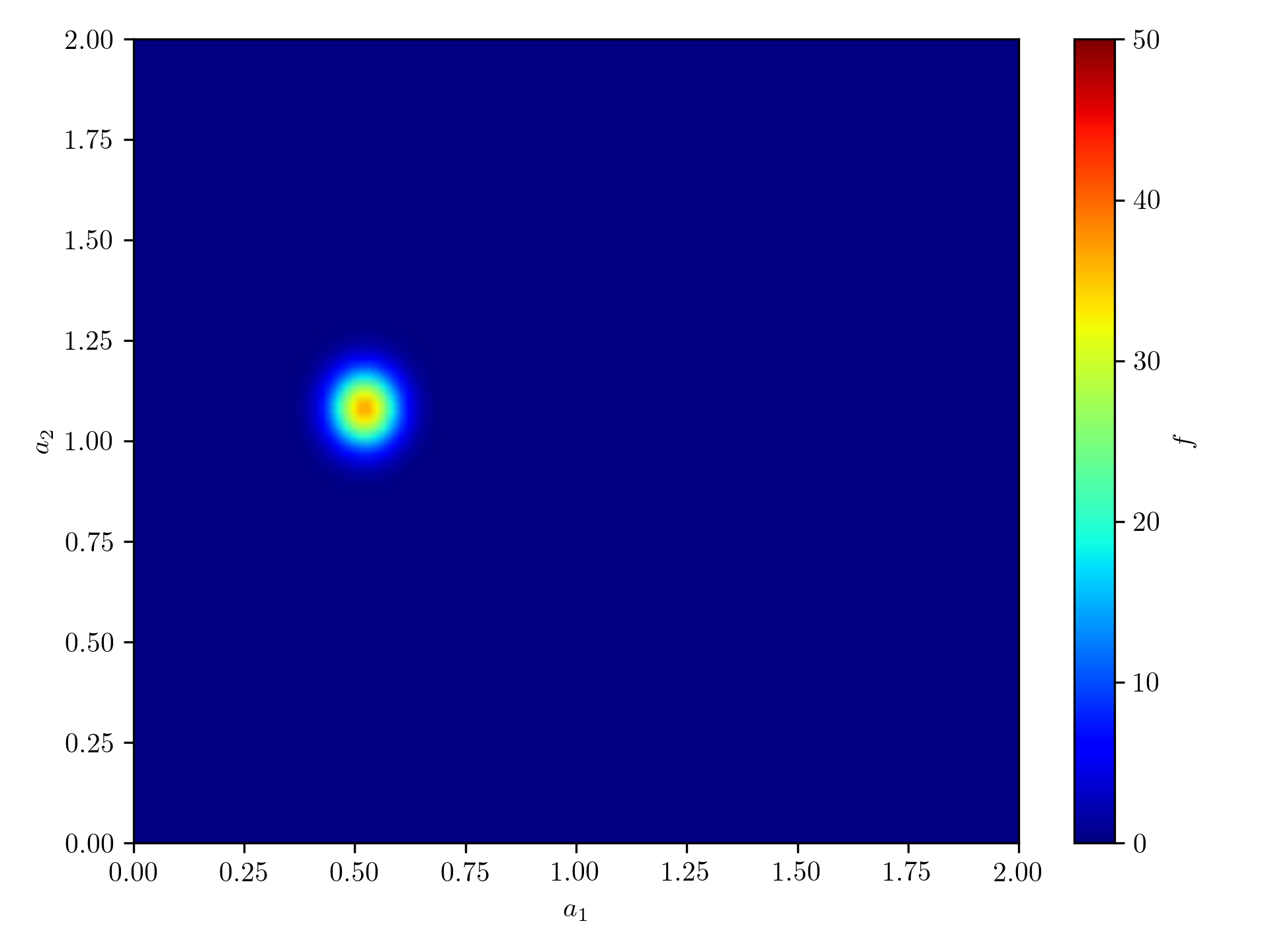}
     \vspace{-0.2cm}\caption{Exact(Numerical)}
    \end{subfigure}
    \caption{Simulation results for Case 2 at $t=1.0$ using the various schemes. 101 grid points are used in both the $a_{1}$ and $a_{2}$ directions for the Upwind schemes on a uniform grid and Exact schemes while 100 cells in both the $a_{1}$ and $a_{2}$ directions for the WENO scheme. The simulations on a nonuniform grid have 277 and 56 grid points in the $a_{1}$ and $a_{2}$ directions respectively.}
    \label{fig:case2simulation}
\end{figure}

\begin{figure}[htbp]
    \centering
    \begin{subfigure}{0.45\textwidth}
        \centering
        \includegraphics[width=\textwidth]{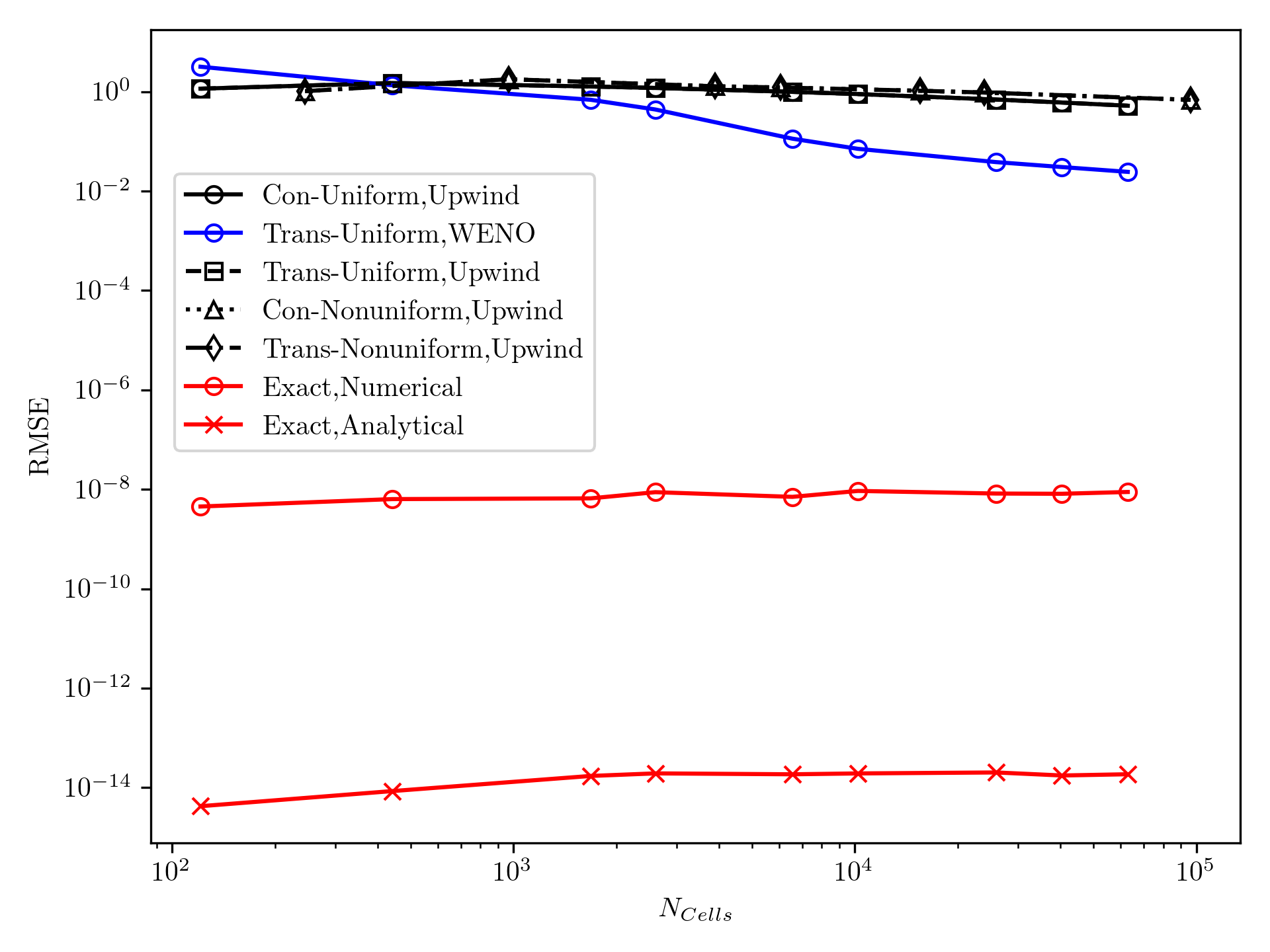}\vspace{-0.2cm}
        \caption{RMSE}
    \end{subfigure}
    \hfill
    \begin{subfigure}{0.45\textwidth}
        \centering
        \includegraphics[width=\textwidth]{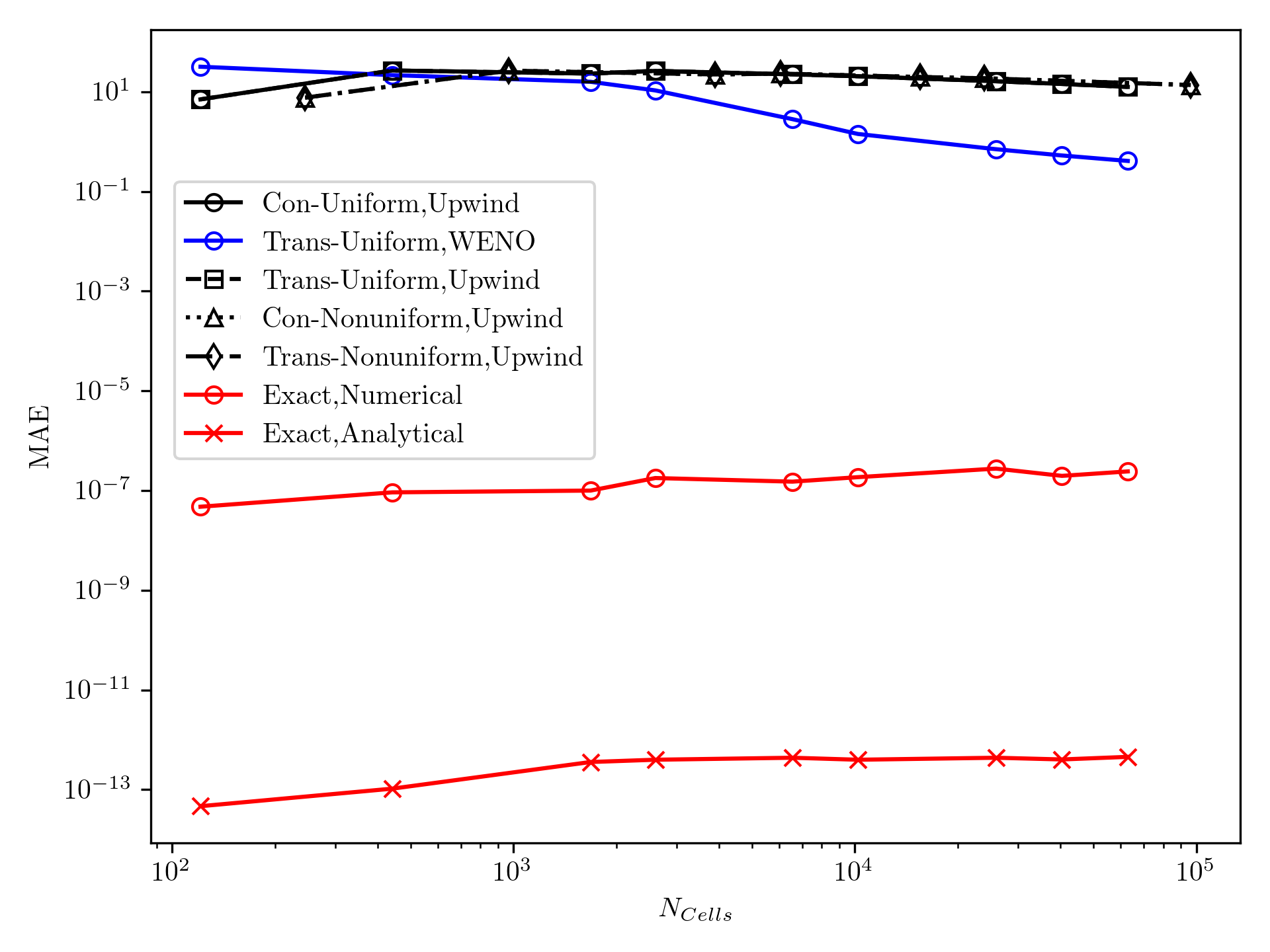}\vspace{-0.2cm}
     \caption{MAE}
    \end{subfigure}
    \caption{Error analysis for Case 2.}
    \label{fig:case2error}
\end{figure}

Consider the PBM adapted from \cite{Gunawan2008},
\begin{equation}
    \frac{\partial f}{\partial t} + \frac{\partial (G_{1}(a_{1})f)}{\partial a_{1}} + \frac{\partial (G_{2}(a_{2})f)}{\partial a_{2}} = 0, \quad f_{0}(a_{1},a_{2}) = 50\exp\!\left( -\frac{(a_{1}-0.4)^{2}}{0.005} - \frac{(a_{2}-0.4)^{2}}{0.005}  \right),
    \label{eq:model2_con}
\end{equation}
with,
\begin{equation}
    G_{1}(a_{1}) = 0.1 + 0.05 a_{1},\quad G_{2}(a_{2}) = 0.5 + 0.25 a_{2},\quad a_{i} \in [0,2].
\end{equation}
Multiplying each term in \eqref{eq:model2_con} by $G_{1}(a_{1})G_{2}(a_{2})$ and defining $\hat{f}(t,a_{1},a_{2}) = G_{1}(a_{1})G_{2}(a_{2})f$ transforms \eqref{eq:model2_con} into
\begin{equation}
    \frac{\partial \hat{f}}{\partial t} + G_{1}(a_{1})\frac{\partial \hat{f}}{\partial a_{1}} + G_{2}(a_{2})\frac{\partial \hat{f}}{\partial a_{2}} = 0, \quad \hat{f}(0,a_{1},a_{2}) = \hat{f}_{0}(a_{1},a_{2}) = G_{1}(a_{1})G_{2}(a_{2})f_{0}(a_{1},a_{2}).
    \label{eq:model2_trans}
\end{equation}
This PBM has an analytical solution,
\begin{equation}
    f(t,a_{1},a_{2}) = \frac{\hat{f}_{0} \left( (a_{1}+2)e^{-0.05t} - 2),(a_{2}+2)e^{-0.25t} - 2)\right)}{G_{1}(a_{1}) G_{2}(a_{2})}
\end{equation}
Exemplar numerical solutions to the PBM can be found in \cref{fig:case2simulation} and convergence analysis in \cref{fig:case2error}. To employ the ``Exact,Analytical" scheme, the functions $\tilde{a}_{i}(a_{i})$ given by \eqref{eq:size1_var_trans} and their inverse need to be separately computed analytically and supplied into the scheme. Another variant of the exact scheme, ``Exact,Numerical" is also considered where $\tilde{a}_{i}(a_{i})$ and its inverse are computed numerically instead. While the accuracy of the ``Exact,Numerical" scheme is constrained by the accuracy of the quadrature step as can be seen in \cref{fig:case2error}, the scheme is still able to perform very well and is more user-friendly as it does not require any pre-computation.

\subsection{Case 3: PBMs with Growth Rate $G_{i} = G_{i}(\mathbf{a})$}

Consider the PBM,
\begin{equation}
    \frac{\partial f}{\partial t} + \frac{\partial (G_{1}(a_{1},a_{2})f)}{\partial a_{1}} + \frac{\partial (G_{2}(a_{1},a_{2})f)}{\partial a_{2}} = 0, \quad f_{0}(a_{1},a_{2}) = 50\exp\!\left( -\frac{(a_{1}-0.4)^{2}}{0.005} - \frac{(a_{2}-0.4)^{2}}{0.005}  \right),
\end{equation}
with,
\begin{equation}
    G_{1}(a_{1},a_{2}) = 0.25 + 0.5(a_{1}+a_{2}) , \quad G_{2}(a_{1},a_{2}) = 0.5 + 0.25(a_{1} + a_{2}).
\end{equation}
The analytical solution to this PBM is
\begin{equation}
    f(t,a_{1},a_{2}) = f_{0}\!\left(\frac{Bc_{2} - Dc_{1}}{BC - AD}, \frac{Cc_{1} - Ac_{2}}{BC - AD} \right)\exp(-0.75t), 
\end{equation}
where
\begin{align}
    c_{1} &= 3a_{1} + 0.75t + 2 - 2e^{0.75t}, \quad 
    c_{2} = 3a_{2} - 0.75t + 1 - e^{0.75t}, \nonumber \\ A &= 1 + 2e^{0.75t}, \quad B = -2 + 2e^{0.75t}, \quad C = -1 + e^{0.75t}, \quad D = 2 + e^{0.75t}.
\end{align}

\begin{figure}[htbp]
    \centering
    \begin{subfigure}{0.45\textwidth}
        \centering
        \includegraphics[width=\textwidth]{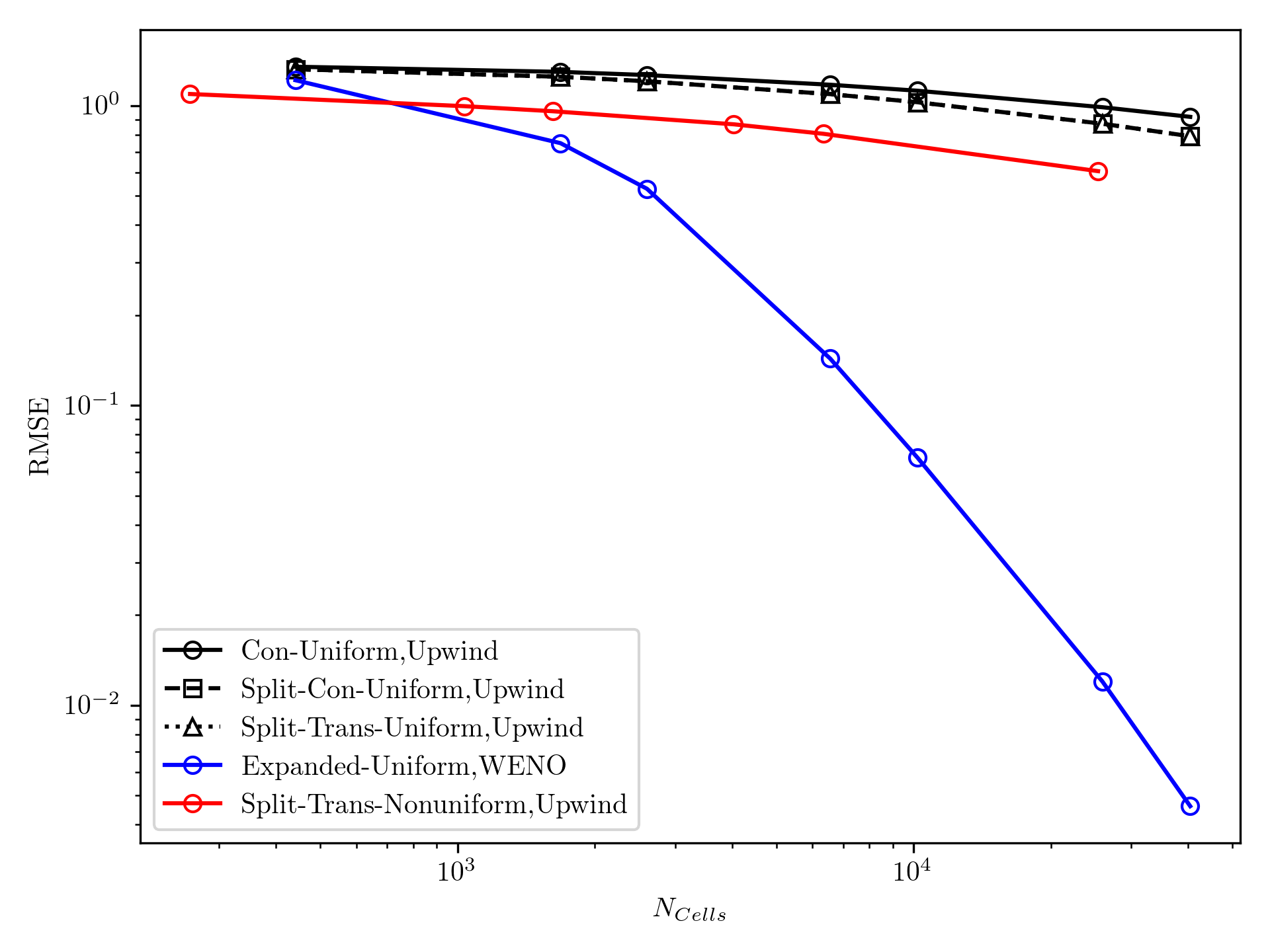}\vspace{-0.2cm}
        \caption{RMSE}
    \end{subfigure}
    \hfill
    \begin{subfigure}{0.45\textwidth}
        \centering
        \includegraphics[width=\textwidth]{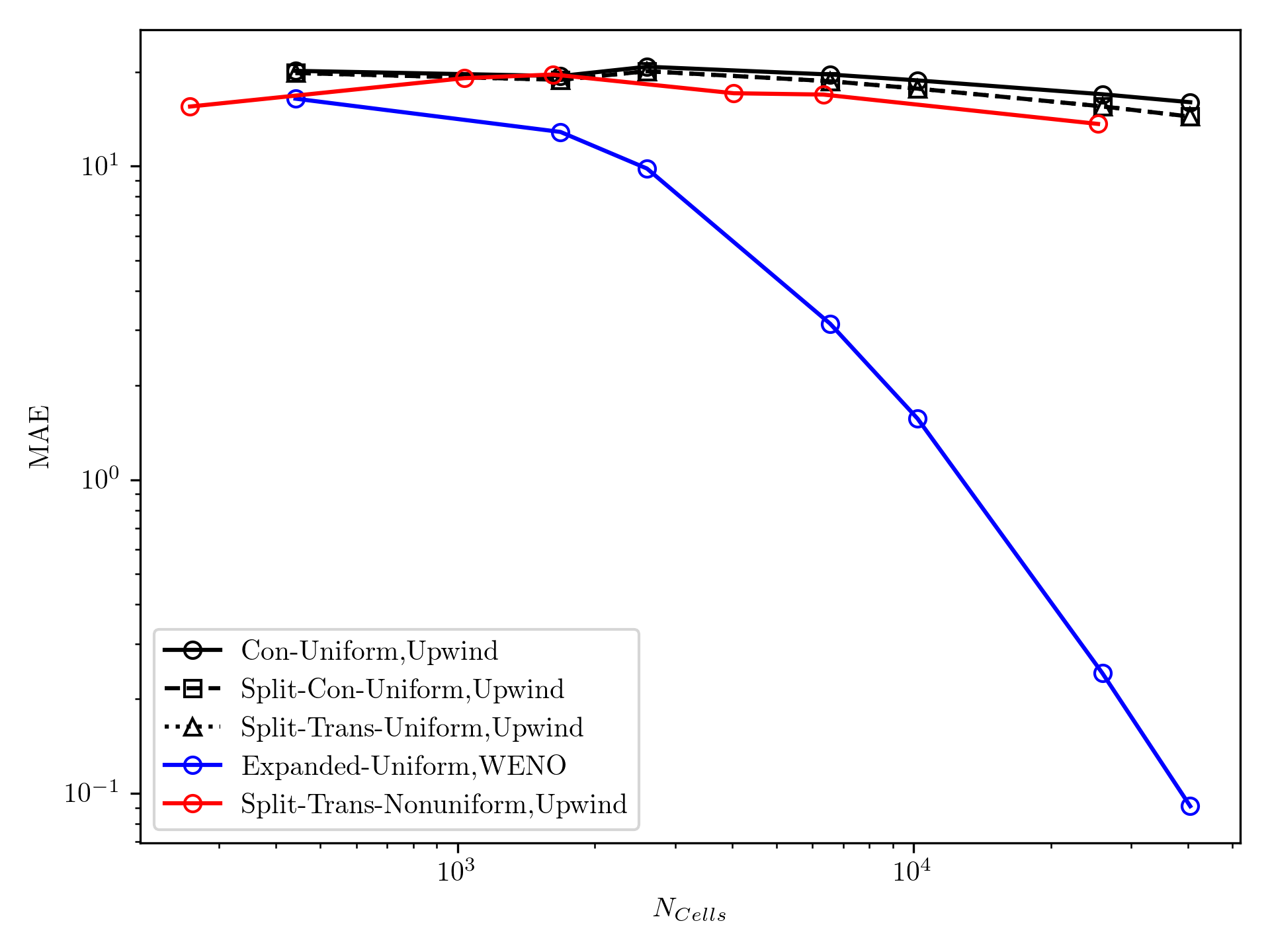}\vspace{-0.2cm}
     \caption{MAE}
    \end{subfigure}
    \hfill
    \begin{subfigure}{0.45\textwidth}
        \centering
        \includegraphics[width=\textwidth]{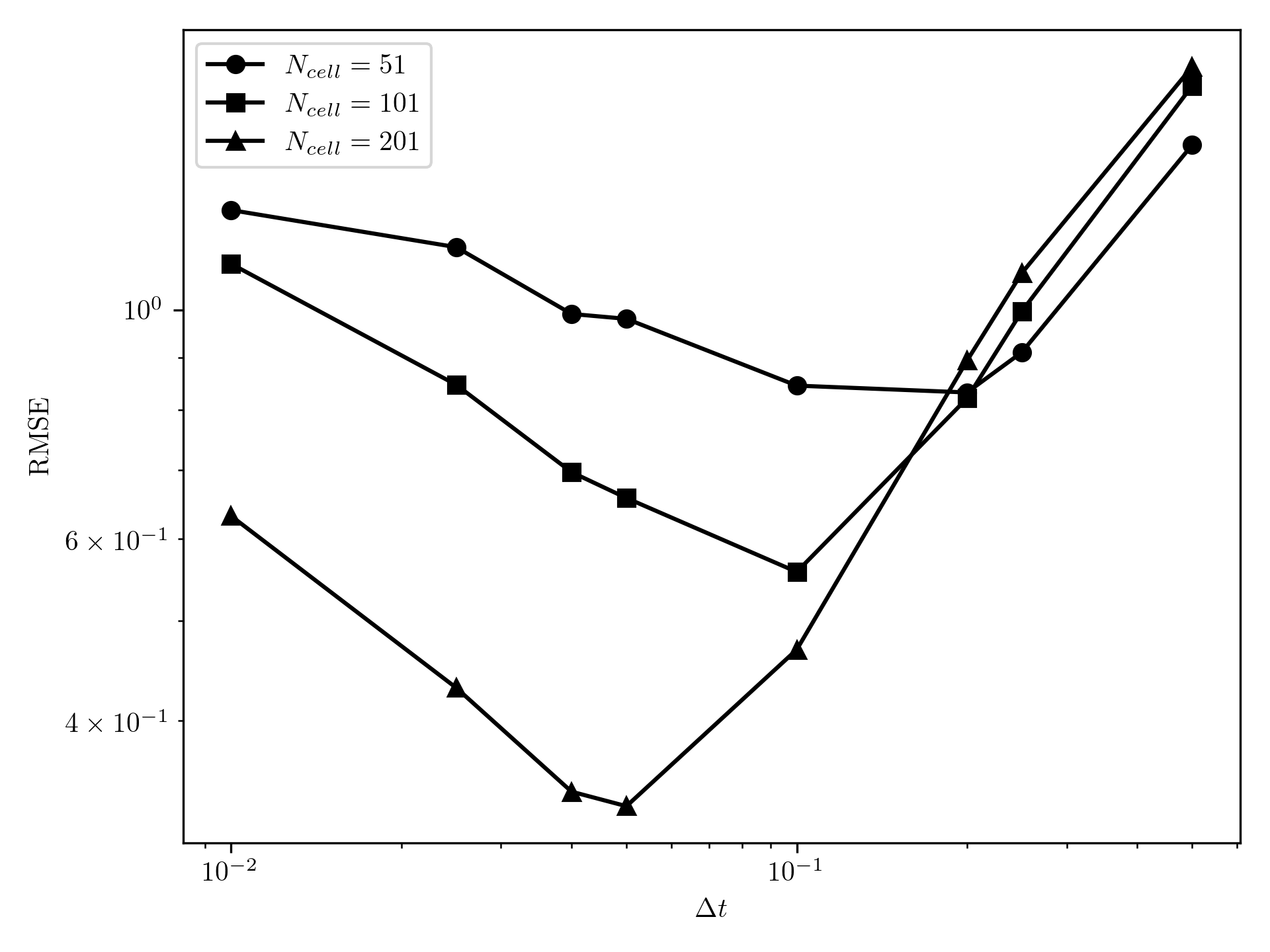}\vspace{-0.2cm}
     \caption{``Split-Exact" RMSE}
    \end{subfigure}
    \hfill
    \begin{subfigure}{0.45\textwidth}
        \centering
        \includegraphics[width=\textwidth]{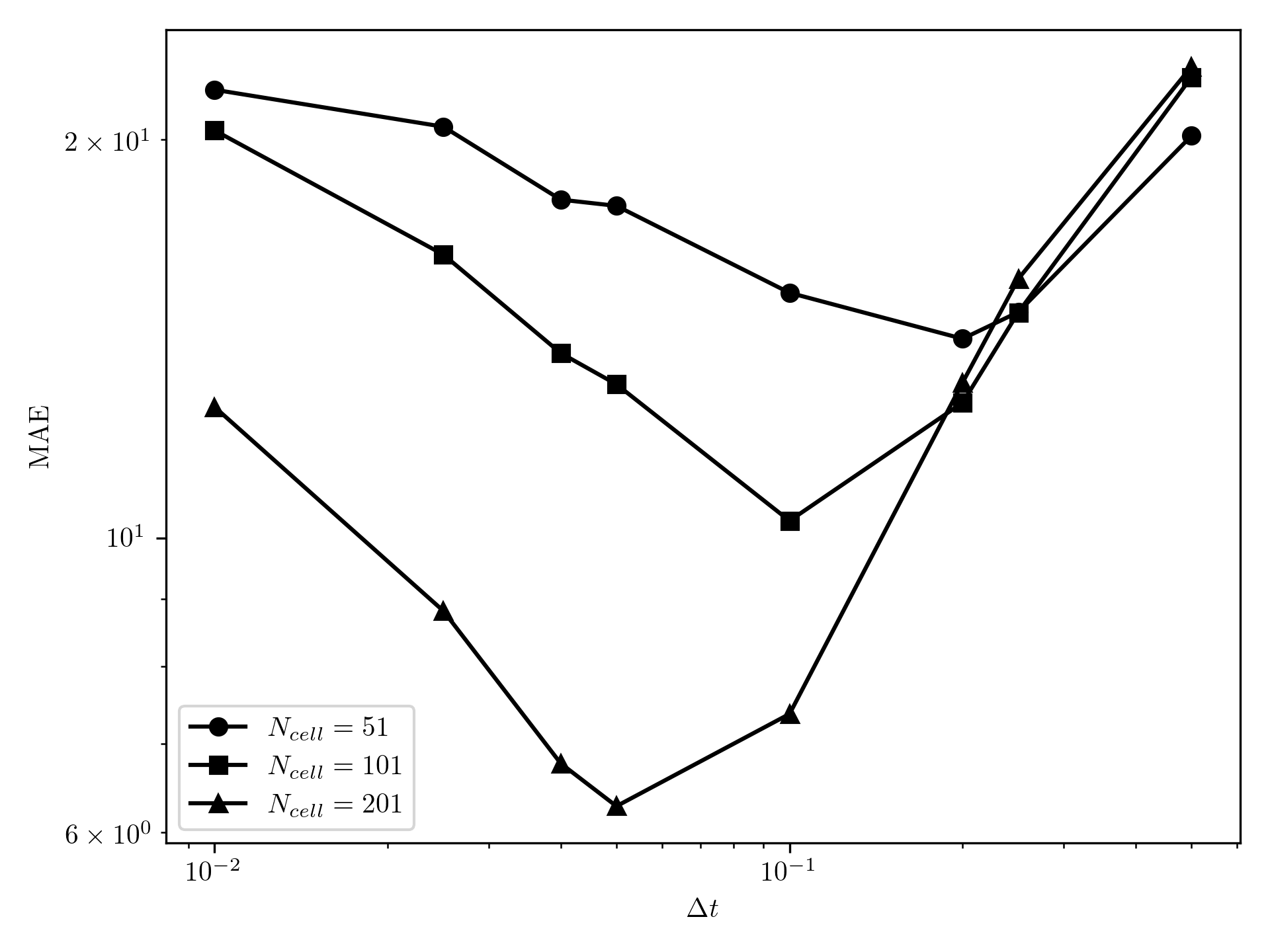}\vspace{-0.2cm}\caption{``Split-Exact" MAE}
    \end{subfigure}
    \vspace{-0.2cm}\caption{Error Analysis for Case 3.}
    \label{fig:case3error}
\end{figure}

\begin{figure}[htbp]
    \centering
    \begin{subfigure}{0.45\textwidth}
        \centering
        \includegraphics[width=0.9\textwidth]{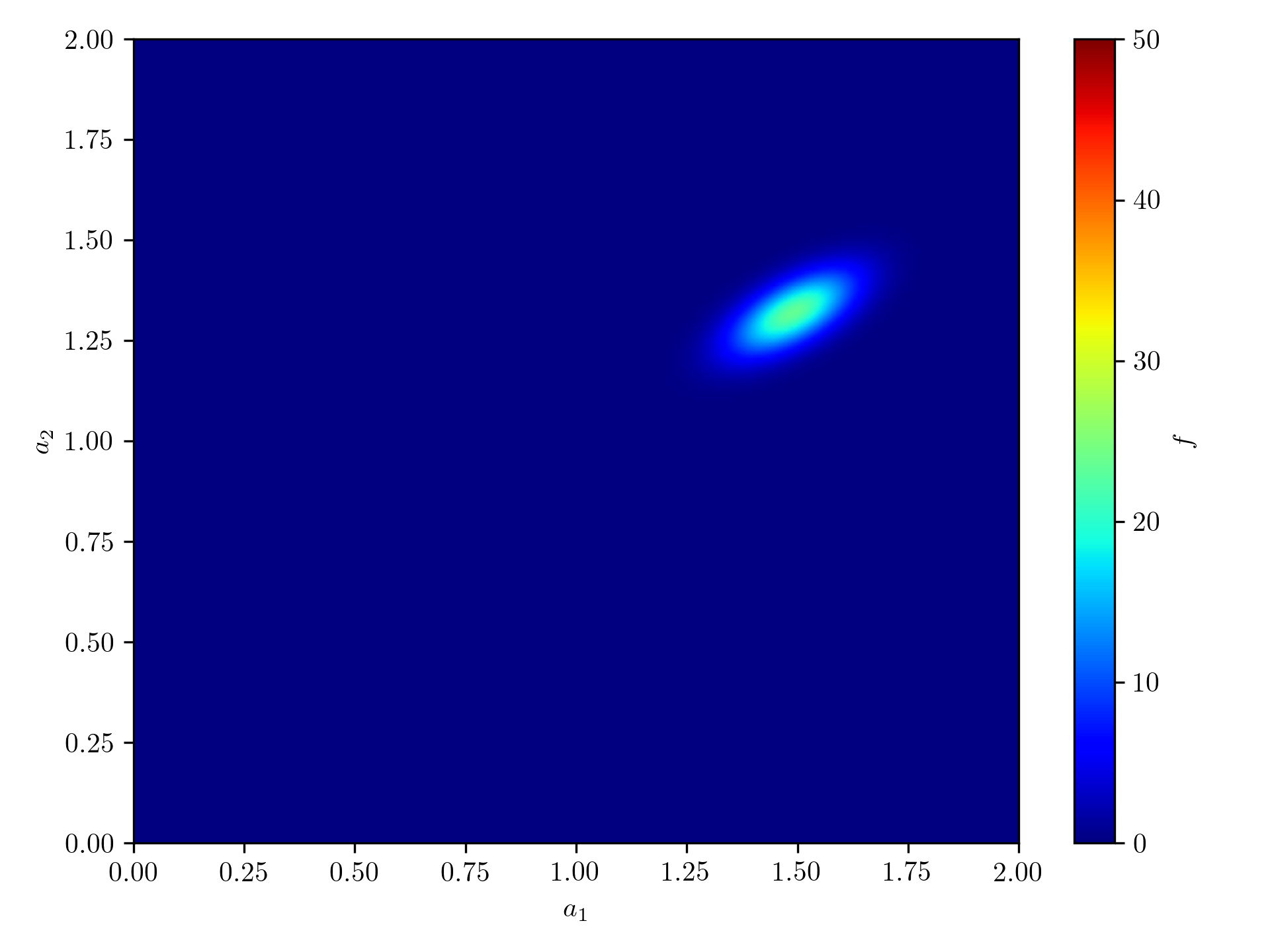}
        \vspace{-0.2cm}\caption{Analytical Solution}
    \end{subfigure}
    \hfill
    \begin{subfigure}{0.45\textwidth}
        \centering
        \includegraphics[width=0.9\textwidth]{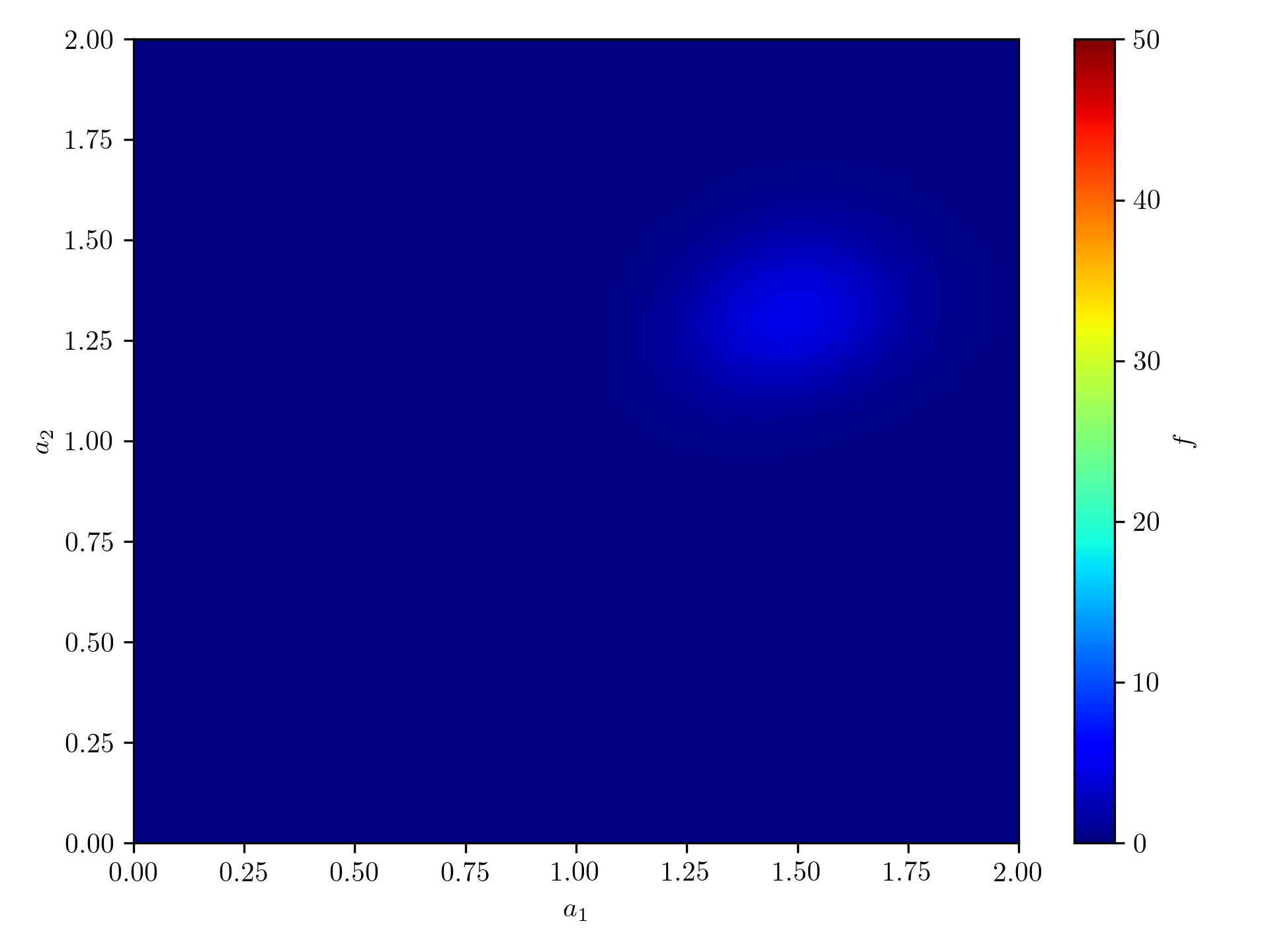}
     \vspace{-0.2cm}\caption{Con-Uniform,Upwind}
    \end{subfigure}
    \hfill
    \begin{subfigure}{0.45\textwidth}
        \centering
        \includegraphics[width=0.9\textwidth]{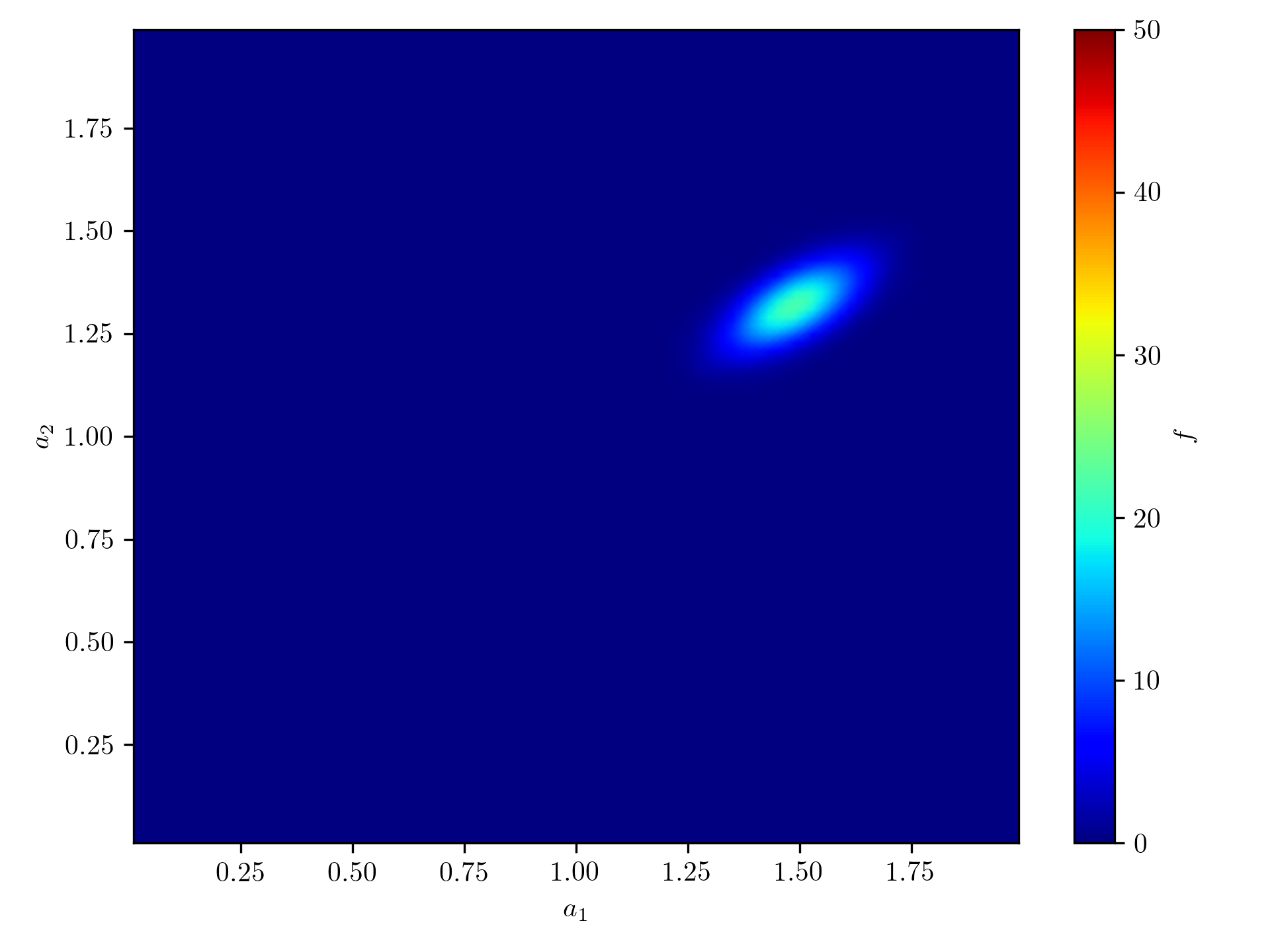}
     \vspace{-0.2cm}\caption{Expanded-Uniform,WENO}
    \end{subfigure}
    \hfill
    \begin{subfigure}{0.45\textwidth}
        \centering
        \includegraphics[width=0.9\textwidth]{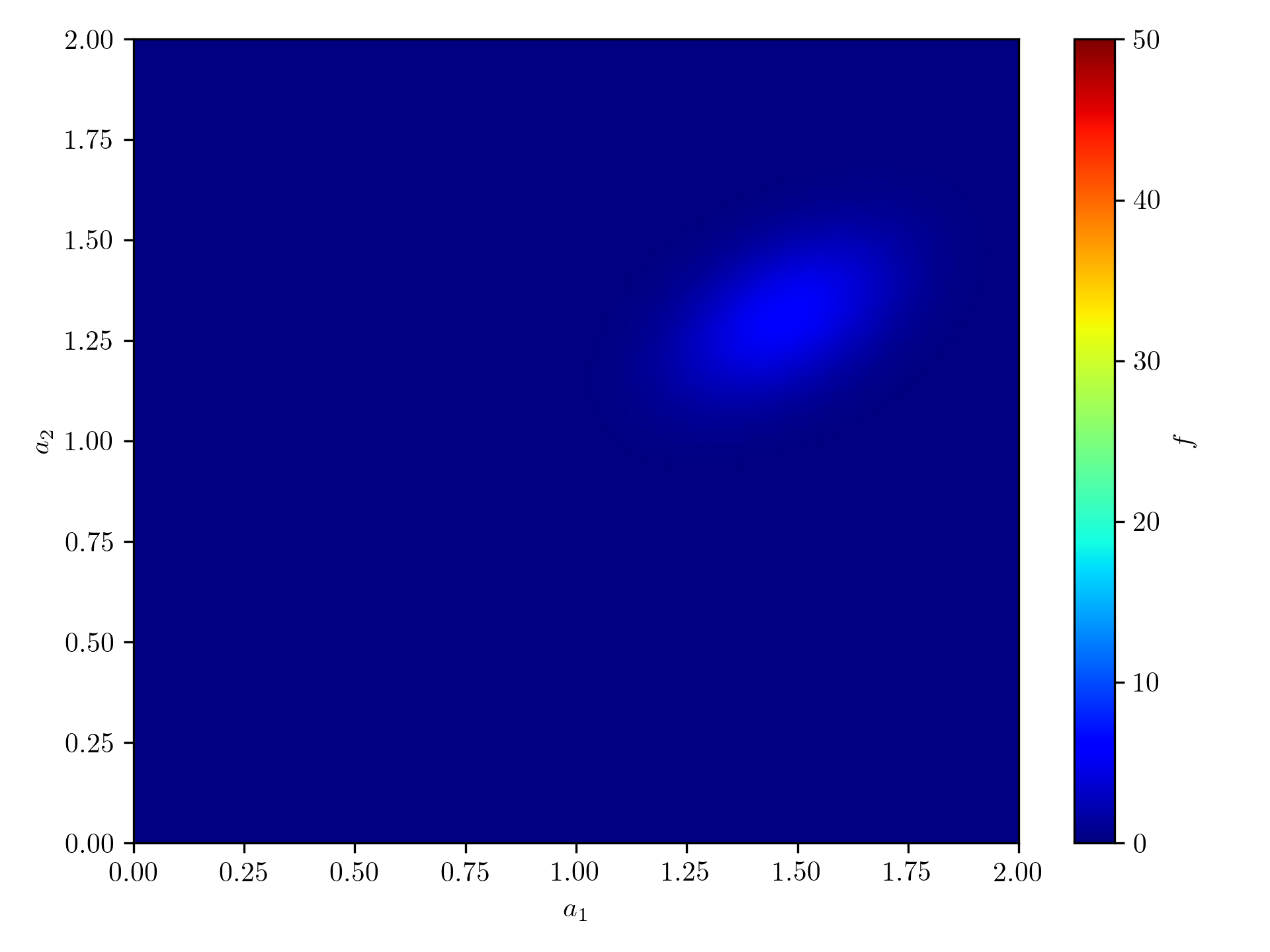}
     \vspace{-0.2cm}\caption{Split-Con-Uniform, Upwind}
    \end{subfigure}
    \hfill
    \begin{subfigure}{0.45\textwidth}
        \centering
        \includegraphics[width=0.9\textwidth]{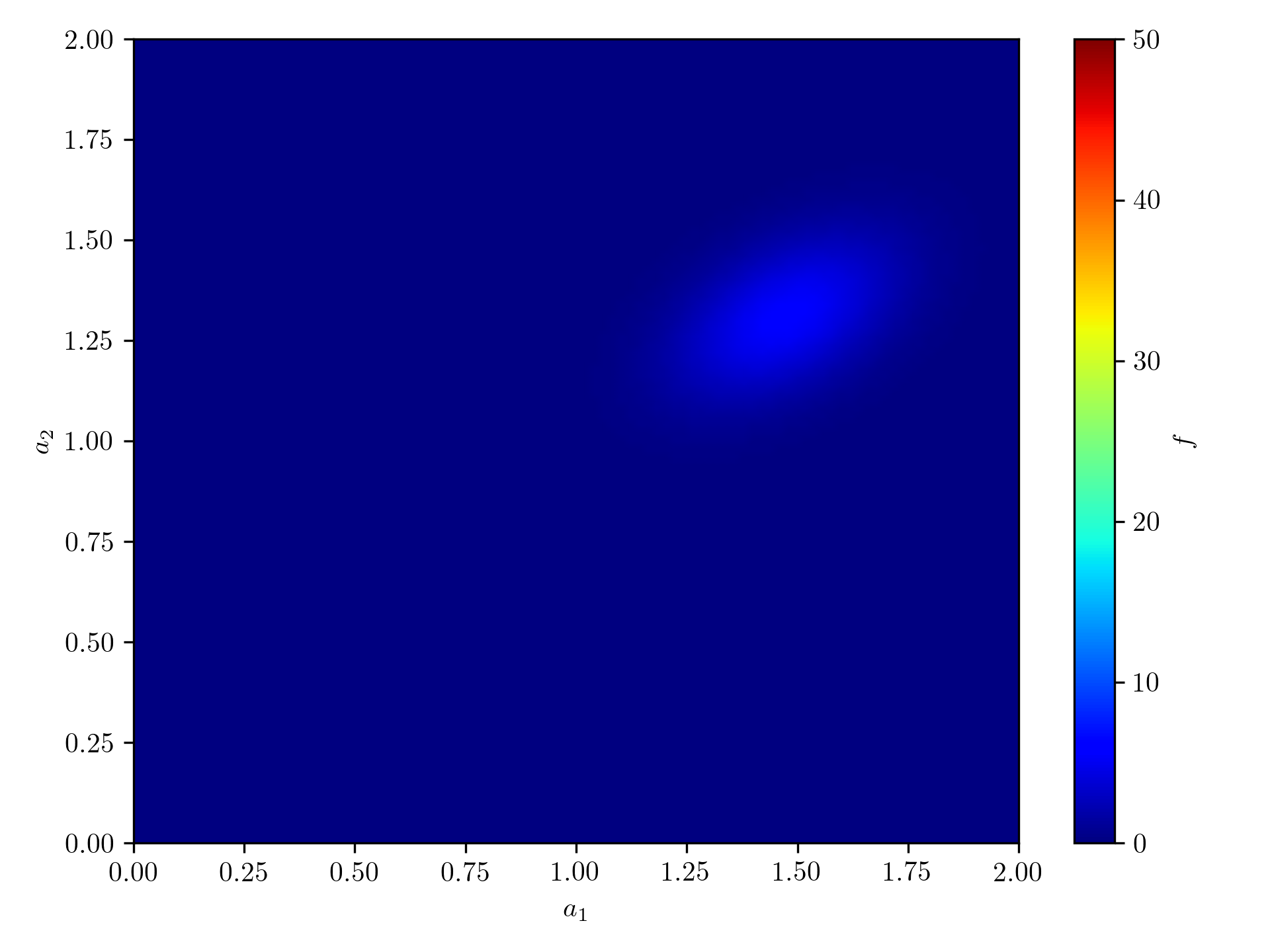}
    \vspace{-0.2cm} \caption{Split-Trans-Uniform,Upwind}
    \end{subfigure}
    \hfill
    \begin{subfigure}{0.45\textwidth}
        \centering
        \includegraphics[width=0.9\textwidth]{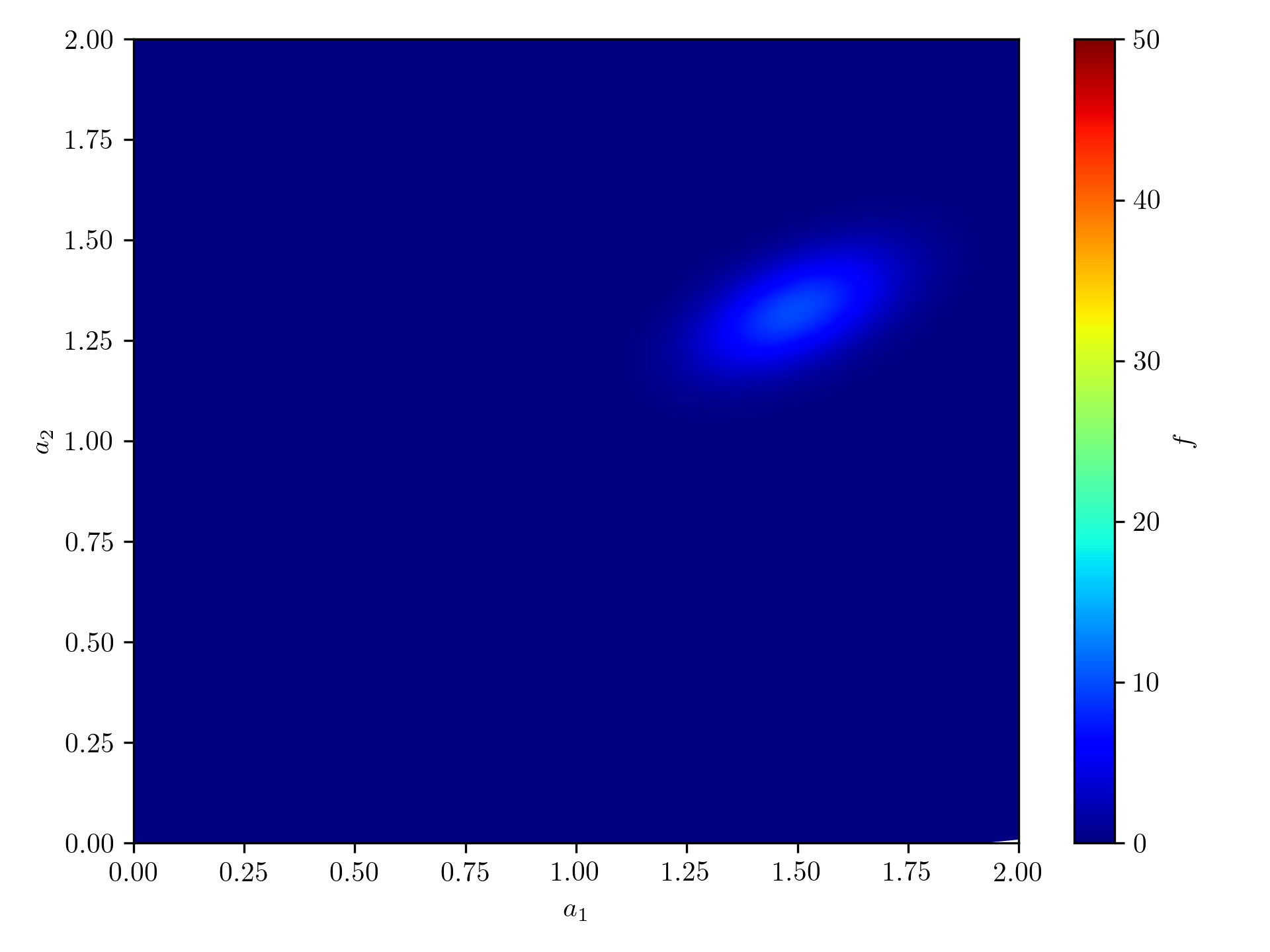}
     \vspace{-0.2cm}\caption{Split-Trans-Nonuniform,Upwind}
    \end{subfigure}
    \hfill
    \begin{subfigure}{0.45\textwidth}
        \centering
        \includegraphics[width=0.9\textwidth]{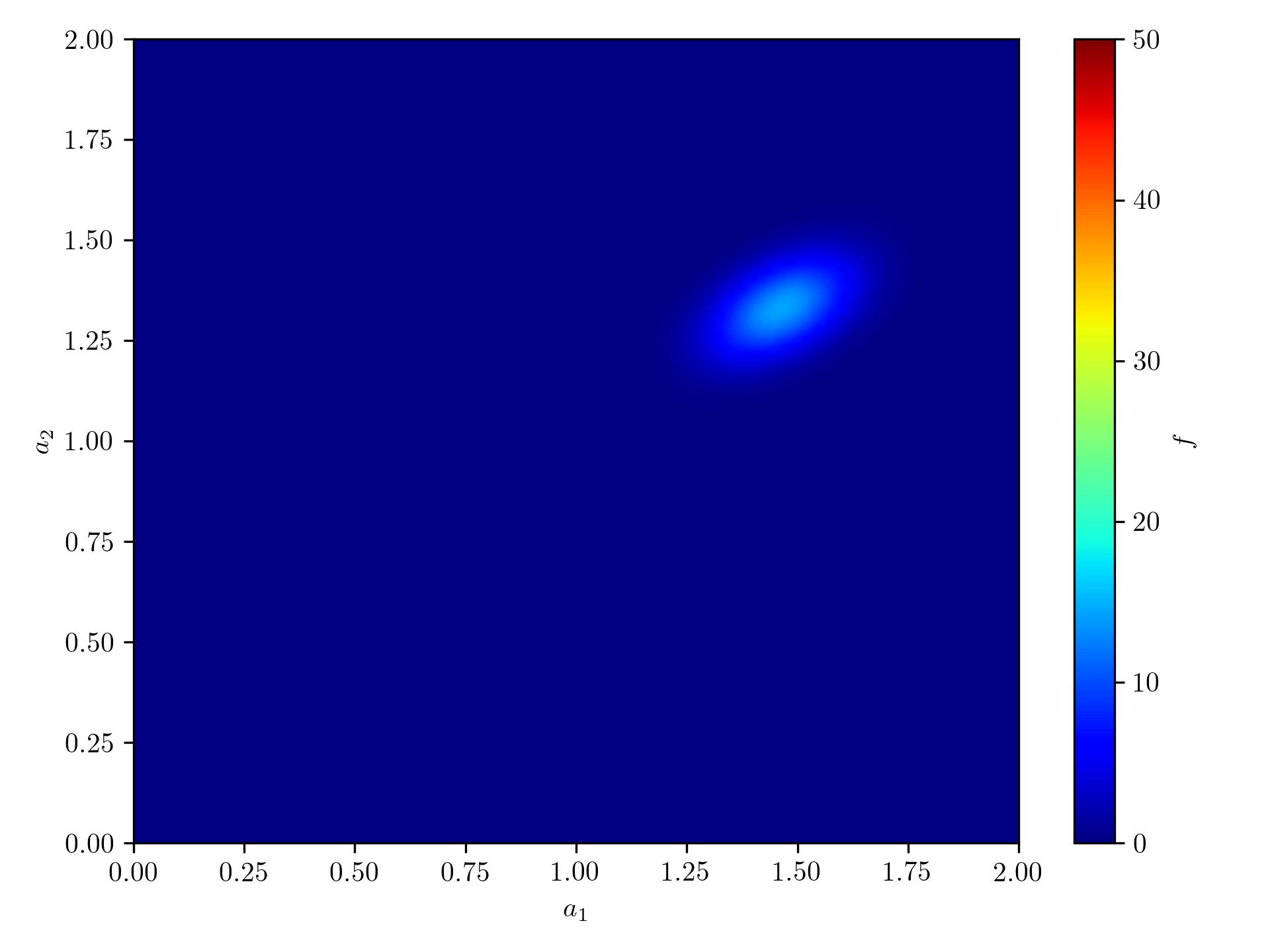}
     \vspace{-0.2cm}\caption{Split-Exact}
    \end{subfigure}
    \caption{Simulation results for Case 3 at $t=1.0$ using the various schemes. 101 grid points are used in both the $a_{1}$ and $a_{2}$ directions for the Upwind schemes on a uniform grid and Split-Exact schemes while 100 cells in both the $a_{1}$ and $a_{2}$ directions for the WENO scheme. The Split-Exact scheme used a value of $\Delta t = 0.1$. The simulation on a nonuniform grid has 25,367 grid points by specifying $\Delta t = 0.01$.}
    \label{fig:case3simulation}
\end{figure}

None of the schemes employed are able to solve the PBM to a high degree of accuracy (Figure~\ref{fig:case3error}). Even though the ``Split-Exact" scheme performs poorer than the WENO scheme (which is regarded as both mathematically involved and computationally expensive), the ``Split-Exact" scheme enables the use of much larger $\Delta t$ values as the sub-problems are effectively solved as a single function call to advance the full $\Delta t$ instead of requiring time-stepping.

The ``Split-Exact scheme" using the method of characteristics can be modified and enhanced to eliminate the need to interpolate between meshes to solve each sub-problem and further speed up the solution. However, this enhancement requires the offline analytical evaluation of two integrals and does not significantly improve the error performance (see Figure A1). This modification is discussed in the Appendix (see Section A1). The lack of improvement in the error performance from using the enhanced scheme indicates that the main source of error with the ``Split-Exact" scheme arises from the operator splitting step rather than the quadrature and interpolation steps. Correspondingly, it might be worth exploring more sophisticated splitting schemes to further improve error performance.

\subsection{Case 4: PBMs with a Nonhomogeneous Term}

Consider the PBM,
\begin{equation}
    \frac{\partial f}{\partial t} + \frac{\partial f}{\partial a_{1}} + \frac{\partial f}{\partial a_{2}} =  1 + a_{1}a_{2}, \quad f_{0}(a_{1},a_{2}) = 10\exp\!\left( -\frac{(a_{1}-0.4)^{2}}{0.005} - \frac{(a_{2}-0.4)^{2}}{0.005}  \right),
\end{equation}
with the same boundary conditions as Case 1. An analytical solution cannot be readily supplied for benchmarking so, to perform the convergence analysis, a ``reference" numerical solution was generated by running the simulation using the presented scheme on a very fine mesh ($\sim 2.56\times 10^{6}$ grid points).

\begin{figure}[htbp]
    \centering
    \begin{subfigure}{0.45\textwidth}
        \centering
        \includegraphics[width=0.9\textwidth]{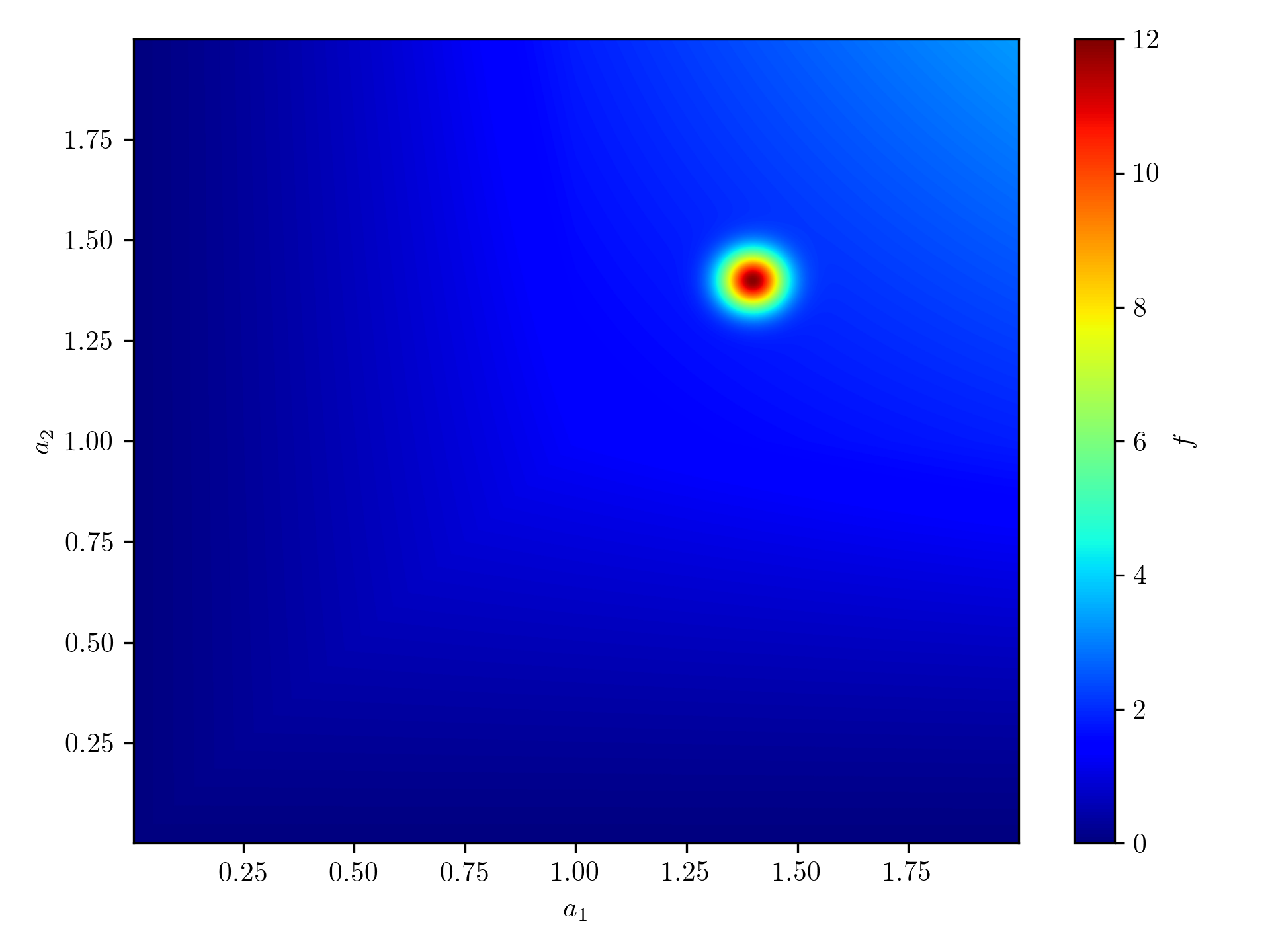}\vspace{-0.2cm}
        \caption{Reference Solution}
    \end{subfigure}
    \hfill
    \begin{subfigure}{0.45\textwidth}
        \centering
        \includegraphics[width=0.9\textwidth]{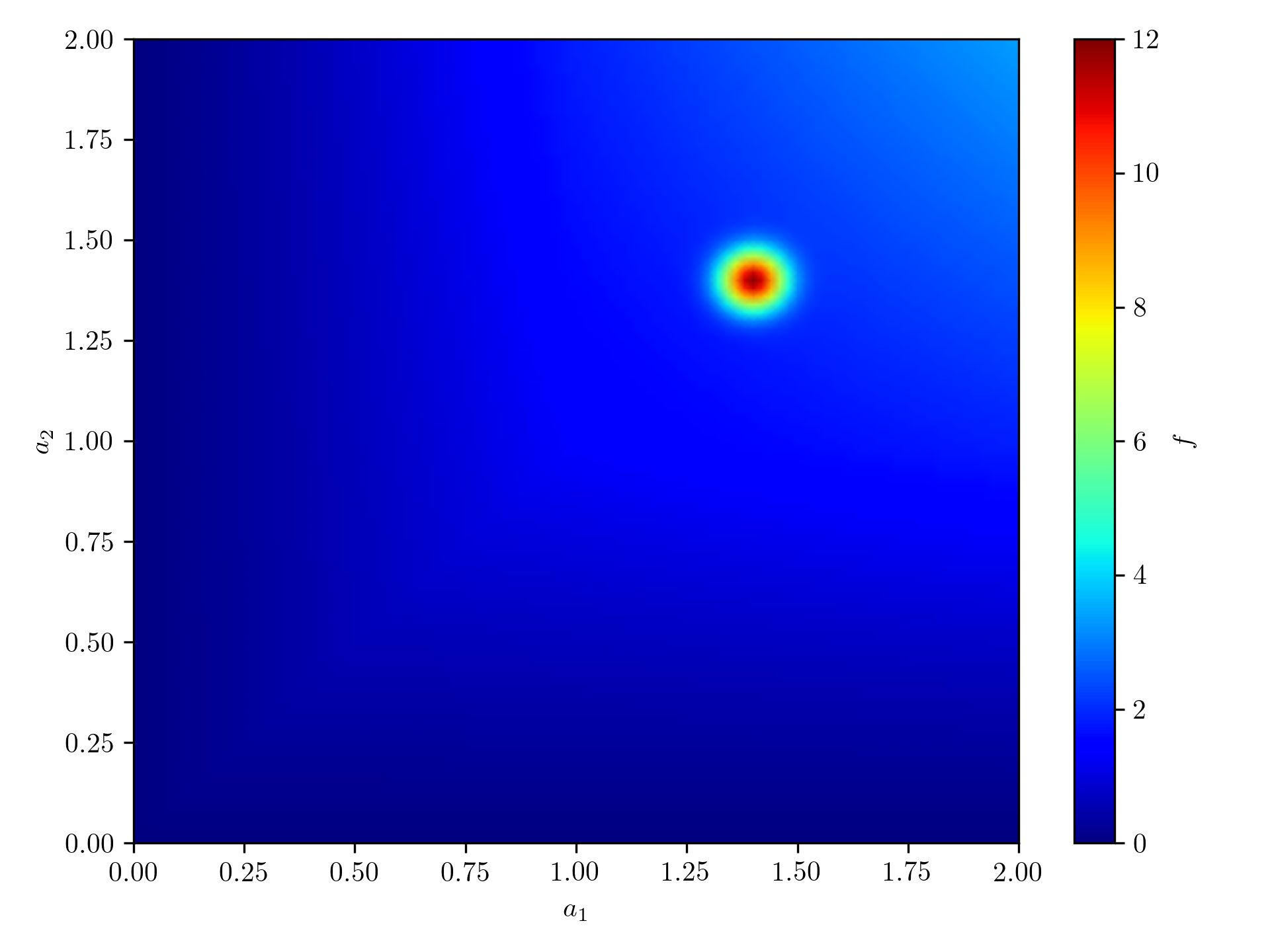}\vspace{-0.2cm}
     \caption{Split Scheme}
    \end{subfigure}
    \hfill
    \begin{subfigure}{0.45\textwidth}
        \centering
        \includegraphics[width=0.9\textwidth]{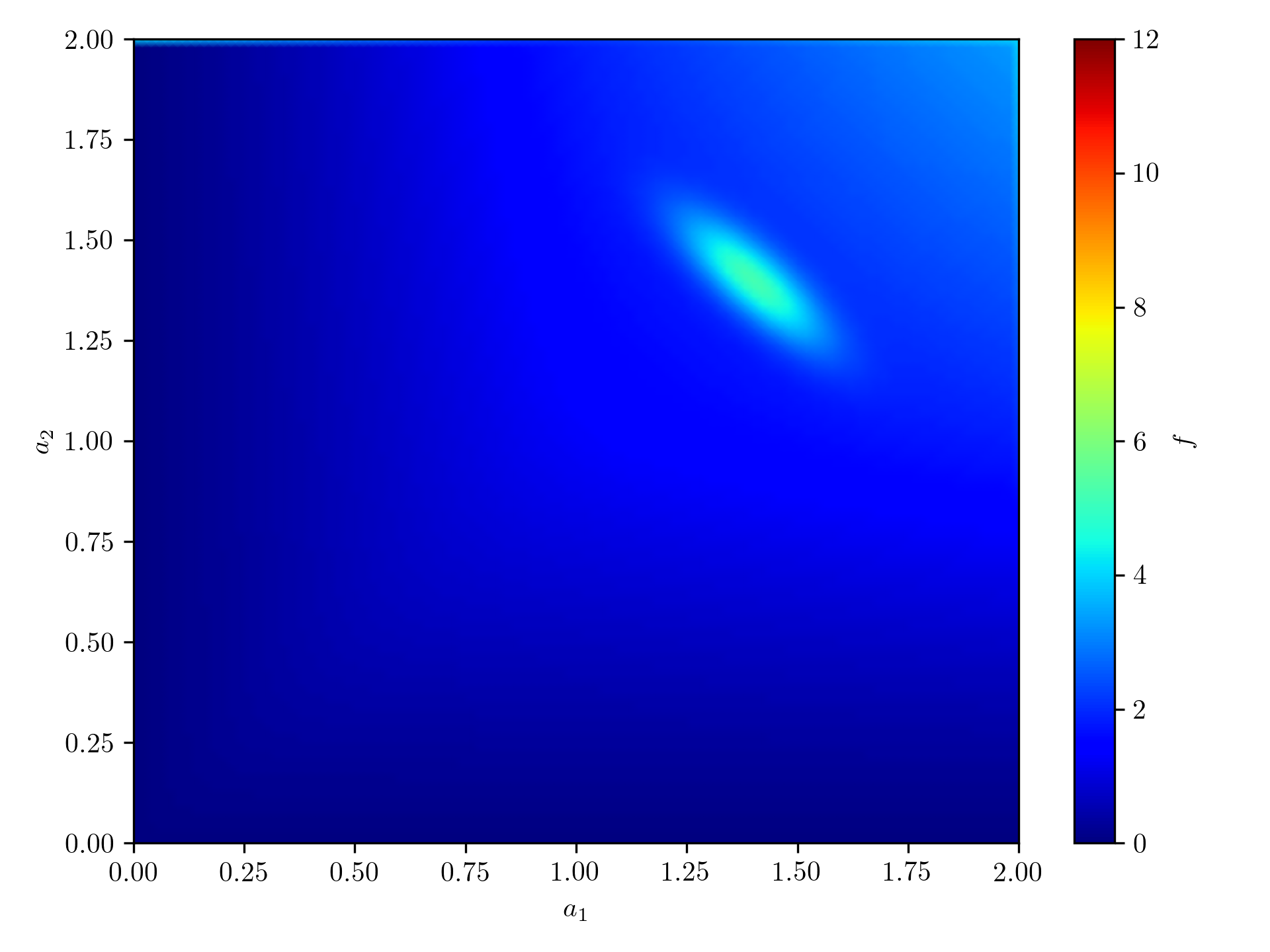}\vspace{-0.2cm}
     \caption{Upwind Scheme}
    \end{subfigure}
    \hfill
    \begin{subfigure}{0.45\textwidth}
        \centering
        \includegraphics[width=0.9\textwidth]{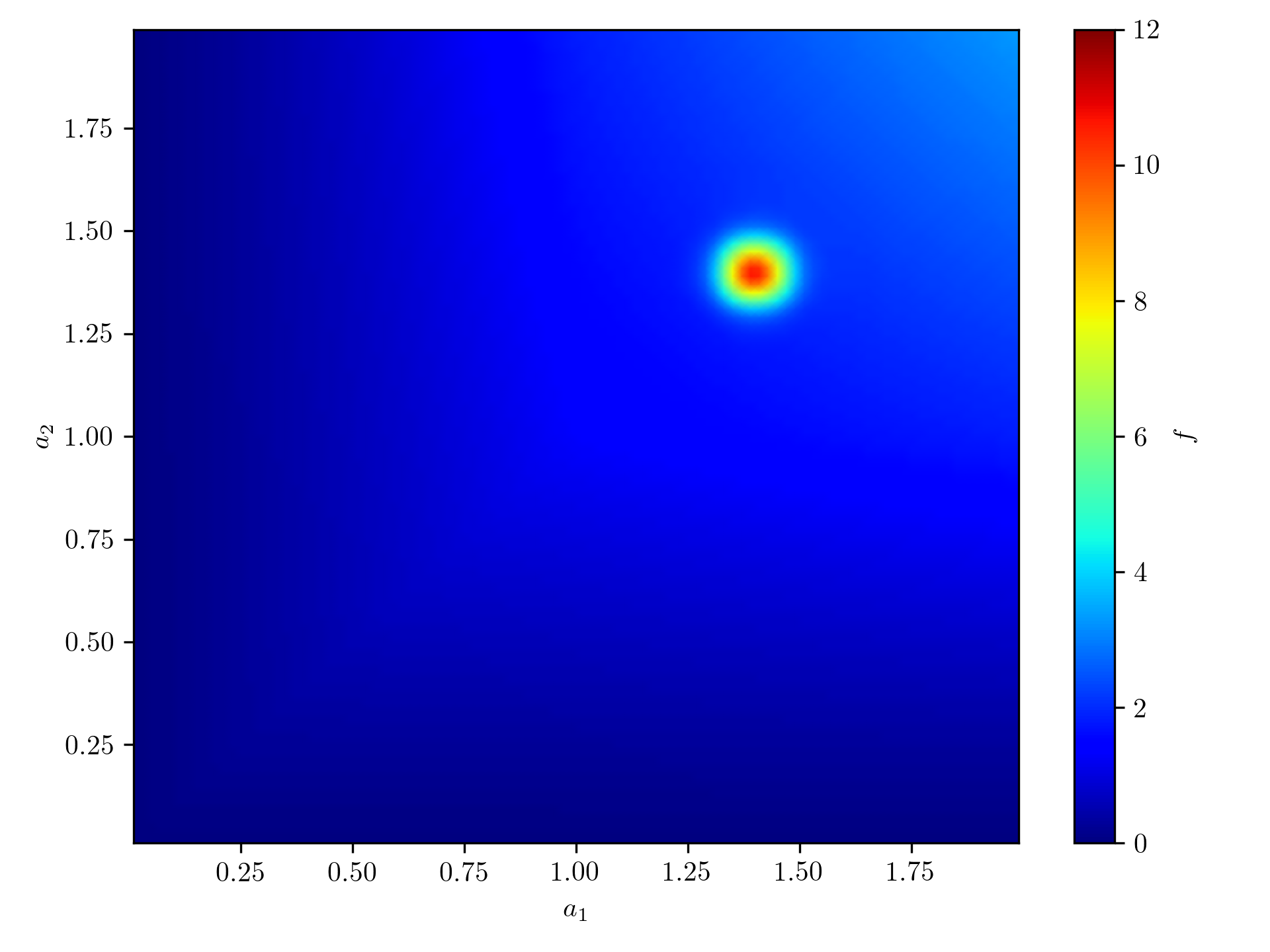}\vspace{-0.2cm}
     \caption{WENO Scheme}
    \end{subfigure}
    \caption{Simulation results for Case 4 at $t=1.0$ using the various schemes. 101 grid points are used in both the $a_{1}$ and $a_{2}$ directions for the Upwind and Split schemes while 100 cells in both the $a_{1}$ and $a_{2}$ directions for the WENO scheme.}
    \label{fig:case4simulation}
\end{figure}

\begin{figure}[htbp]
    \centering
    \begin{subfigure}{0.45\textwidth}
        \centering
        \includegraphics[width=\textwidth]{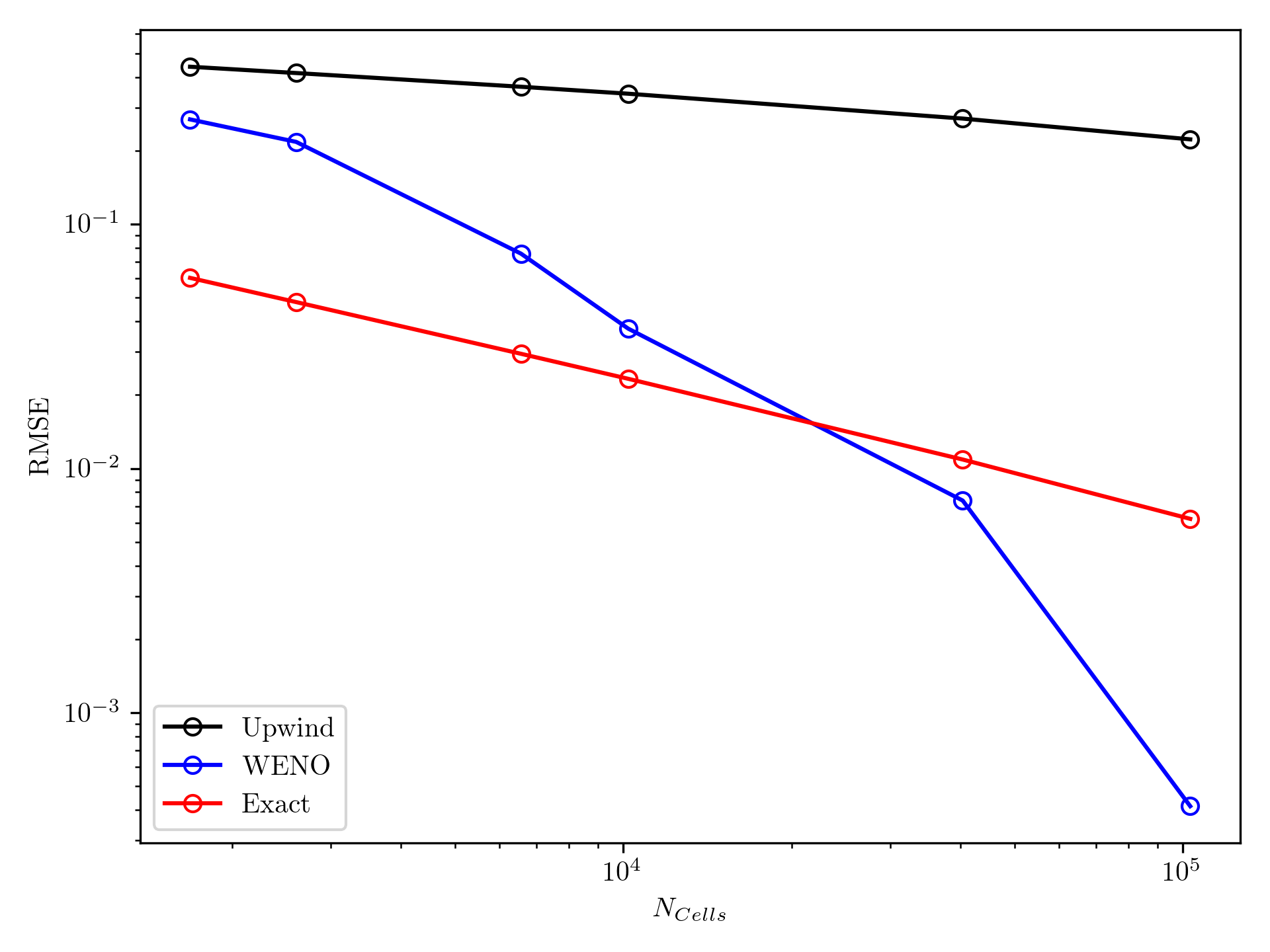}
        \caption{RMSE}
    \end{subfigure}
    \hfill
    \begin{subfigure}{0.45\textwidth}
        \centering
        \includegraphics[width=\textwidth]{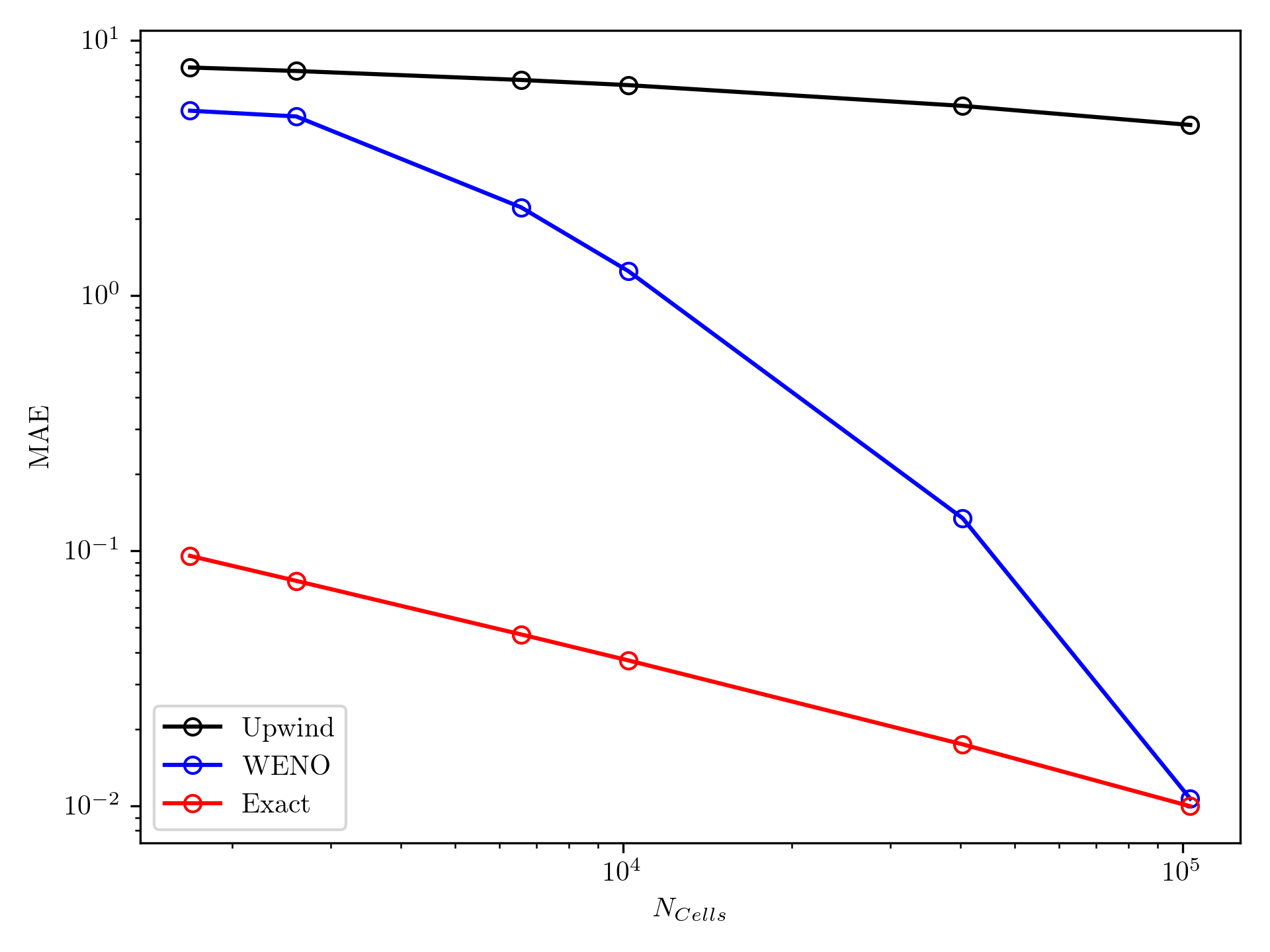}
     \caption{MAE}
    \end{subfigure}
    \caption{Error analysis for Case 4.}
    \label{fig:case4error}
\end{figure}

The numerical diffusion is much smaller for the Split and WENO than the Upwind scheme (Figure~\ref{fig:case4simulation}). At low resolution, the proposed Split scheme has much higher accuracy than the WENO scheme (Figure~\ref{fig:case4error}). Employing operator splitting and solving the ``advection" component of the Split problem exactly results in much less numerical diffusion on coarser meshes compared to WENO scheme.

\subsection{Case 5: PBMs with a Linear Nonhomogeneous Term}

Consider an example of the 2D von Foerster equation,
\begin{equation}
    \frac{\partial f}{\partial t} + \frac{\partial f}{\partial a_{1}} + \frac{\partial f}{\partial a_{2}} = -(a_{1} + a_{2})f, \quad f_{0}(a_{1},a_{2}) = 50\exp\!\left( -\frac{(a_{1}-0.4)^{2}}{0.005} - \frac{(a_{2}-0.4)^{2}}{0.005}  \right),
\end{equation}
with the boundary conditions,
\begin{equation}
    f(t,a_{1} = 0, a_{2}) = f(t, a_{1}, a_{2} = 0) = 0.
\end{equation}
This PBM has the analytical solution,
\begin{equation}
    f(t,a_{1},a_{2}) = 
    \begin{cases} f_{0}(a_{1}-t, a_{2}-t) \exp\!{\left(-(a_{1}+a_{2})t + t^{2} \right)}, & a_{1},a_{2} \geq t, \\
    0, & a_{1},a_{2} < t.
    \end{cases}
\end{equation}

\begin{figure}[htbp]
    \centering
    \begin{subfigure}{0.45\textwidth}
        \centering
        \includegraphics[width=\textwidth]{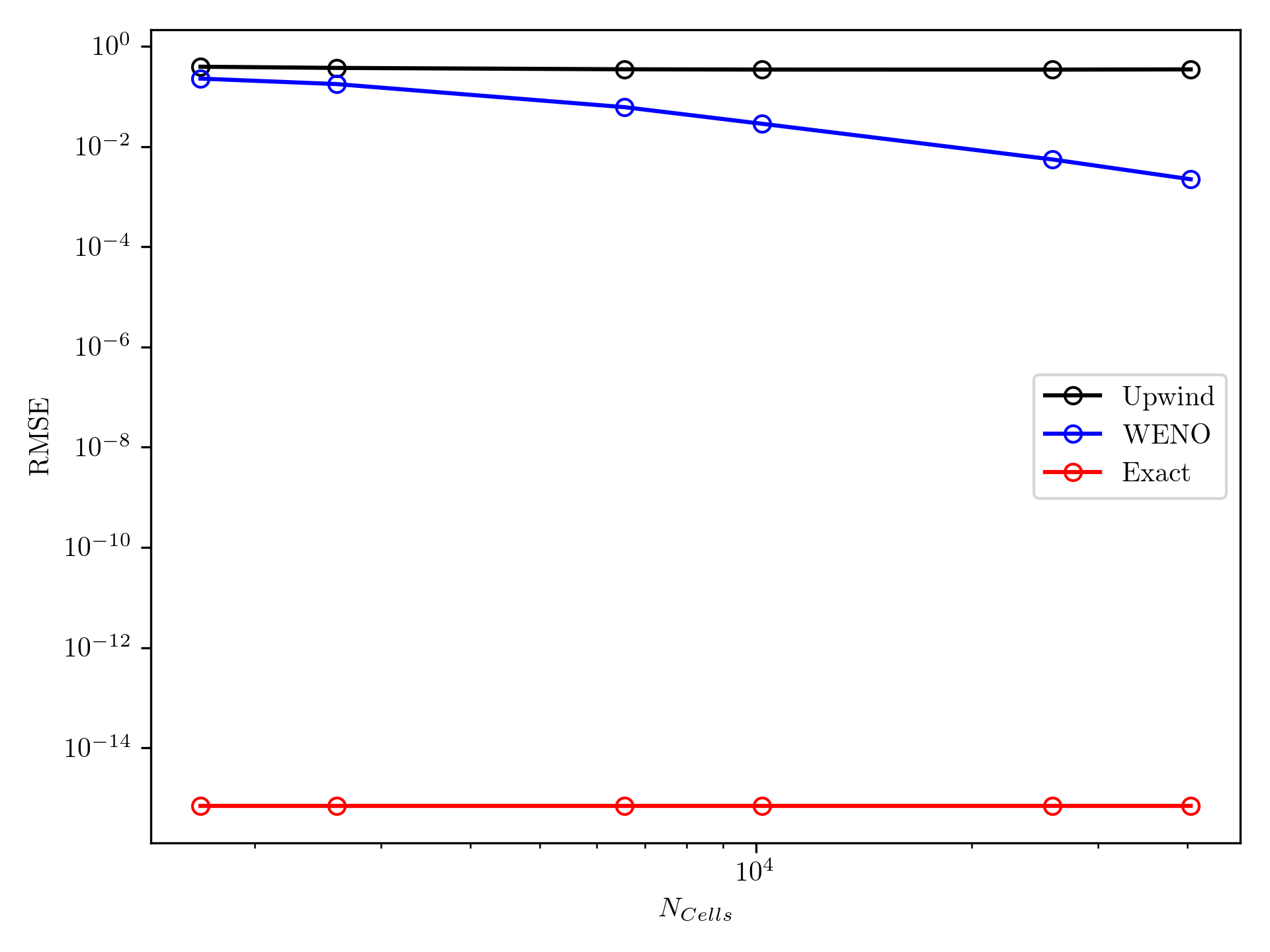}
        \caption{RMSE}
    \end{subfigure}
    \hfill
    \begin{subfigure}{0.45\textwidth}
        \centering
        \includegraphics[width=\textwidth]{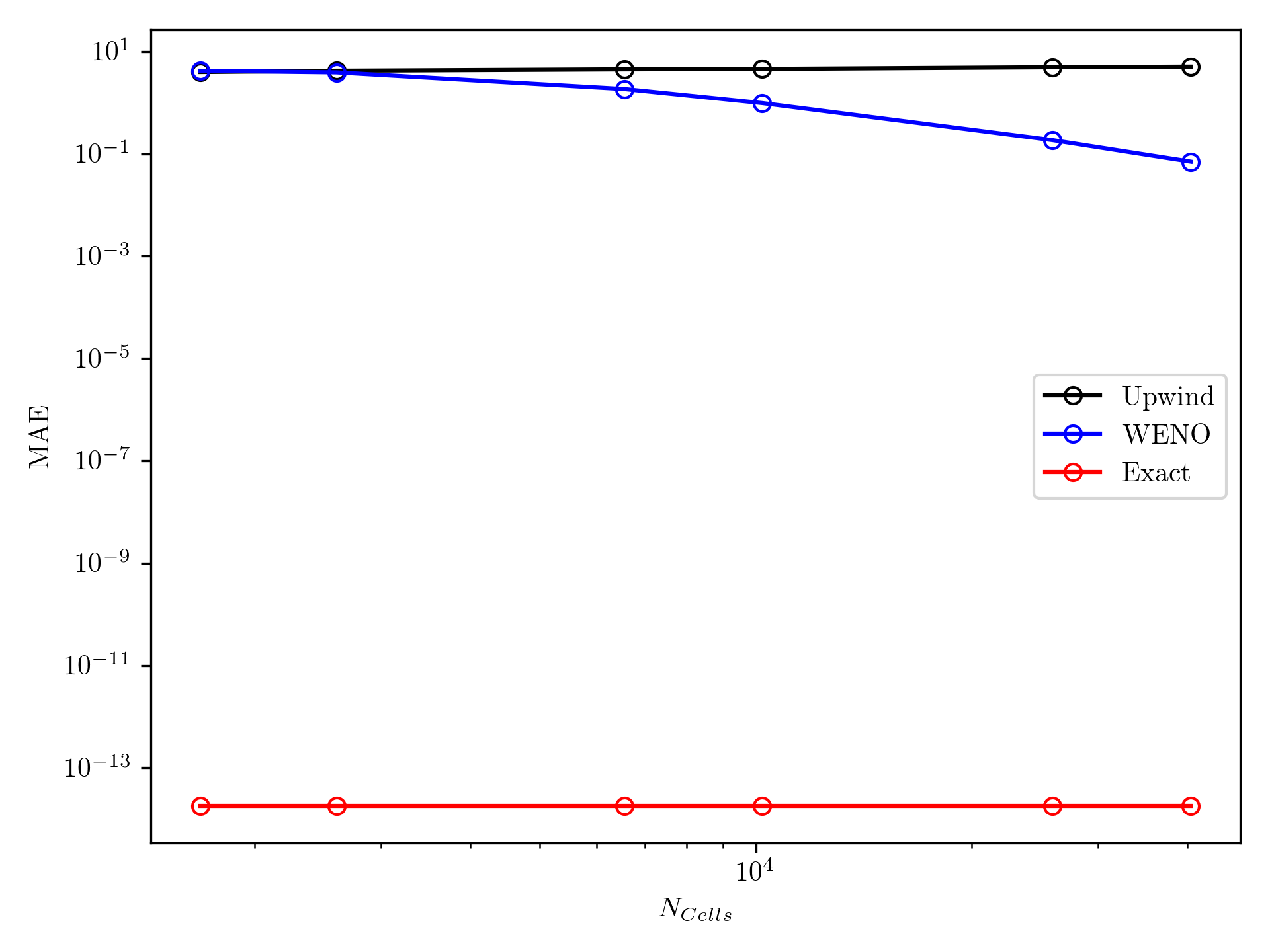}
     \caption{MAE}
    \end{subfigure}
    \caption{Error analysis for Case 5.}
    \label{fig:case5error}
\end{figure}

The simulation and error results (see Figures~\ref{fig:case4simulation} and \ref{fig:case5error}) are qualitatively similar to Cases 1 and 2, with the proposed exact scheme solving the PBM to machine precision, and the upwind and WENO schemes having much lower numerical accuracy.

\begin{figure}[htbp]
    \centering
    \begin{subfigure}{0.45\textwidth}
        \centering
        \includegraphics[width=0.9\textwidth]{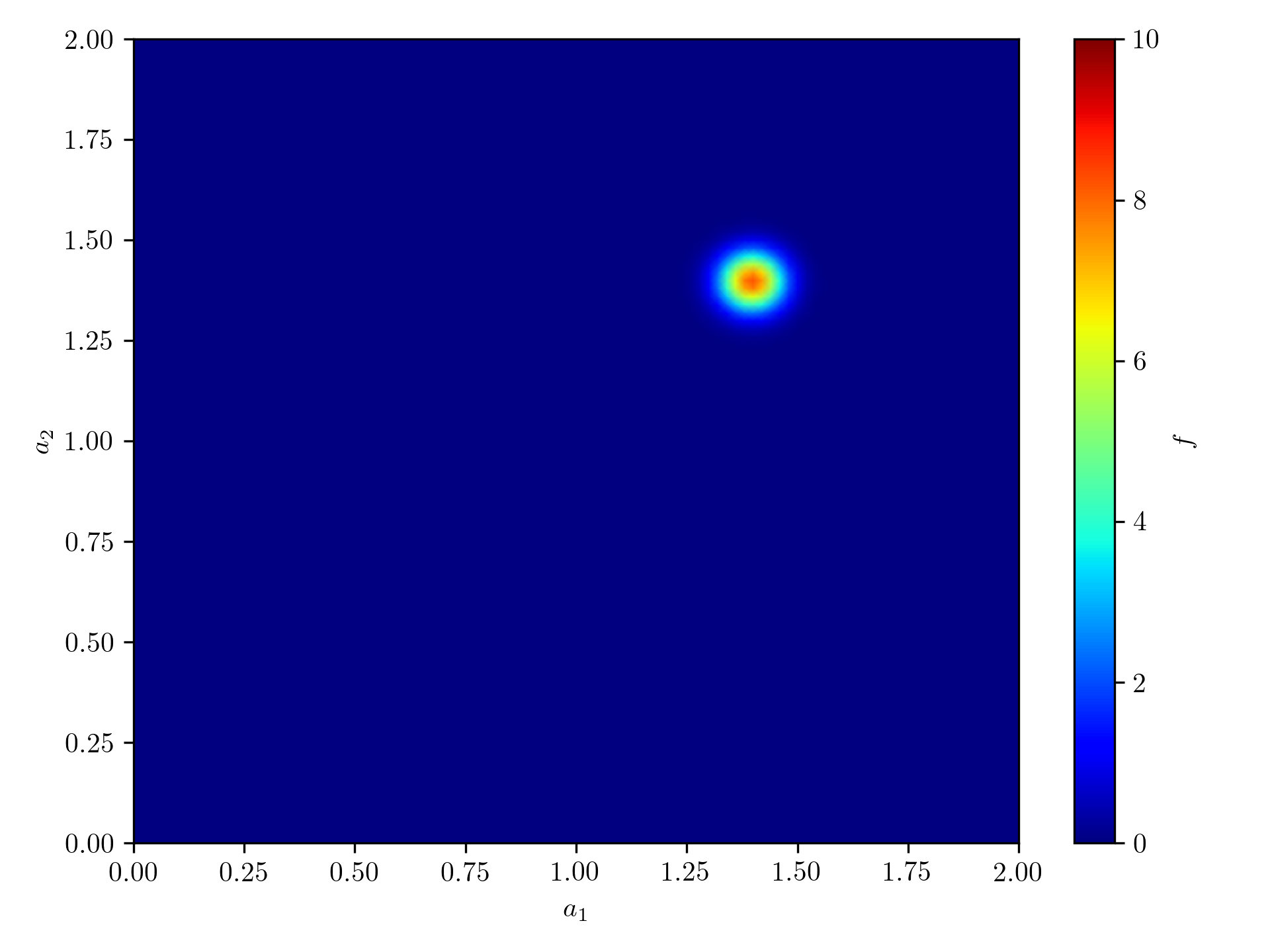}\vspace{-0.2cm}
        \caption{Analytical Solution}
    \end{subfigure}
    \hfill
    \begin{subfigure}{0.45\textwidth}
        \centering
        \includegraphics[width=0.9\textwidth]{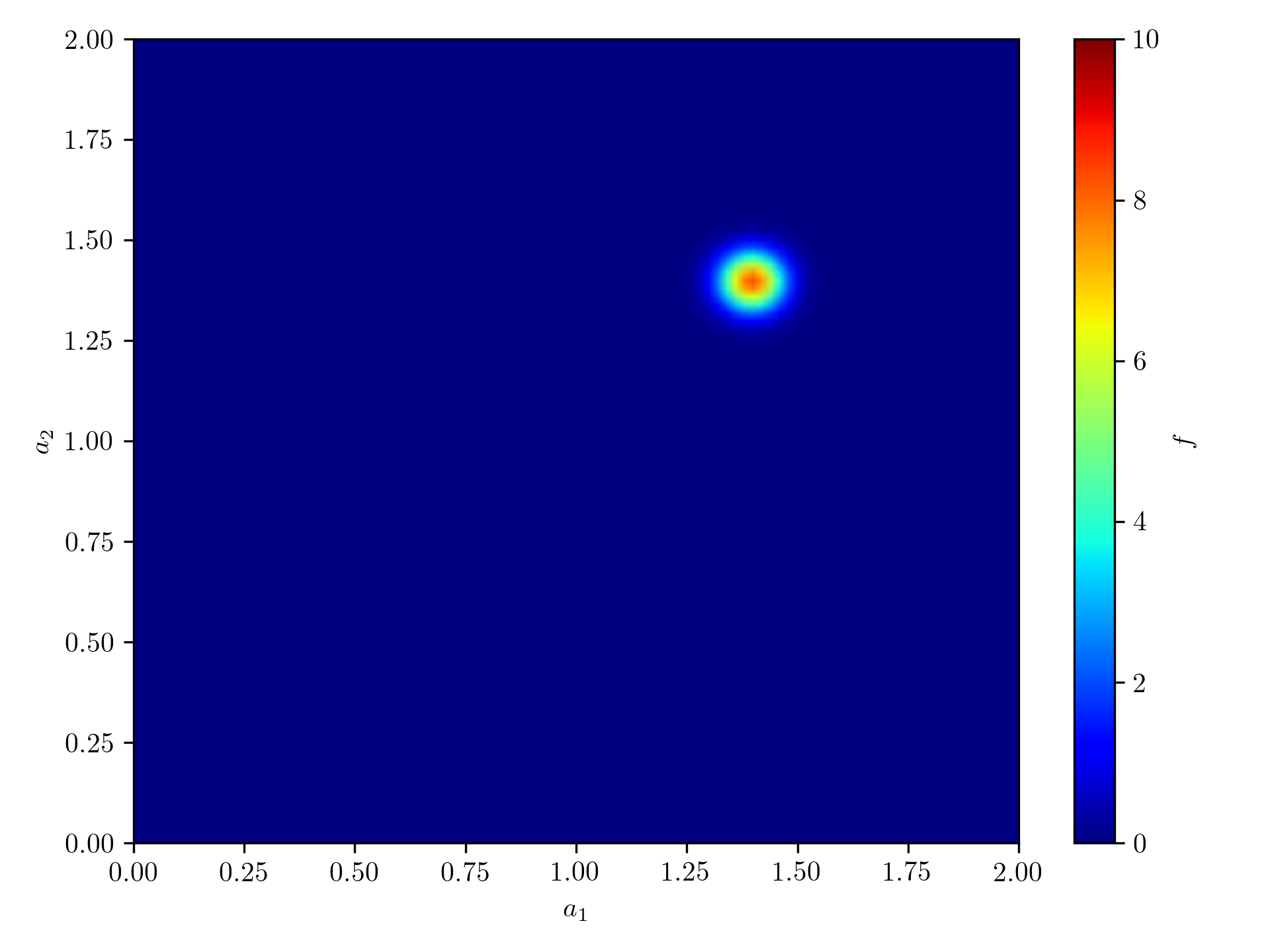}\vspace{-0.2cm}
     \caption{Exact Scheme}
    \end{subfigure}
    \hfill
    \begin{subfigure}{0.45\textwidth}
        \centering
        \includegraphics[width=0.9\textwidth]{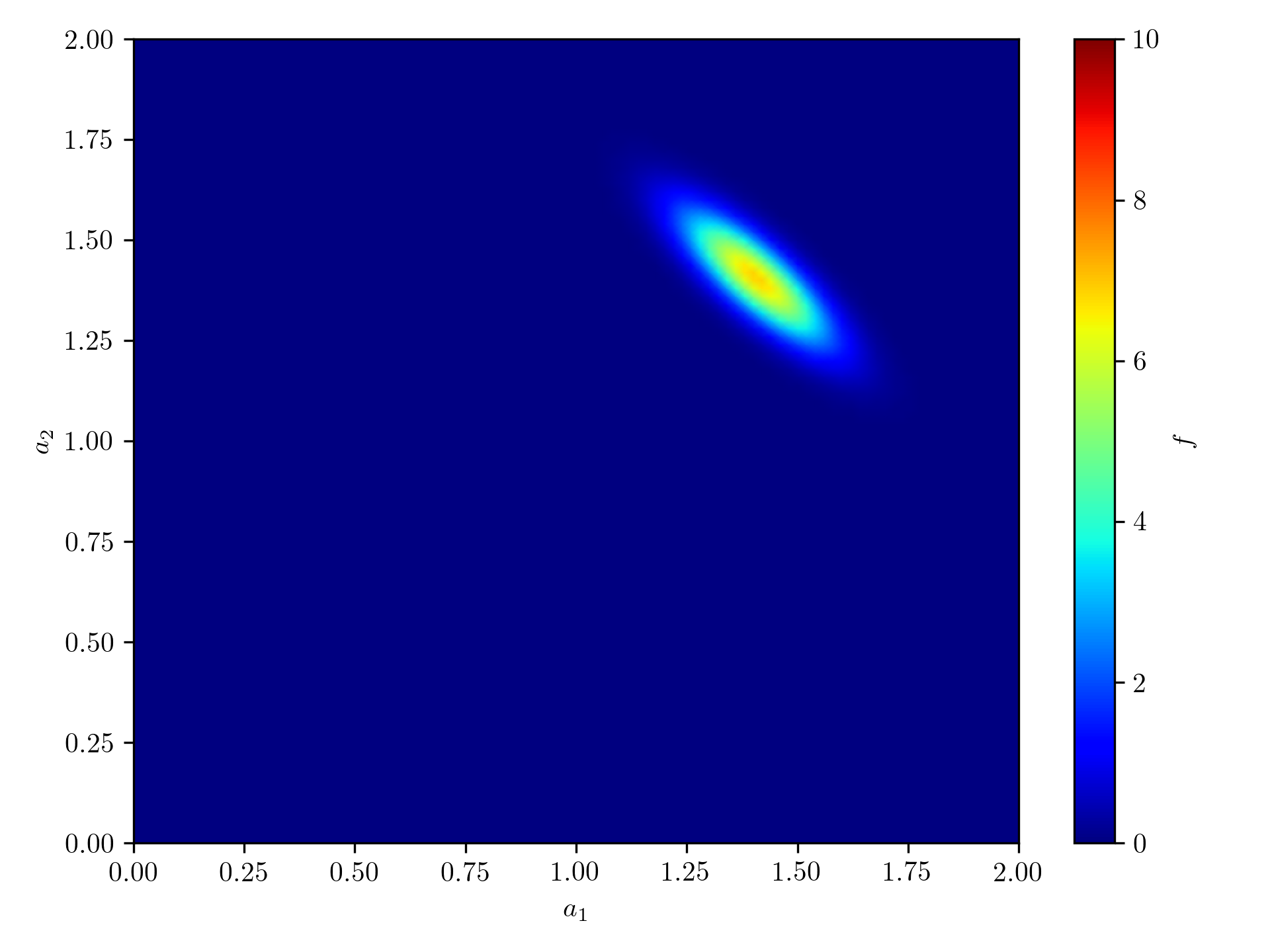}\vspace{-0.2cm}
     \caption{Upwind Scheme}
    \end{subfigure}
    \hfill
    \begin{subfigure}{0.45\textwidth}
        \centering
        \includegraphics[width=0.9\textwidth]{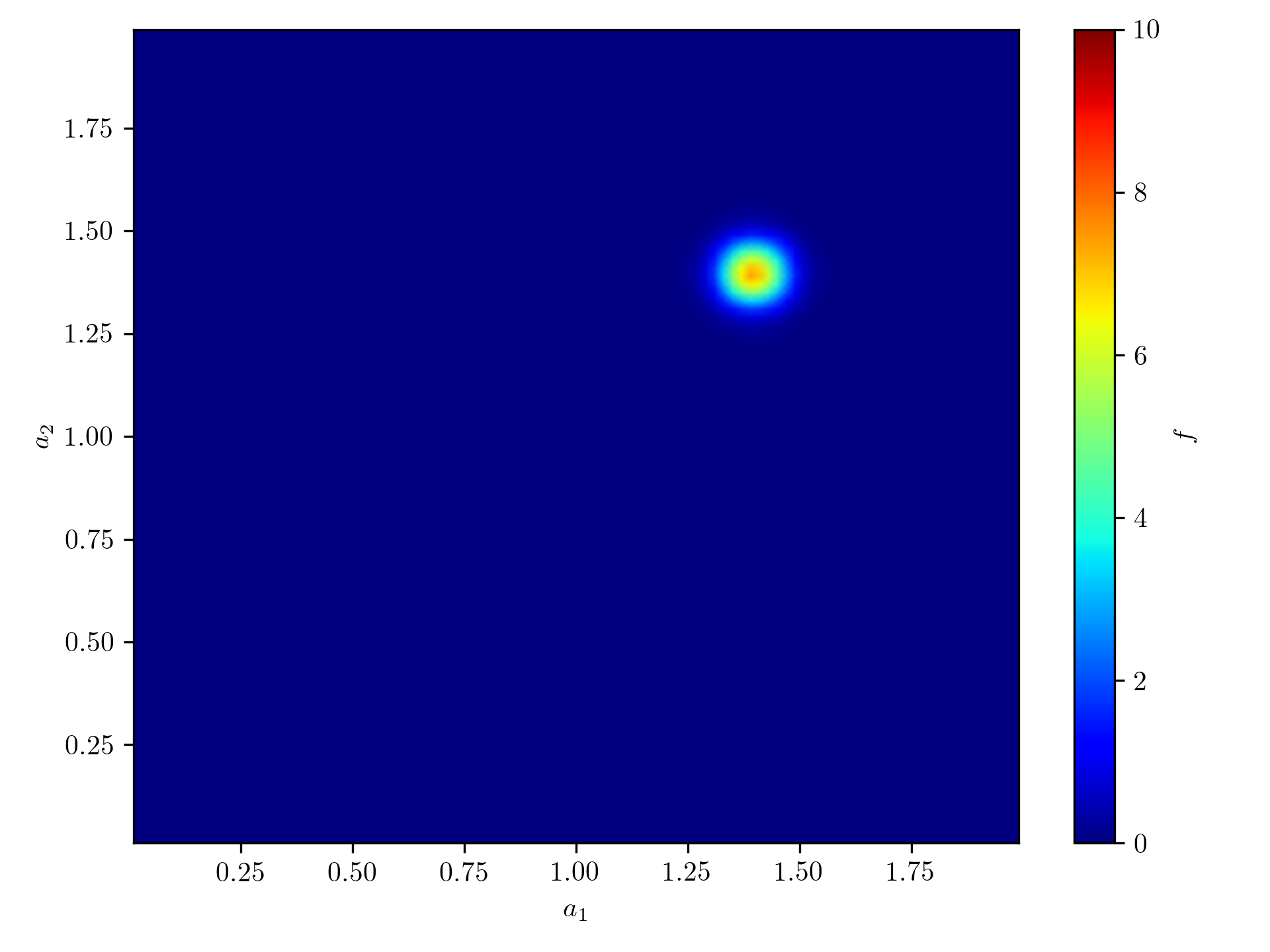}\vspace{-0.2cm}
     \caption{WENO Scheme}
    \end{subfigure}
    \caption{Simulation results for Case 5 at $t=1.0$ using the various schemes. 101 grid points are used in both the $a_{1}$ and $a_{2}$ directions for the Upwind and Exact schemes while 100 cells in both the $a_{1}$ and $a_{2}$ directions for the WENO scheme. The ability of the proposed scheme to solve the PBM to machine precision is demonstrated.}
    \label{fig:case5simulation}
\end{figure}

\section{Conclusions}

With a combination of dimensional splitting, variable transformations, and operating at the limit of numerical stability, the upwind finite difference scheme is able to solve many classes of multidimensional PBMs either to machine precision or with sufficiently high accuracy for most applications. One of the most significant features of the proposed numerical scheme is its high computational efficiency, requiring only memory reallocation or, in some instances, a minimal number of floating point operations and function calls during the time-stepping. Even in Case 3 in Section~\ref{sec:results}, where the accuracy of all of the tested numerical schemes is comparatively poor, the proposed scheme can be more efficient due to the larger time-steps made possible by the scheme effectively transforming each sub-problem into a function call. The low computational cost enables the direct incorporation of the multidimensional population balance model into on-line optimization-based control, aka model predictive control. 

\section{Availability of Code}

A Python implementation of the various schemes discussed in this work is available at \url{https://github.com/pavaninguva/PBM_Schemes}. 

\section*{Acknowledgments}
Financial support is acknowledged from the U.S. Food and Drug Administration (75F40121C00090) and the Agency for Science, Technology and Research (A*STAR), Singapore.

\printbibliography

\end{document}


\maketitle

\section{PBMs with Growth Rate $G_{i} = G_{i}(\mathbf{a})$}

Consider the PBM,
\begin{equation}
    \frac{\partial f}{\partial t} + \frac{\partial (G_{1}(a_{1},a_{2})f)}{\partial a_{1}} + \frac{\partial (G_{2}(a_{1},a_{2})f)}{\partial a_{2}} = 0.
    \label{eq:size2}
\end{equation}
Applying first-order splitting as discussed in the manuscript to \eqref{eq:size2} gives
\begin{align}
    &\frac{\partial f^{*}}{\partial t} + \frac{\partial (G_{1}(a_{1},a_{2})f^{*})}{\partial a_{1}} = 0, \quad f^{*}(t,a_{1}, a_{2}) = f(t,a_{1},a_{2}), \nonumber \\
    &\frac{\partial f^{**}}{\partial t} + \frac{\partial (G_{2}(a_{1},a_{2})f^{**})}{\partial a_{2}} = 0, \quad f^{**}(t,a_{1}, a_{2}) = f^{*}(t+\Delta t,a_{1},a_{2}), \nonumber \\
    &f(t+\Delta t, a_{1},a_{2}) = f^{**}(t+\Delta t, a_{1},a_{2}).
    \label{eq:size2_split}
\end{align}
Applying the variable transformation $\hat{f} = G_{i}(a_{1},a_{2})f$ for each sub-problem in \eqref{eq:size2_split} transforms \eqref{eq:size2_split} into
\begin{align}
    &\frac{\partial \hat{f}^{*}}{\partial t} + G_{1}(a_{1},a_{2})\frac{\partial \hat{f}^{*}}{\partial a_{1}} = 0, \quad \hat{f}^{*}(t,a_{1}, a_{2}) = G_{1}(a_{1},a_{2})f(t,a_{1},a_{2}), \nonumber \\
    &\frac{\partial \hat{f}^{**}}{\partial t} + G_{2}(a_{1},a_{2})\frac{\partial \hat{f}^{**}}{\partial a_{2}} = 0, \quad \hat{f}^{**}(t,a_{1}, a_{2}) = \hat{f}^{*}(t+\Delta t,a_{1},a_{2})\frac{G_{2}(a_{1},a_{2})}{G_{1}(a_{1},a_{2})}, \nonumber \\
    &f(t+\Delta t, a_{1},a_{2}) = \frac{\hat{f}^{**}(t+\Delta t, a_{1},a_{2})}{G_{2}(a_{1},a_{2})}.
    \label{eq:size2_split_trans}
\end{align}
To implement the enhancement to the ``Split-Exact" scheme discussed in the main text, it is important to first recognize that each sub-problem in \eqref{eq:size2_split_trans} can be solved analytically using the method of characteristics. Consider the first sub-problem in \eqref{eq:size2_split_trans},
\begin{equation}
    \frac{\partial \hat{f}^{*}}{\partial t} + G_{1}(a_{1},a_{2})\frac{\partial \hat{f}^{*}}{\partial a_{1}} = 0, \quad \hat{f}^{*}(t,a_{1}, a_{2}) = G_{1}(a_{1},a_{2})f(t,a_{1},a_{2}).
    \label{eq:size2_split_trans_1}
\end{equation}
Equation~\ref{eq:size2_split_trans_1} has the analytical solution,
\begin{equation}
    \hat{f}^{*}(t+\Delta t, a_{1},a_{2}) = \hat{f}^{*}(t, a_{1,0},a_{2}),
\end{equation}
where $a_{1,0}$ is evaluated by computing the integral,
\begin{equation}
    \int_{a_{1,0}}^{a_{1}} \frac{1}{G_{1}(a_{1},a_{2}}\partial a_{1} = t.
\end{equation}
In the example considered, $G_{1}(a_{1},a_{2}) = 0.25 + 0.5a_{1} + 0.5a_{2}$ which gives
\begin{equation}
    a_{1,0} = e^{-t/2}(0.5+ a_{1} + a_{2}) - 0.5 - a_{2}.
\end{equation}
Another way of conceptualizing the mesh construction step for the ``Split-Exact" scheme is that the scheme numerically computes this transformation which is reflected on the mesh. Hence, by evaluating this integral offline prior to the simulation, it is possible to specify a single mesh such as a uniformly spaced mesh on which the solution for both sub-problems can be evaluated on as opposed to requiring one mesh for each sub-problem. This not only eliminates the initial step of computing the mesh (either analytically which has no error penalty or by using quadrature which imposes a minor error penalty as shown in the second case study), it also eliminates the need to interpolate the solution between both meshes at each time-step. For 201 grid points in both the $a_{1}$ and $a_{2}$ directions using a $\Delta t = 0.05$, the ``Split-Exact" scheme took $\sim$56s while this enhanced scheme took $\sim$40s. These times are provided to illustrate the difference in relative CPU time and should not be taken as a metric for the absolute performance of the schemes as the schemes are implemented in Python (which can be relatively slow) and the code is not optimized for speed. 

\begin{figure}[htbp]
    \centering
    \begin{subfigure}{0.45\textwidth}
        \centering
        \includegraphics[width=\textwidth]{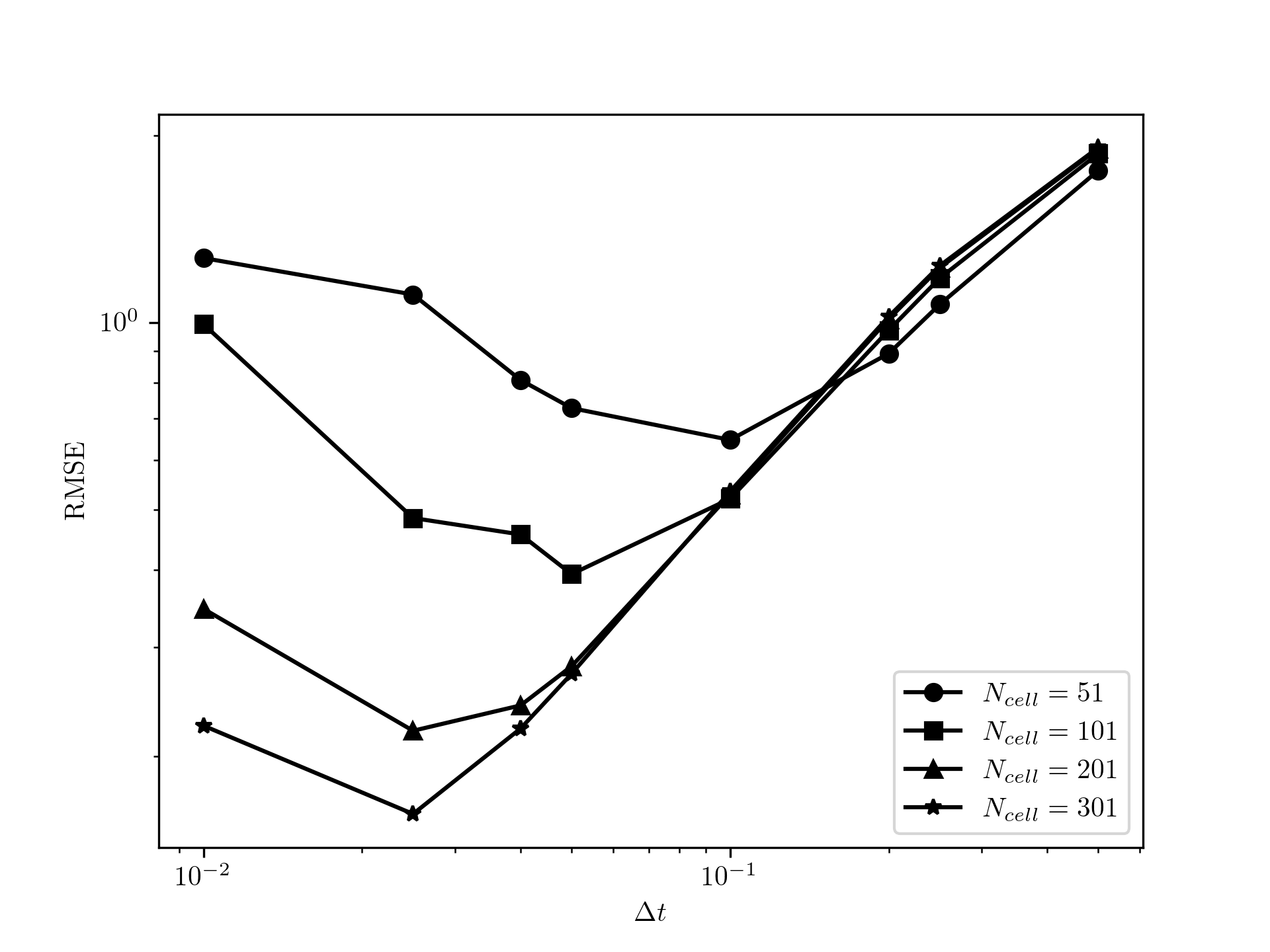}
        \caption{RMSE}
    \end{subfigure}
    \hfill
    \begin{subfigure}{0.45\textwidth}
        \centering
        \includegraphics[width=\textwidth]{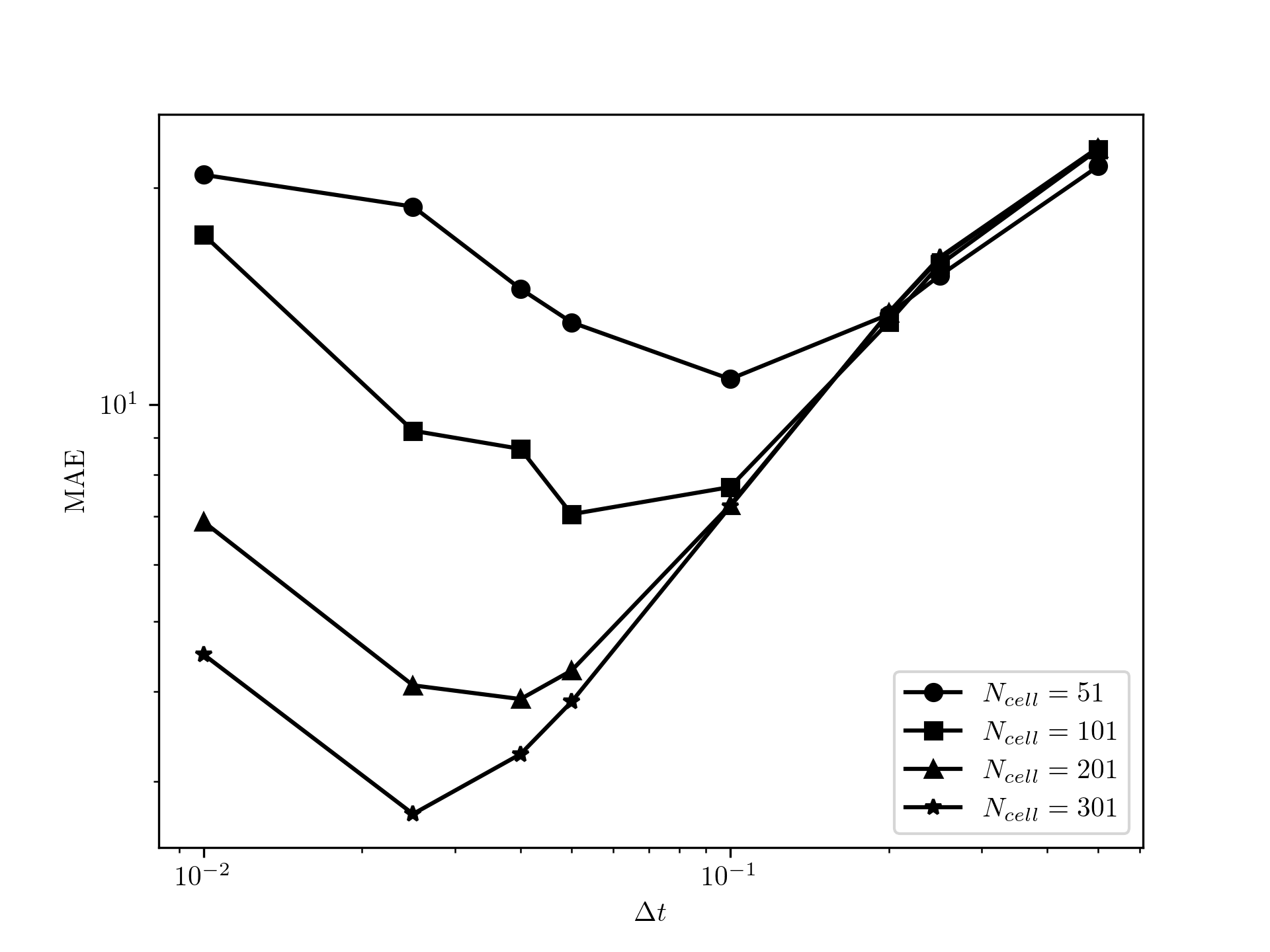}
     \caption{MAE}
    \end{subfigure}
    \caption{Error analysis for case 3 using the enhanced ``Split-Exact" scheme.}
    \label{fig:case3_scratch_error}
\end{figure}

\begin{figure}[htbp]
    \centering
    \begin{subfigure}{0.45\textwidth}
        \centering
        \includegraphics[width=\textwidth]{Figures/Case3/case3_analytical.png}
        \caption{Analytical Solution}
    \end{subfigure}
    \hfill
    \begin{subfigure}{0.45\textwidth}
        \centering
        \includegraphics[width=\textwidth]{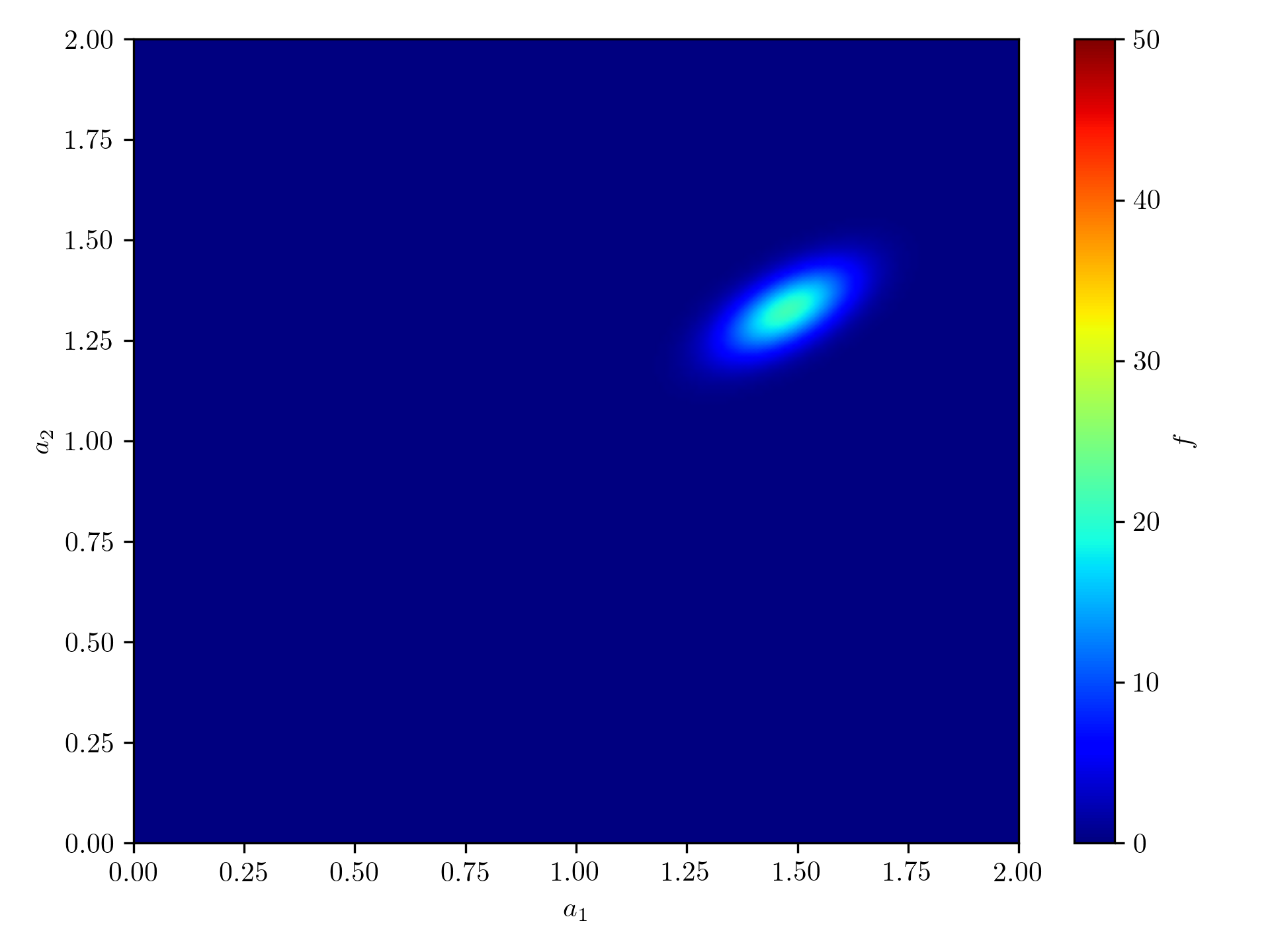}
     \caption{Enhanced ``Split-Exact"}
    \end{subfigure}
    \caption{Simulation results using the enhanced ``Split-Exact" scheme with 201 grid points in both $a_{1}$ and $a_{2}$ directions and $\Delta t = 0.05$.}
    \label{fig:case3_scratch}
\end{figure}
